\documentclass[aps,prb,reprint,showpacs,superscriptaddress,nofootinbib,floatfix]{revtex4-1}
\usepackage{color}
\usepackage{amsmath}
\usepackage{amssymb}
\usepackage{graphicx}

\begin{document}

\title{The Generalized Langevin Equation: An efficient approach to non-equilibrium
molecular dynamics of open systems}

\author{L. Stella}
\email{l.stella@qub.ac.uk}
\affiliation{Atomistic Simulation Centre, School of Mathematics and Physics, Queen's
University Belfast, University Road, Belfast BT7 1NN, Northern Ireland,
UK}

\author{C.D. Lorenz}
\affiliation{Department of Physics, School of Natural and Mathematical Sciences,
King's College London, the Strand, London WC2R 2LS, UK}

\author{L. Kantorovich}
\affiliation{Department of Physics, School of Natural and Mathematical Sciences,
King's College London, the Strand, London WC2R 2LS, UK}

\begin{abstract}
The Generalized Langevin Equation (GLE) has been recently suggested
to simulate the time evolution of classical solid and molecular systems
when considering general non-equilibrium processes. In this approach,
a part of the whole system (an open system), which interacts and exchanges
energy with its dissipative environment, is studied. Because the GLE
is derived by projecting out exactly the harmonic environment, the
coupling to it is realistic, while the equations of motion are non-Markovian.
Although the GLE formalism has already found promising applications,
e.g., in nanotribology and as a powerful thermostat for equilibration
in classical molecular dynamics simulations, efficient algorithms
to solve the GLE for realistic memory kernels are highly non-trivial,
especially if the memory kernels decay non-exponentially. This is
due to the fact that one has to generate a colored noise and take
account of the memory effects in a consistent manner. In this contribution,
we present a simple, yet efficient, algorithm for solving the GLE
for practical memory kernels and we demonstrate its capability for
the exactly solvable case of a harmonic oscillator coupled to a Debye
bath.
\end{abstract}

\pacs{05.10.Gg, 05.70.Ln, 02.70.−c, 63.70.+h}

\maketitle

\section{Introduction\label{sec:intro}}

Nanoscale devices and materials are becoming increasingly important
in the development of novel technologies. In many of the application
areas of these new nanotechnologies, the materials and devices are
part of a driven system in which understanding their non-equilibrium
properties is of utmost importance. Of particular interest in many
applications is understanding the thermal conductivity of materials
(i.e., molecular junctions,
\cite{Segal2002,Dubi2011}
nanotubes,
\cite{Berber2000,Kim2001,Shi2002,Padgett2004,Hu2008}
nanorods,
\cite{Padgett2006}
nanowires,
\cite{Yang2008}
semiconductors
\cite{Estreicher2009})
and the heat transport within nanodevices.
\cite{Cahill2002,Pop2010,Zebarjadi2012}
Other applications in which the non-equilibrium properties of materials
are of interest include: (a) the bulk energy dissipation in crystals
due to an excited point defect
\cite{West2006}
or crack propagation;
\cite{Kermode2008}
(b) interfacial chemical reactions between adsorbed molecules and
the surface that generate excess energy which is dissipated into the
surface;
\cite{Backus2005,Ueba2005}
(c) surfaces interacting with
energetic lasers,
\cite{Mak1987}
atomic/ionic
\cite{Chan2007,Toussaint2009}
or molecular beams
\cite{Wucher2010}
when substantial energy is released
along the particles trajectory into the surface; (d) in tribology,
where two surfaces shear upon each other with bonds between them forming
and breaking that results in consuming and releasing a considerable
amounts of energy; 
\cite{Szlufarska2008,Benassi10,Benassi12} 
and (e) molecules which
are driven by a heat gradient.
\cite{Lohrasebi2011}

Over the years molecular dynamics (MD) simulations have proven to
be a powerful and yet simple tool for investigating the vibrational
energy dissipation of atoms. There are several thermostats that can
be used in MD simulations, which have been described in great detail
in
\cite{Toton2010} 
to sample a canonical distribution of the system
at a given volume and temperature: Andersen, 
\cite{Andersen1980}
Nos\'{e},
\cite{Nose1984,Nose1984a} 
Hoover,
\cite{Hoover1985}
Langevin
\cite{Schneider1978} 
and other stochastic thermostats.
\cite{Bussi2007}
However, these methods can only enable the modeling systems of interest
in thermodynamic equilibrium corresponding to the given volume, temperature
and number of particles. 

At the same time, these equilibrium thermostats are increasingly being
applied to simulations studying non-equilibrium processes including
tribology,
\cite{Szlufarska2007,Barry2009} 
energy dissipation,
\cite{Trevethan2004}
crack propagation,
\cite{Kermode2008} 
heat transport
\cite{Mazyar2006,Hu2008,Hu2009,Guo2010,Hu2011,Manikandan2011}
and irradiation.
\cite{Hsu2007} 
In some instances,
\cite{Szlufarska2007,Kermode2008,Hsu2007}
the equilibrium thermostats are applied to all atoms of the system
in order to impose a specific temperature, while in other studies,
\cite{Trevethan2004,Mazyar2006,Hu2008,Barry2009,Hu2009,Guo2010,Hu2011,Manikandan2011}
the Nos\'{e},
\cite{Nose1984},
Hoover 
\cite{Hoover1985}
or Berendsen
\cite{Berendsen1984}
thermostats were used to thermostat only certain regions of the systems,
although, strictly speaking, they were only proven to work if applied
to the whole system (and additionally the Berendsen thermostat is
not truly canonical). When these equilibrium thermostats are applied
to non-equilibrium MD simulations, they introduce artifacts into the
resulting trajectories in these simulations. For example, in nanotribology
MD simulations, the most commonly used method to thermostat the system
is by applying a Langevin thermostat only in the direction perpendicular
to the shear plane of the system,
\cite{Thompson1990}
but this method
has limitations at high shear rates
\cite{Thompson1990}
which are
required to study friction of low viscosity fluids.
\cite{Lorenz2010}

An exact and elegant solution to this problem is provided by the Generalized
Langevin Equation (GLE). 
\cite{Zwanzig2001}
Under rather general
assumptions concerning the classical Hamiltonian of the open system
and its interaction with its surroundings, assumed to be harmonic,
one arrives at non-Markovian dynamics of the open system with multivariate
Gaussian distributed random force and the memory kernel that is shown
to be exactly proportional to the random force autocorrelation function.
\cite{Kantorovich2008a}
Although the GLE has been around for a while (see Ref.~\onlinecite{Kantorovich2008a}
and references therein), its application to interesting simulated
systems has only recently become realized. Ceriotti and collaborators
\cite{Ceriotti2009a,Ceriotti2011,Morrone2011} 
have utilized the GLE
approach to develop an efficient equilibrium thermostat for improving
the convergence during the advanced sampling of the degrees of freedom
(DoFs) within a system. Others have used a similar approach to generate
quantum heat baths that can be utilized in MD simulations of
both equilibrium 
\cite{Dammak2009,Barrat2011}
and out-of-equilibrium systems.
\cite{Lu12,Biele13}

In this manuscript, we present a very efficient algorithm which enables
one to solve the GLE numerically taking into account both of its fundamental
features, namely its non-Markovian character and the colored noise.
Moreover, the proposed algorithm allows one to solve a realistic GLE
with the noise and the memory kernel entering the memory term, which
can be calculated from a realistic Hamiltonian of the entire system
consisting of both the open system and the environment. The aim in
developing this method is so that we will be able to apply it to MD
simulations of driven systems that are out-of-equilibrium and therefore
provide a fundamentally sound non-equilibrium thermostat.

The remainder of the paper will present in Section \ref{sec:GLE}
the underlying mathematical development of the GLE equations and the
algorithm itself, while the example of a harmonic oscillator coupled
to a harmonic bath on which we have tested the algorithm is given
in Section \ref{sec:harmonic}. Finally, conclusions are presented
in Section \ref{sec:conclusions}.

\section{GLE for solids\label{sec:GLE}}

Let us start by considering a solid divided into two regions: \emph{the
open system} --- hereafter referred to simply as the \emph{system}
--- consisting of a finite, possibly small, portion of the solid,
and the rest of the solid --- hereafter the \emph{bath} --- which
is assumed to be large enough to be faithfully described in terms
of its thermodynamic properties, e.g. its temperature, $T$.

\subsection{Equations of motion for a system coupled to the bath\label{sub:eoms}}

Let us consider a system-bath interaction modeled by the classical
Lagrangian $\mathcal{L}\equiv\mathcal{L}_{{\rm sys}}+\mathcal{L}_{{\rm bath}}+\mathcal{L}_{{\rm int}}$,
where
\begin{equation}
\mathcal{L}_{sys}\left(\mathbf{r},\dot{\mathbf{r}}\right)=\sum_{i\alpha}\frac{1}{2}m_{i}\dot{r}_{i\alpha}^{2}-V\left({\bf r}\right)
\label{eq:lagrangian-1}
\end{equation}
\begin{multline}
\mathcal{L}_{bath}\left(\mathbf{u},\dot{\mathbf{u}}\right)=\sum_{l\gamma}\frac{1}{2}\mu_{l}\dot{u}_{l\gamma}^{2}\\
-\frac{1}{2}\sum_{l\gamma}\sum_{l^{\prime}\gamma^{\prime}}\sqrt{\mu_{l}\mu_{l^{\prime}}}u_{l\gamma}D_{l\gamma,l^{\prime}\gamma^{\prime}}u_{l^{\prime}\gamma^{\prime}}
\label{eq:lagrangian-2}
\end{multline}
\begin{equation}
\mathcal{L}_{int}\left(\mathbf{r},\mathbf{u}\right)=-\sum_{l\gamma}\mu_{l}f_{l\gamma}\left(\mathbf{r}\right)u_{l\gamma}
\label{eq:lagrangian-3}
\end{equation}
Here index $i=1,\dots,N$ labels the system atoms, their masses being
$m_{i}$. The positions of the system atoms are given by vectors $\mathbf{r}_{i}=\left(r_{i\alpha}\right)$
with the Greek index $\alpha$ indicating the appropriate Cartesian
components, i.e., $r_{i\alpha}$ gives the Cartesian component $\alpha$
of the position of atom $i$. $\mathcal{L}_{sys}$ is the Lagrangian
of the system with potential energy $V(\mathbf{r})$, and the vector
$\mathbf{r}$ collects the Cartesian components of all the positions
of the system atoms. Similarly, the vector $\dot{\mathbf{r}}$ collects
the velocities of all the system atoms. The Lagrangian $\mathcal{L}_{bath}$
describes a harmonic bath and the index $l=1\dots,L$ labels the bath
atoms, their masses being $\mu_{l}$. The displacements of the bath
atoms from their equilibrium positions are given by vectors $\mathbf{u}_{l}=\left(u_{l\gamma}\right)$,
with the Greek index $\gamma$ indicating the appropriate Cartesian
components. The vector $\mathbf{u}$ collects the Cartesian components
of all the displacements of the bath atoms. Similarly, the vector
$\dot{\mathbf{u}}$ collects all the velocities of the bath atoms.
As the bath is described in the harmonic approximation, the potential
energy of the bath is quadratic in the atomic displacements, the matrix
$\mathbf{D}=\left(D_{l\gamma,l^{\prime}\gamma^{\prime}}\right)$ being
the dynamic matrix of the bath. The system-bath interaction defined
in $\mathcal{L}_{{\rm int}}$ has been chosen to be linear in $\mathbf{u}$
in order to have $\mathcal{L}_{{\rm bath}}+\mathcal{L}_{{\rm int}}$
harmonic in the bath DoFs. Note, however, that the dependence of the
interaction term on the system DoFs (via $f_{l\gamma}\left(\mathbf{r}\right)$)
remains arbitrary.

From the Lagrangian, Eqs. (\ref{eq:lagrangian-1}) - (\ref{eq:lagrangian-3}),
the following equations of motion (EoMs) for the system and bath DoFs
are derived:
\begin{equation}
m_{i}\ddot{r}_{i\alpha}=-\frac{\partial V\left(\mathbf{r}\right)}{\partial r_{i\alpha}}-\sum_{l\gamma}\mu_{l}g_{i\alpha,l\gamma}\left(\mathbf{r}\right)u_{l\gamma}
\label{eq:sys_eoms}
\end{equation}
\begin{equation}
\mu_{l}\ddot{u}_{l\gamma}=-\sum_{l^{\prime}\gamma^{\prime}}\sqrt{\mu_{l}\mu_{l^{\prime}}}D_{l\gamma,l^{\prime}\gamma^{\prime}}u_{l^{\prime}\gamma^{\prime}}-\mu_{l}f_{l\gamma}\left(\mathbf{r}\right)
\label{eq:bath_eom}
\end{equation}
where $g_{i\alpha,l\gamma}\left(\mathbf{r}\right)=\partial f{}_{l\gamma}\left({\bf r}\right)/\partial r_{i\alpha}$.
Eqs. (\ref{eq:bath_eom}) can be solved analytically
\cite{Kantorovich2008a}
to give 
\begin{multline}
u_{l\gamma}\left(t;\mathbf{r}\right)=\sum_{l^{\prime}\gamma^{\prime}}\sqrt{\frac{\mu_{l^{\prime}}}{\mu_{l}}}\left[\dot{\Omega}_{l\gamma,l^{\prime}\gamma^{\prime}}\left(t\right)u_{l^{\prime}\gamma^{\prime}}\left(-\infty\right)\right.\\
\left. +\Omega_{l\gamma,l^{\prime}\gamma^{\prime}}\left(t\right)\dot{u}_{l^{\prime}\gamma^{\prime}}\left(-\infty\right)\right.\\
\left.-\int_{-\infty}^{t}\Omega_{l\gamma,l^{\prime}\gamma^{\prime}}\left(t-t^{\prime}\right)f_{l^{\prime}\gamma^{\prime}}\left(\mathbf{r}\left(t^{\prime}\right)\right)\mbox{d}t^{\prime}\right]
\label{eq:bath_evolution}
\end{multline}
where $u_{l\gamma}\left(-\infty\right)$ and $\dot{u}_{l\gamma}\left(-\infty\right)$
are the initial positions and velocities of the bath atoms, which,
at variance with Ref.~\onlinecite{Kantorovich2008a}, 
are set at $t\to-\infty$
for numerical convenience (see Sec. \ref{sub:integration}). In Eq.
(\ref{eq:bath_evolution}) we have made use of the resolvent 
\begin{equation}
\Omega_{l\gamma,l^{\prime}\gamma^{\prime}}\left(t-t^{\prime}\right)=\sum_{\lambda}\frac{v_{l\gamma}^{\left(\lambda\right)}v_{l^{\prime}\gamma^{\prime}}^{\left(\lambda\right)}}{\omega_{\lambda}}\sin\left(\omega_{\lambda}\left(t-t^{\prime}\right)\right)
\label{eq:resolvent}
\end{equation}
where the bath normal modes $\mathbf{v}^{\left(\lambda\right)}=\left(v_{l\gamma}^{(\lambda)}\right)$
and frequencies $\omega_{\lambda}$ are defined via the usual vibration
eigenproblem: 
\begin{equation}
\sum_{l^{\prime}\gamma^{\prime}}D_{l\gamma,l^{\prime}\gamma^{\prime}}v_{l^{\prime}\gamma^{\prime}}^{\left(\lambda\right)}=\omega_{\lambda}^{2}v_{l\gamma}^{\left(\lambda\right)}
\label{eq:vibr-problem}
\end{equation}

By first substituting Eq. (\ref{eq:bath_evolution}) into Eq. (\ref{eq:sys_eoms})
and then performing an integration by parts
\cite{Kantorovich2008a},
the following EOMs for the system are found
\begin{multline}
m_{i}\ddot{r}_{i\alpha}=-\frac{\partial\bar{V}\left(\mathbf{r}\right)}{\partial r_{i\alpha}}\\
-\int_{-\infty}^{t}\sum_{i^{\prime}\alpha^{\prime}}K_{i\alpha,i^{\prime}\alpha^{\prime}}\left(t,t^{\prime};\mathbf{r}\right)\dot{r}_{i^{\prime}\alpha^{\prime}}(t^{\prime})\mbox{d}t^{\prime}+{\bf \eta}_{i\alpha}\left(t;{\bf r}\right)
\label{eq:gle}
\end{multline}
There are three terms in the right hand side. The first term is a
conservative force from the effective potential energy of the system
defined as 
\begin{multline}
\bar{V}\left(\mathbf{r}\right)=V\left(\mathbf{r}\right)\\
-\frac{1}{2}\sum_{l\gamma}\sum_{l^{\prime}\gamma^{\prime}}\sqrt{\mu_{l}\mu_{l^{\prime}}}f_{l\gamma}\left(\mathbf{r}\right)\Pi_{l\gamma,l^{\prime}\gamma^{\prime}}\left(0\right)f_{l^{\prime}\gamma^{\prime}}\left(\mathbf{r}\right)
\label{eq:potential_pol}
\end{multline}
which includes a polaronic correction (the second term in Eq. (\ref{eq:potential_pol}))
as the equilibrium positions of the bath atoms are modified by the
linear system-bath interaction defined in Eq. (\ref{eq:lagrangian-3}). 
The second term in Eq. (\ref{eq:gle}) describes
the friction forces acting on the atoms in the system; this term depends
on the whole trajectory of system atoms \emph{prior} to the current
time $t$, i.e., this term explicitly contains memory effects. The
corresponding memory kernel is given by
\begin{widetext}
\begin{equation}
K_{i\alpha,i^{\prime}\alpha^{\prime}}\left(t,t^{\prime};\mathbf{r}\right)=\sum_{l\gamma}\sum_{l^{\prime}\gamma^{\prime}}\sqrt{\mu_{l}\mu_{l^{\prime}}}g_{i\alpha,l\gamma}\left(\mathbf{r}(t)\right)\Pi_{l\gamma,l^{\prime}\gamma^{\prime}}\left(t-t^{\prime}\right)g_{i^{\prime}\alpha^{\prime},l^{\prime}\gamma^{\prime}}\left(\mathbf{r}\left(t^{\prime}\right)\right)
\label{eq:kernel-ini}
\end{equation}
Finally, the last term in the right hand side of Eq. (\ref{eq:gle})
describes the stochastic (and hence non-conservative) forces given
by
\begin{equation}
\eta_{i\alpha}\left(t;\mathbf{r}\right)=-\sum_{l\gamma}\sum_{l^{\prime}\gamma^{\prime}}\sqrt{\mu_{l}\mu_{l^{\prime}}}g_{i\alpha,l\gamma}\left(\mathbf{r}\left(t\right)\right)\left(\dot{\Omega}_{l\gamma,l^{\prime}\gamma^{\prime}}\left(t\right)u_{l^{\prime}\gamma^{\prime}}\left(-\infty\right)+\Omega_{l\gamma,l^{\prime}\gamma^{\prime}}\left(t\right)\dot{u}_{l^{\prime}\gamma^{\prime}}\left(-\infty\right)\right)
\label{eq:forces_ini}
\end{equation}
\end{widetext}
Both the memory kernel and the dissipative forces are (causal) functionals
of the open system atomic trajectories, ${\bf r}\left(t\right)$.
The bath polarization matrix used in Eqs. (\ref{eq:potential_pol})
and (\ref{eq:kernel-ini}) is defined as the integral of the resolvent,
Eq. (\ref{eq:resolvent}), so that
\begin{equation}
\Pi_{l\gamma,l^{\prime}\gamma^{\prime}}\left(t-t^{\prime}\right)=\sum_{\lambda}\frac{v_{l\gamma}^{\left(\lambda\right)}v_{l^{\prime}\gamma^{\prime}}^{\left(\lambda\right)}}{\omega_{\lambda}^{2}}\cos\left(\omega_{\lambda}\left(t-t^{\prime}\right)\right)
\label{eq:pol_ini}
\end{equation}
For an infinite bath possessing a continuum phonon spectrum, the polarization
matrix decays to zero in the limit of $t-t^{\prime}\rightarrow\infty$.
Note that since $t>t^{\prime}$ in Eq. (\ref{eq:kernel-ini}), the
polarization matrix can be defined just for $t-t^{\prime}\geq0$.
To define its Fourier transform (FT) 
\begin{equation}
\Pi_{l\gamma,l^{\prime}\gamma^{\prime}}\left(\omega\right)=\int_{-\infty}^{\infty}\Pi_{l\gamma,l^{\prime}\gamma^{\prime}}\left(s\right)e^{-i\omega s}\mbox{d}s\;,
\label{eq:FT-pol}
\end{equation}
where $s=t-t^{\prime},$ it is convenient to extend the definition
of the polarization matrix also to the negative times $t-t^{\prime}<0$.
In that respect, various choices are possible. One possibility is
that the polarization matrix is defined by Eq. (\ref{eq:pol_ini})
for all times and is therefore an even function of time decaying to
zero at the $\left|t-t^{\prime}\right|\rightarrow\infty$ limit. Another
possibility is to impose the causality condition on the polarization
matrix by requiring that it is equal to zero for $t-t^{\prime}<0$,
i.e., one can introduce the\emph{ causal} polarization matrix $\widetilde{\Pi}_{l\gamma,l^{\prime}\gamma^{\prime}}\left(t-t^{\prime}\right)=\theta\left(t-t^{\prime}\right)\Pi_{l\gamma,l^{\prime}\gamma^{\prime}}\left(t-t^{\prime}\right)$,
where $\theta(t)$ is the Heaviside step function. In that case the
real and imaginary parts of the polarization matrix $\widetilde{\Pi}_{l\gamma,l^{\prime}\gamma^{\prime}}\left(\omega\right)$
satisfy the Kramers-Kronig relationships. This choice has an advantage
as the corresponding memory kernel will be also causal. Hence, the
upper limit in the time integral in the GLE, Eq. (\ref{eq:gle}),
can be extended to infinity which facilitates using the FT when required.
We shall use a tilde hereafter to indicate causal quantities. 

The polarization matrix and the memory kernel satisfy the obvious
symmetry identities:
\begin{align}
&\Pi_{l\gamma,l^{\prime}\gamma^{\prime}}\left(t-t^{\prime}\right)=\Pi_{l^{\prime}\gamma^{\prime},l\gamma}\left(t-t^{\prime}\right)
\label{eq:symmetry-pol-matrix}\\
&K_{i\alpha,i^{\prime}\alpha^{\prime}}\left(t,t^{\prime};\mathbf{r}\right)=K_{i^{\prime}\alpha^{\prime},i\alpha}\left(t^{\prime},t;\mathbf{r}\right)
\label{eq:symmetry-kernel}
\end{align}

As it follows from Eq. (\ref{eq:pol_ini}), to calculate the exact
memory kernel, the bath vibration eigenproblem, Eq. (\ref{eq:vibr-problem}),
must be solved first as the bath dynamics is encoded in its polarization
matrix, $\Pi_{l\gamma,l^{\prime}\gamma^{\prime}}\left(t-t^{\prime}\right)$;
the latter is the central factor in both the memory kernel and the
polaronic correction in Eq. (\ref{eq:potential_pol}). 

The system-bath coupling has three important effects: (i) it modifies
the equilibrium configuration of the system atoms due to the polaronic
correction in Eq. (\ref{eq:potential_pol}) (the polaronic effect); (ii)
the memory term is responsible for the system energy dissipation (i.e.,
friction) by draining energy from the system; (iii) finally, atoms
of the system experience stochastic forces (\ref{eq:forces_ini})
due to the last term Eq. (\ref{eq:gle}) which on average bring energy
into the system. The last two effects are better understood by looking
at the time derivative of the system energy:
\begin{widetext}
\begin{equation}
\frac{d}{dt}\left(\frac{1}{2}\sum_{i\alpha}m_{i}\dot{r}_{i\alpha}^{2}+\bar{V}(\mathbf{r})\right)=-\int_{-\infty}^{t}\sum_{i\alpha}\sum_{i^{\prime}\alpha^{\prime}}\dot{r}_{i\alpha}(t)K_{i\alpha,i^{\prime}\alpha^{\prime}}\left(t,t^{\prime};\mathbf{r}\right)\dot{r}_{i^{\prime}\alpha^{\prime}}\left(t^{\prime}\right)\mbox{d}t^{\prime}+\sum_{i\alpha}\dot{r}_{i\alpha}(t){\bf \eta}_{i\alpha}(t;\mathbf{r})
\label{eq:en_dt}
\end{equation}
\end{widetext}
which depends on two apparently uncorrelated contributions: the first
one describes the energy drain, while the second one describes the
work on the system atoms by the random forces.

The dissipative forces defined in Eq. (\ref{eq:forces_ini}) depend
on a large number of unknown initial positions, $u_{l\gamma}\left(-\infty\right)$,
and velocities, $\dot{u}_{l\gamma}\left(-\infty\right)$, of the bath
atoms. Given that the bath is assumed to be much larger than the system
(in fact, macroscopically large), and hence the number of bath DoFs
is infinite, it is impossible to specify all of them explicitly and
hence, a statistical approach is in order to describe the bath
\cite{Zwanzig2001}.
Assuming the bath (described by the combined Lagrangian $\mathcal{L}_{{\rm bath}}+\mathcal{L}_{{\rm int}}$)
is in thermodynamic equilibrium at temperature $T$, the stochastic
forces $\eta_{i\alpha}\left(t;\mathbf{r}\right)$ can be treated as
random variables. Indeed, it has been demonstrated in Ref.~\onlinecite{Kantorovich2008a}
that from this assumption the dissipative forces are well described
by a multi-dimensional Gaussian stochastic process with correlation
functions
\begin{align}
&\left\langle \eta_{i\alpha}\left(t;{\bf r}\right)\right\rangle =0
\label{eq:noise-av-1}\\
&\left\langle \eta_{i\alpha}\left(t;{\bf r}\right)\eta_{i^{\prime}\alpha^{\prime}}\left(t^{\prime};\mathbf{r}\right)\right\rangle =k_{B}TK_{i\alpha,i^{\prime}\alpha^{\prime}}\left(t,t^{\prime};\mathbf{r}\right)
\label{eq:Fluct-Diss-Theorem}
\end{align}
The last equation (\ref{eq:Fluct-Diss-Theorem}) is equivalent to
the (second) fluctuation-dissipation theorem.
\cite{Zwanzig2001}
As a consequence, Eq. (\ref{eq:gle}) becomes a stochastic integro-differential
equation for the system DoFs, which is in essence what the GLE actually
is: it describes dynamics of a (classical) open system which interacts
and exchanges energy with its environment (i.e., the bath), however,
the bath DoFs are not explicitly present in the formulation. In particular,
if $\left\langle \eta_{i\alpha}(t;\mathbf{r})\eta_{i^{\prime}\alpha^{\prime}}(t^{\prime};\mathbf{r})\right\rangle \propto\delta\left(t-t'\right)$
the dissipative forces provide a multi-dimensional Wiener process
(or \emph{white noise}), while in the general case, the dissipative
forces are said to give a \emph{colored noise}. 

We also note here that in the case of the white noise the GLE goes
over into the ordinary Langevin dynamics. Indeed, assuming that the
memory kernel decays with time much faster than the characteristic
change in the velocities of the system atoms, the velocity $\dot{r}_{i^{\prime}\alpha^{\prime}}(t)$
can be taken out of the integral; the integral of the memory kernel
then becomes the friction constant $\Gamma_{i\alpha,i^{\prime}\alpha^{\prime}}\left({\bf r}\left(t\right)\right)$
multiplying the velocity in the EoMs as in an ordinary Langevin equation.
This transformation is formally obtained by writing the memory kernel
as
\begin{equation}
K_{i\alpha,i^{\prime}\alpha^{\prime}}\left(t,t^{\prime};{\bf r}\right)=2\Gamma_{i\alpha,i^{\prime}\alpha^{\prime}}\left({\bf r}\left(t\right)\right)\delta\left(t-t^{\prime}\right)
\label{eq:kernel_inf}
\end{equation}
with the friction constant possibly depending on the positions of
system atoms in a non-trivial way. In this case, the GLE reduces to
the Langevin equation
\begin{equation}
m_{i}\ddot{r}_{i\alpha}=-\frac{\partial\bar{V}}{\partial r_{i\alpha}}-\sum_{i^{\prime}\alpha^{\prime}}\Gamma_{i\alpha,i^{\prime}\alpha^{\prime}}\left({\bf r}\right)\dot{r}_{i^{\prime}\alpha^{\prime}}+{\bf \eta}_{i\alpha}\left(t;{\bf r}\right)
\label{eq:Langevin-eq}
\end{equation}
with the \emph{white noise} ${\bf \eta}_{i\alpha}(t;{\bf r})$ thanks
to the (second) fluctuation-dissipation theorem (\ref{eq:Fluct-Diss-Theorem}):
\begin{equation*}
\left\langle \eta_{i\alpha}\left(t;{\bf r}\right)\eta_{i^{\prime}\alpha^{\prime}}\left(t^{\prime};\mathbf{r}\right)\right\rangle =2k_{B}T\Gamma_{i\alpha,i^{\prime}\alpha^{\prime}}\left({\bf r}\left(t\right)\right)\delta\left(t-t^{\prime}\right)
\end{equation*}

Non-trivial numerical issues must be faced when solving the GLE, namely:
(i) the integral containing the memory kernel computed at time $t$
is a functional of the system history (i.e., atomic trajectories at
all previous times $t^{\prime}<t$); (ii) the colored noise has to
be properly generated, on-the-fly when possible. Approximations can
be introduced  to avoid the calculation of the integral containing
the memory kernel at each time-step.
\cite{Barrat2011,Biele13,Baczewski13}
Although in practice they narrow the scope of the GLE, the analytic 
on-the-fly colored noise generation is possible in just a very few 
cases,
\cite{Luczka05}
and in some cases the noise cannot be generated 
\emph{a priori} for the duration of the whole simulation.
\cite{Rice44,Billah90,Mannella92}

In the following section, we shall present a convenient alternative
which --- at the price of introducing some auxiliary DoFs --- yields
a simple and general algorithm to: (i) generate the Gaussian stochastic
forces on-the-fly using well-established algorithms for Wiener stochastic
processes; (ii) model non-stationary correlations, when the memory
kernel has the exact structure given by Eq. (\ref{eq:kernel-ini})
and hence are not definite positive
\cite{Baczewski13} 
and can depend
on \emph{both} $t$ and $t^{\prime}$ separately, not just on their
difference; (iii) avoid the explicit calculation of the integral containing
the memory kernel so to circumvent this formidable computation bottleneck.

\subsection{Mapping the GLE onto complex Langevin dynamics in an extended phase
space\label{sub:Mapping-of-GLE}}

Our goal is to generate \emph{on-the-fly} a stochastic process associated
with a non-trivial colored noise such as in Eqs. (\ref{eq:noise-av-1})
and (\ref{eq:Fluct-Diss-Theorem}). We shall now show that the GLE
equations (\ref{eq:gle}) can be solved in an appropriately extended
phase by introducing auxiliary DoFs which satisfy stochastic equations
of the Langevin type (i.e., without the integral of the memory kernel
and with the white noise). We shall demonstrate that, by choosing
appropriately the dynamics of the auxiliary DoFs (i.e., their EoMs),
it is possible to provide an approximate, yet converging, mapping
to the original GLE. This strategy is convenient because there are
efficient numerical approaches to integrate Langevin equations with
the white noise.
\cite{Allen}
According to this strategy, after the
MD trajectories have been simulated in the extended phase space, the
GLE evolution is obtained by tracing out the auxiliary DoFs. Our approach
has been inspired by a similar, yet more efficiency led, algorithm
devised by Ceriotti \emph{et al.} 
\cite{Ceriotti2009a,Ceriotti2011,Morrone2011}
to provide a GLE thermostat. The major difference between the two
approaches is that we are constrained by the specific form of the
noise and the memory kernel derived from the actual dynamics of the
realistic system and bath which interact with each other, where the
compound system is a solid, while in Refs.~\onlinecite{Ceriotti2009a,Ceriotti2011,Morrone2011}
the authors were mostly preoccupied with the efficient, yet unphysical,
thermalisation of the system. In addition, our scheme leads to a rather
natural interpretation of the auxiliary DoFs as effective collective
modes of the bath.

Let us introduce $2(K+1)$ real auxiliary DoFs, $s_{1}^{(k)}(t)$
and $s_{2}^{(k)}(t)$, (where $k=0,1,2,\ldots,K$) which satisfy the
following EoMs:
\begin{align}
&\dot{s}_{1}^{(k)}=-s_{1}^{(k)}/\tau_{k}+\omega_{k}s_{2}^{(k)}+A_{k}(t)+B_{k}{\bf \xi}_{1}^{(k)}
\label{eq:EoM-s1}\\
&\dot{s}_{2}^{(k)}=-s_{2}^{(k)}/\tau_{k}-\omega_{k}s_{1}^{(k)}+B_{k}{\bf \xi}_{2}^{(k)}
\label{eq:EoM-s2}
\end{align}
Here a number of parameters have been introduced: $\tau_{k}$ sets
the relaxation time for a pair of auxiliary DoFs, $\omega_{k}$ provides
the coupling between a pair of auxiliary DoFs $s_{1}^{(k)}$ and $s_{2}^{(k)}$,
and finally $\xi_{1}^{(k)}(t)$ and $\xi_{2}^{(k)}(t)$ are independent
Wiener stochastic processes with correlation functions
\begin{equation}
\begin{aligned}
&\left\langle \xi_{1}^{(k)}(t)\right\rangle =\left\langle \xi_{2}^{(k)}(t)\right\rangle =0\\
&\left\langle \xi_{1}^{(k)}(t)\xi_{1}^{(k^{\prime})}\left(t{}^{\prime}\right)\right\rangle =\left\langle \xi_{2}^{(k)}(t)\xi_{2}^{(k^{\prime})}\left(t^{\prime}\right)\right\rangle =\delta_{kk^{\prime}}\delta\left(t-t^{\prime}\right)\\
&\left\langle \xi_{1}^{(k)}(t)\xi_{2}^{(k^{\prime})}\left(t^{\prime}\right)\right\rangle =0
\end{aligned}
\label{eq:gnoise}
\end{equation}
The function $A_{k}(t)$ and the parameter $B_{k}$ for each $k$
will be determined later on. The idea is to emulate the collective
dynamics of the realistic bath by appropriately setting the free parameters
in the definition of $A_{k}\left(t\right)$ and $B_{k}$. More explicitly,
we shall approximate the displacements $u_{l\gamma}(t)$ as a linear
combination of the auxiliary DoFs. This is not a straightforward change
of co-ordinates, as the number of auxiliary DoFs, namely $2\left(K+1\right),$
will be always kept much smaller than the number of the bath DoFs,
i.e. $K\ll L$. The goal is to achieve a satisfactory approximation
of the bath dynamics through a minimum of possible number of auxiliary
DoFs. 

Since the EoMs of the system atoms (\ref{eq:sys_eoms}) contain the
contribution from the bath in a form of the linear combination of
the bath atoms displacements with the prefactor $\mu_{l}g_{i\alpha,l\gamma}\left(\mathbf{r}\right)$,
we introduce the auxiliary DoFs into the EoMs (\ref{eq:sys_eoms})
for the system (i.e., physical) DoFs linearly as well:
\begin{equation}
m_{i}\ddot{r}_{i\alpha}=-\frac{\partial\bar{V}}{\partial r_{i\alpha}}+\sum_{l\gamma}\mu_{l}g_{i\alpha,l\gamma}\left(\mathbf{r}\right)\left(\sum_{k}\theta_{l\gamma}^{\left(k\right)}s_{1}^{(k)}\right)
\label{eq:EoM-r}
\end{equation}
where we introduced some yet unknown rectangular matrix $\theta_{l\gamma}^{(k)}$.
We have also included the polaronic correction to the potential (see
Eq. (\ref{eq:potential_pol})) to match Eq. (\ref{eq:gle}). Note
that only $s_{1}^{(k)}(t)$ enter the dynamics of the physical DoFs,
the reason for this will become apparent later. 

We shall now find the appropriate forms for the parameters $\theta_{l\gamma}^{(k)}$
and $B_{k}$ and the functions $A_{k}(t)$ which would map the auxiliary
dynamics given by Eqs. (\ref{eq:EoM-s1}), (\ref{eq:EoM-s2}) and
(\ref{eq:EoM-r}) onto the real dynamics of the physical variables
given by the GLE (\ref{eq:gle}).

To this end, we first notice that the Langevin dynamics of the auxiliary
DoFs given by Eqs. (\ref{eq:EoM-s1}) and (\ref{eq:EoM-s2}) possess
a natural complex structure which is revealed by defining the complex
DoF, $s^{(k)}=s_{1}^{(k)}+is_{2}^{(k)}$, satisfying the EoM:
\begin{equation*}
\dot{s}^{(k)}=-\left(\frac{1}{\tau_{k}}+i\omega_{k}\right)s^{(k)}-A_{k}(t)+B_{k}{\bf \xi}^{(k)}
\end{equation*}
where $\xi^{(k)}=\xi_{1}^{(k)}+i\xi_{2}^{(k)}$ is now a complex Wiener
stochastic process. The above equation has the following solution
(vanishing at $t=-\infty$):
\begin{multline*}
s^{(k)}(t)=-\int_{-\infty}^{t}\mbox{d}t^{\prime}\left[A_{k}\left(t^{\prime}\right)-B_{k}\xi^{(k)}\left(t^{\prime}\right)\right]\\
\exp\left[-\left(\frac{1}{\tau_{k}}+i\omega_{k}\right)\left(t-t^{\prime}\right)\right]\mbox{d}t^{\prime}
\end{multline*}
Substituting the real part of the solution, $s_{1}^{(k)}(t)=\mbox{Re}\left[s^{(k)}(t)\right]$,
back into Eq. (\ref{eq:EoM-r}), we obtain:
\begin{multline}
m_{i}\ddot{r}_{i\alpha}=-\frac{\partial\bar{V}}{\partial r_{i\alpha}}+\sum_{l\gamma}\mu_{l}g_{i\alpha,l\gamma}\left(\mathbf{r}\right)\sum_{k}\theta_{l\gamma}^{(k)}\\
\int_{-\infty}^{t}A_{k}\left(t^{\prime}\right)\phi_{k}\left(t-t^{\prime}\right)\mbox{d}t^{\prime}+\eta_{i\alpha}(t)
\label{eq:interm1}
\end{multline}
where
\begin{equation}
\eta_{i\alpha}(t)=\sum_{l\gamma}\mu_{l}g_{i\alpha,l\gamma}\left(\mathbf{r}\right)\sum_{k}\theta_{l\gamma}^{(k)}B_{k}\chi_{k}(t)
\label{eq:stochastic-force}
\end{equation}
and, for the sake of notation, we have also introduced
\begin{equation}
\phi_{k}(t)=e^{-\left|t\right|/\tau_{k}}\cos\left(\omega_{k}t\right)
\label{eq:function-phi}
\end{equation}
and
\begin{multline}
\chi_{k}(t)=\int_{-\infty}^{t}e^{-\left(t-t^{\prime}\right)/\tau_{k}}\left[\xi_{1}^{(k)}\left(t^{\prime}\right)\cos\left(\omega_{k}\left(t-t^{\prime}\right)\right)\right.\\
\left. +\xi_{2}^{(k)}\left(t^{\prime}\right)\sin\left(\omega_{k}\left(t-t^{\prime}\right)\right)\right]\mbox{d}t^{\prime}
\label{eq:auxiliary-fun}
\end{multline}
Since the force $\eta_{i\alpha}$ is related directly to the Wiener
stochastic processes and hence must be the only one responsible for
the stochastic forces in Eq. (\ref{eq:gle}), the second term in the
right hand side of Eq. (\ref{eq:interm1}) must then have exactly
the same form as the memory term in the GLE (\ref{eq:gle}). This
is only possible with the following choice of the function $A_{k}(t)$:
\begin{equation*}
A_{k}(t)=\sum_{l\gamma}\vartheta_{l\gamma}^{(k)}\left[\sum_{i\alpha}g_{i\alpha,l\gamma}\left(\mathbf{r}\left(t\right)\right)\dot{r}_{i\alpha}(t)\right]
\end{equation*}
with some additional parameters $\vartheta_{l\gamma}^{(k)}$. This
choice leads to the memory kernel having the same structure as in
Eq. (\ref{eq:kernel-ini}), but with the polarization matrix
\begin{equation*}
\Pi_{l\gamma,l^{\prime}\gamma^{\prime}}\left(t-t^{\prime}\right)=\sqrt{\frac{\mu_{l}}{\mu_{l^{\prime}}}}\sum_{k}\theta_{l\gamma}^{(k)}\vartheta_{l^{\prime}\gamma^{\prime}}^{(k)}\phi_{k}\left(t-t^{\prime}\right)
\end{equation*}
Since the polarization matrix must be symmetric, see Eq. (\ref{eq:symmetry-pol-matrix}),
one has to choose $\vartheta_{l^{\prime}\gamma^{\prime}}^{(k)}=\zeta_{k}\mu_{l^{\prime}}\theta_{l^{\prime}\gamma^{\prime}}^{(k)}$.
The proportionality constant $\zeta_{k}$ can be chosen arbitrarily;
it is convenient to choose it such that $\zeta_{k}$ does not depend
on $k$. We shall denote the proportionality constant by $\bar{\mu}$
which can be thought of as the mass of the auxiliary DoFs (see below)
and hence, $\vartheta_{l^{\prime}\gamma^{\prime}}^{(k)}=\bar{\mu}\mu_{l^{\prime}}\theta_{l^{\prime}\gamma^{\prime}}^{(k)}$.
Finally, we set $\theta_{l\gamma}^{(k)}=c_{l\gamma}^{(k)}/\sqrt{\bar{\mu}\mu_{l}}$,
where $c_{l\gamma}^{(k)}$ are new parameters. These definitions finally
bring the polarization matrix into the form:
\begin{multline}
\Pi_{l\gamma,l^{\prime}\gamma^{\prime}}\left(t-t^{\prime}\right)=\sum_{k}c_{l\gamma}^{(k)}c_{l^{\prime}\gamma^{\prime}}^{(k)}\\
e^{-\left(t-t^{\prime}\right)/\tau_{k}}\cos\left(\omega_{k}\left(t-t^{\prime}\right)\right)
\label{eq:pol}
\end{multline}
and the original EoMs for the physical DoFs, Eq. (\ref{eq:EoM-r}),
can now be written as: 
\begin{equation}
m_{i}\ddot{r}_{i\alpha}=-\frac{\partial\bar{V}}{\partial r_{i\alpha}}+\sum_{l\gamma}\sqrt{\frac{\mu_{l}}{\bar{\mu}}}g_{i\alpha,l\gamma}\left(\mathbf{r}\right)\sum_{k}c_{l\gamma}^{\left(k\right)}s_{1}^{(k)}
\label{eq:EoM-physical-cvariable-final}
\end{equation}
which when compared with Eq. (\ref{eq:sys_eoms}) yield
\begin{equation}
u_{l,\gamma}\;\Longrightarrow\;\frac{1}{\sqrt{\mu_{l}\bar{\mu}}}\sum_{k}c_{l\gamma}^{\left(k\right)}s_{1}^{\left(k\right)}
\label{eq:collective}
\end{equation}
that is, new variables provide an approximate linear representation
for the actual displacements of the bath atoms.

We now need to make sure that the stochastic force (\ref{eq:stochastic-force})
satisfies Eqs. (\ref{eq:noise-av-1}) and (\ref{eq:Fluct-Diss-Theorem})
which is necessary for the dynamics of the auxiliary DoFs to mimic
correctly that of the actual bath DoFs. Using the definitions (\ref{eq:gnoise})
for the Wiener stochastic processes, Eq. (\ref{eq:noise-av-1}) follows
immediately. To check the (second) fluctuation-dissipation theorem,
Eq. (\ref{eq:Fluct-Diss-Theorem}), we first note that from the properties
of the Wiener stochastic processes, $\xi_{1}^{(k)}(t)$ and $\xi_{2}^{(k)}(t)$,
it follows that the correlation function of the auxiliary function
(\ref{eq:auxiliary-fun}),
\begin{multline*}
\left\langle \chi_{k}(t)\chi_{k^{\prime}}\left(t^{\prime}\right)\right\rangle =\delta_{kk^{\prime}}e^{-\left(t-t^{\prime}\right)/\tau_{k}}\cos\left(\omega_{k}\left(t-t^{\prime}\right)\right)\\
\int_{-\infty}^{\min\left(t,t^{\prime}\right)}e^{2x/\tau_{k}}\mbox{d}x=\delta_{kk^{\prime}}\frac{\tau_{k}}{2}\phi_{k}\left(t-t^{\prime}\right)
\end{multline*}
depends only on the absolute value of the time difference, $\left|t-t^{\prime}\right|$,
via $\phi_{k}\left(t-t^{\prime}\right)$ defined by Eq. (\ref{eq:function-phi}).
This in turn results in the following correlation function of the
noise (\ref{eq:stochastic-force}):
\begin{multline*}
\left\langle \eta_{i\alpha}(t)\eta_{i^{\prime}\alpha^{\prime}}\left(t^{\prime}\right)\right\rangle =\sum_{l\gamma}\sum_{l^{\prime}\gamma^{\prime}}\sqrt{\mu_{l}\mu_{l^{\prime}}}g_{i\alpha,l\gamma}\left(\mathbf{r}\left(t\right)\right)\\
\left[\frac{1}{\bar{\mu}}\sum_{k}\frac{\tau_{k}B_{k}^{2}}{2}c_{l\gamma}^{(k)}c_{l^{\prime}\gamma^{\prime}}^{(k)}\phi_{k}\left(t-t^{\prime}\right)\right]g_{i^{\prime}\alpha^{\prime},l^{\prime}\gamma^{\prime}}\left(\mathbf{r}\left(t^{\prime}\right)\right)
\end{multline*}
To satisfy the (second) fluctuation-dissipation theorem (\ref{eq:Fluct-Diss-Theorem}),
one has to choose $B_{k}=\sqrt{2k_{B}T\bar{\mu}/\tau_{k}}$ which
would make the correlation function above to be exactly equal to the
$k_{B}T$ times the memory kernel (\ref{eq:kernel-ini}) with the
polarization matrix given by expression (\ref{eq:pol}). Therefore,
as both the functions $A_{k}(t)$ and the constants $B_{k}$ are determined,
we can now fully define the EoMs for the auxiliary DoFs, Eqs. (\ref{eq:EoM-s1})
and (\ref{eq:EoM-s2}), as: 
\begin{multline}
\dot{s}_{1}^{(k)}=-s_{1}^{(k)}/\tau_{k}+\omega_{k}s_{2}^{(k)}-\sum_{l\gamma}\sqrt{\bar{\mu}\mu_{l}}c_{l\gamma}^{(k)}\\
\sum_{i\alpha}g_{ia,l\gamma}\left(\mathbf{r}\left(t\right)\right)\dot{r}_{i\alpha}(t)+\sqrt{\frac{2k_{B}T\bar{\mu}}{\tau_{k}}}{\bf \xi}_{1}^{(k)}
\label{eq:EoM-s1-final}
\end{multline}
\begin{equation}
\dot{s}_{2}^{(k)}=-s_{2}^{(k)}/\tau_{k}-\omega_{k}s_{1}^{(k)}+\sqrt{\frac{2k_{B}T\bar{\mu}}{\tau_{k}}}{\bf \xi}_{2}^{(k)}
\label{eq:EoM-s2-final}
\end{equation}
Equations (\ref{eq:EoM-physical-cvariable-final}), (\ref{eq:EoM-s1-final})
and (\ref{eq:EoM-s2-final}) together define a set of complex Langevin
equations
\begin{widetext}
\begin{equation}
\begin{cases}
m_{i}\ddot{r}_{i\alpha} & =-\frac{\partial\bar{V}}{\partial r_{i\alpha}}+\sum_{l\gamma}\sum_{k}\sqrt{\frac{\mu_{l}}{\bar{\mu}}}g_{i\alpha,l\gamma}\left(\mathbf{r}\right)c_{l\gamma}^{\left(k\right)}s_{1}^{(k)}\\
\dot{s}_{1}^{(k)} & =-\frac{s_{1}^{(k)}}{\tau_{k}}+\omega_{k}s_{2}^{(k)}-\sum_{i\alpha}\sum_{l\gamma}\sqrt{\mu_{l}\bar{\mu}}g_{i\alpha,l\gamma}\left(\mathbf{r}\right)c_{l\gamma}^{\left(k\right)}\dot{r}_{i\alpha}+\sqrt{\frac{2k_{B}T\bar{\mu}}{\tau_{k}}}{\bf \xi}_{1}^{(k)}\\
\dot{s}_{2}^{(k)} & =-\frac{s_{2}^{(k)}}{\tau_{k}}-\omega_{k}s_{1}^{(k)}+\sqrt{\frac{2k_{B}T\bar{\mu}}{\tau_{k}}}{\bf \xi}_{2}^{(k)}
\end{cases}
\label{eq:complex_LEs}
\end{equation}
\end{widetext}
which defines the required mapping: the introduction of a finite number
of auxiliary DoFs ($s_{1}^{(k)}$ and $s_{2}^{(k)}$), as discussed
above, allows one to obtain the EoMs for the physical variables that
are the same as the exact GLE (\ref{eq:gle}), provided that the polarization
matrix (\ref{eq:pol_ini}) is replaced by that shown in expression
(\ref{eq:pol}).

The polarization matrix entering the memory kernel and defined in
Eq. (\ref{eq:pol}) is formally different from the GLE counterpart
defined in Eq. (\ref{eq:pol_ini}). In practice, by properly choosing
the values of the parameters $\omega_{k}$, $\tau_{k}$, and $c_{l\gamma}^{\left(k\right)}$,
one can ensure that the matrix (\ref{eq:pol}) yields a satisfactory
approximation of the original one. As the mass, $\bar{\mu}$, does
not appear in Eq. (\ref{eq:pol}), it can be freely adjusted to improve
the efficiency of the algorithm. In principle, this approximation
is not trivial as we would like to represent the bath dynamics through
a much smaller set of auxiliary DoFs, as $K\ll L$. However, the agreement
is expected to improve as $K$ is increased as more fitting parameters
for the polarization matrix will become available. 

Instead of a straightforward fit of the free parameters to ensure
that Eqs. (\ref{eq:pol}) and (\ref{eq:pol_ini}) agree as much as
possible in the time domain, we prefer a scheme which takes full advantage
of the functional form of the bath polarization matrix. In fact, we
find that it is more convenient to ensure that the two polarization
matrices agree in the frequency domain. Assuming that the polarization
matrix (\ref{eq:pol}) is defined as an even function of its time
argument (see the discussion at the end of Sec. \ref{sub:eoms}),
this method is facilitated by the fact that the FT of the polarization
matrix
\begin{multline}
\Pi_{l\gamma,l^{\prime}\gamma^{\prime}}(\omega)=\sum_{k}c_{l\gamma}^{\left(k\right)}c_{l^{\prime}\gamma^{\prime}}^{\left(k\right)}\left[\frac{\tau_{k}}{1+(\omega-\omega_{k})^{2}\tau_{k}^{2}}\right.\\
\left. +\frac{\tau_{k}}{1+(\omega+\omega_{k})^{2}\tau_{k}^{2}}\right]
\label{eq:kernel_ft}
\end{multline}
is real and proportional to the weighted sum of $2(K+1)$ Lorentzians
centered at $\omega=\pm\omega_{k}$ and with full width at half maximum
$2/\tau_{k}$. Therefore, after computing independently the polarization
matrix using the bath eigenvectors, Eq. (\ref{eq:pol_ini}), one chooses
the fitting parameters $\omega_{k}$, $\tau_{k}$, and $c_{l\gamma}^{\left(k\right)}$
(where $k=0,1,\ldots,K$) in Eq. (\ref{eq:pol}) to provide a good
fit for it in the frequency space. Once the appropriate set of the
parameters is selected, the dynamics of the physical and auxiliary
DoFs is fully defined and should represent the dynamics of our system
surrounded by the realistic bath.

We also note that simple generalization of the above scheme exists
which allows one constructing a mapping whereby the noise correlation
function of the GLE is no longer proportional to the memory kernel,
\cite{Ceriotti2009b,Dammak2009,Ceriotti2010b}
i.e., could be a different function also decaying with time. This
point is briefly addressed in Appendix \ref{sec:quantum}.

\subsection{Fokker-Plank equation and equilibrium properties\label{sec:fp}}

In this section we start the derivation of our numerical algorithm
for solving the stochastic differential equations (\ref{eq:EoM-physical-cvariable-final})-(\ref{eq:EoM-s2-final})
with the white noise. The idea of the method is based on establishing
a Fokker-Planck (FP) equation which is equivalent to our equations
(see, e.g., Refs.~\onlinecite{Gillespie96a,Gillespie96b}) 
and it is similar to
the algorithm proposed by Ceriotti \emph{et al}.
\cite{Ceriotti2010a}
The FP equation is rewritten in the Liouville form which then allows
one constructing the required numerical algorithm. In this section,
we focus on the functional form of the FP equation itself, while the
integration algorithm will be discussed in the next section. In this
way, we can: (i) demonstrate that the Langevin dynamics defined by
our EoMs for the extended set (i.e., physical and auxiliary) of DoFs
can describe the thermalisation of the actual system to the correct
equilibrium Maxwell-Boltzmann distribution and (ii) devise an efficient
algorithm to integrate our equations. As the general idea of this
derivation is well known, only the final results will be stated here
with some details given in Appendix \ref{sec:Derivation-of-the-FP}.

The FP equation corresponding to Eqs. (\ref{eq:EoM-physical-cvariable-final})-(\ref{eq:EoM-s2-final})
is a \emph{deterministic} EoM for the probability density function
(PDF), $P\left(\mathbf{r},\mathbf{p},{\bf s}_{1},{\bf s}_{2},t\right)$,
where the vectors $\mathbf{s}_{1}$ and $\mathbf{s}_{2}$ collect
all auxiliary DoFs $s_{1}^{(k)}$ and $s_{2}^{(k)}$, and the vector
$\mathbf{p}$ collects the Cartesian components of all the momenta
of the system atoms. The PDF satisfies the appropriate FP equation
which we shall write in a form reminiscent of the Liouville equation:
\cite{Tuckerman90,Donnelly05}
\begin{multline}
\dot{P}\left(\mathbf{r},\mathbf{p},{\bf s}_{1},{\bf s}_{2},t\right)=-\hat{\mathfrak{L}}_{FP}P\left(\mathbf{r},\mathbf{p},{\bf s}_{1},{\bf s}_{2},t\right)\\
=-\left(\hat{\mathfrak{L}}_{cons}+\hat{\mathfrak{L}}_{diss}\right)P\left(\mathbf{r},\mathbf{p},{\bf s}_{1},{\bf s}_{2},t\right)
\label{eq:fp}
\end{multline}
where we have split the FP Liouvillian operator, $\hat{\mathfrak{L}}{}_{FP}$,
into its conservative, $\hat{\mathfrak{L}}_{cons}$, and dissipative,
$\hat{\mathcal{\mathfrak{L}}}_{diss}$, parts, see Appendix \ref{sec:Derivation-of-the-FP}
for some details of the derivation. (The minus sign is conventionally
used to stress that the $\hat{\mathfrak{L}}{}_{FP}$ is a positive
semi-definite operator.) 

Based on the Liouville theorem in the extended phase space, the conservative
part of the Liouvillian can be written as
\cite{Tuckerman90,Donnelly05}
\begin{multline}
\hat{\mathfrak{L}}_{cons}=\sum_{i\alpha}\left(\dot{r}_{i\alpha}\frac{\partial}{\partial r_{i\alpha}}+\dot{p}_{i\alpha}\frac{\partial}{\partial p_{i\alpha}}\right)\\
+\sum_{k}\left(\dot{s}_{1}^{(k)}\frac{\partial}{\partial s_{1}^{(k)}}+\dot{s}_{2}^{(k)}\frac{\partial}{\partial s_{2}^{(k)}}\right)
\label{eq:continuity}
\end{multline}
where the dynamics associated with this part of the Liouvillian is
given by the following EoMs:
\begin{align}
&\dot{r}_{i\alpha}=\frac{p_{i\alpha}}{m_{i}}
\label{eq:eoms_cons-r}\\
&\dot{p}_{i\alpha}=-\frac{\partial\bar{V}}{\partial r_{i\alpha}}+\sum_{l\gamma}\sum_{k}\sqrt{\frac{\mu_{l}}{\bar{\mu}}}g_{i\alpha,l\gamma}\left(\mathbf{r}\right)c_{l\gamma}^{\left(k\right)}s_{1}^{(k)}
\label{eq:eoms_cons-p}\\
&\dot{s}_{1}^{(k)}=\omega_{k}s_{2}^{(k)}-\sum_{l\gamma}\sqrt{\bar{\mu}\mu_{l}}c_{l\gamma}^{(k)}\sum_{i\alpha}g_{ia,l\gamma}\left(\mathbf{r}(t)\right)\frac{p_{i\alpha}(t)}{m_{i}}
\label{eq:eoms_cons-s1}\\
&\dot{s}_{2}^{(k)}=-\omega_{k}s_{1}^{(k)}
\label{eq:eoms_cons-s2}
\end{align}
These EoMs correspond to the conservative part of the dynamics. Indeed,
their dynamics conserves the pseudo-energy 
\begin{multline}
\varepsilon_{ps}\left(\mathbf{r},\mathbf{p},{\bf s}_{1},{\bf s}_{2}\right)=\sum_{i\alpha}\frac{p_{i\alpha}^{2}}{2m_{i}}+\bar{V}({\bf r})\\
+\frac{1}{2\bar{\mu}}\sum_{k}\left[\left(s_{1}^{(k)}\right)^{2}+\left(s_{2}^{(k)}\right)^{2}\right]
\label{eq:pse}
\end{multline}
as, by using the EoMs of the conservative dynamics written above,
it is easily verified that $\dot{\epsilon}_{ps}=0$. Remarkably, this
pseudo-energy consists of two terms, the first being the total energy
of the physical system and the second one just ``harmonically''
depending on the auxiliary DoFs and their masses $\bar{\mu}$. 

The remaining dissipative part of the FP operator 
\begin{multline}
\hat{\mathcal{\mathfrak{L}}}_{diss}=-\sum_{k}\frac{1}{\tau_{k}}\left[\frac{\partial}{\partial s_{1}^{(k)}}\left(s_{1}^{(k)}+k_{B}T\bar{\mu}\frac{\partial}{\partial s_{1}^{(k)}}\right)\right.\\\left.+\frac{\partial}{\partial s_{2}^{(k)}}\left(s_{2}^{(k)}+k_{B}T\bar{\mu}\frac{\partial}{\partial s_{2}^{(k)}}\right)\right]
\label{eq:ldiss}
\end{multline}
describes $K+1$ pairs of non-interacting FP processes in the phase
space of the auxiliary DoFs which are equivalent to the Langevin dynamics
governed by the EoMs
\cite{Gillespie96a,Gillespie96b}:
\begin{equation}
\begin{aligned}
&\dot{s}_{1}^{(k)} & =-s_{1}^{(k)}/\tau_{k}+\sqrt{2k_{B}T\bar{\mu}/\tau_{k}}\xi_{1}^{(k)}\\
&\dot{s}_{2}^{(k)} & =-s_{2}^{(k)}/\tau_{k}+\sqrt{2k_{B}T\bar{\mu}/\tau_{k}}\xi_{2}^{(k)}
\end{aligned}
\label{eq:eoms_diss}
\end{equation}
Note that combining the right hand sides of equations (\ref{eq:eoms_cons-r})
- (\ref{eq:eoms_cons-s2}) and (\ref{eq:eoms_diss}) gives the corresponding
right hand sides of the full EoMs (\ref{eq:complex_LEs}), as required.

As a result of the mapping from the complex Langevin equations (\ref{eq:complex_LEs})
to the correspondent FP equation (\ref{eq:fp}), it is now straightforward
to verify that 
\begin{equation}
P^{({\rm eq})}\left(\mathbf{r},\mathbf{p},{\bf s}_{1},{\bf s}_{2}\right)\propto\exp\left(-\varepsilon_{{\rm ps}}/k_{B}T\right)
\label{eq:pdf_eq}
\end{equation}
is a stationary solution of Eq. (\ref{eq:fp}) since $\hat{\mathfrak{L}}_{cons}P^{({\rm eq})}=0$
and $\hat{\mathfrak{L}}_{diss}P^{({\rm eq})}=0$ hold separately and
hence also $\hat{\mathfrak{L}}_{FP}P^{({\rm eq})}=0$. In addition,
it can also be proven that Eq. (\ref{eq:pdf_eq}) corresponds to the
equilibrium PDF, i.e., the solution of the FP equation (\ref{eq:fp})
always converges to $P^{(eq)}\left(\mathbf{r},\mathbf{p},{\bf s}_{1},{\bf s}_{2}\right)$
at $t\rightarrow\infty$ (see Appendix \ref{sec:equilibrium}).

Finally, as stated at the beginning of Sec. \ref{sub:Mapping-of-GLE},
the physical dynamics defined by the solution of Eq. (\ref{eq:gle})
is obtained by tracing the auxiliary DoFs out of the solution of Eqs.
(\ref{eq:EoM-physical-cvariable-final})-(\ref{eq:EoM-s2-final}).
Accordingly, the physical equilibrium PDF is obtained by tracing out
the auxiliary DoFs from Eq. (\ref{eq:pdf_eq}): 
\begin{multline*}
P^{({\rm eq})}\left(\mathbf{r},\mathbf{p}\right)\equiv\int\prod_{k}\mbox{d}s_{1}^{\left(k\right)}\mbox{d}s_{2}^{\left(k\right)}P^{({\rm eq})}\left(\mathbf{r},\mathbf{p},{\bf s}_{1},{\bf s}_{2}\right)\\
\propto\exp\left[-\frac{1}{k_{B}T}\left(\sum_{i\alpha}\frac{p_{i\alpha}^{2}}{2m_{i}}+\bar{V}({\bf r})\right)\right]
\end{multline*}
which is indeed the expected Maxwell-Boltzmann distribution (see also
discussion in Ref.~\onlinecite{Kantorovich2008a}).

\subsection{The integration algorithm\label{sub:integration}}

Eq. (\ref{eq:fp}) can be formally integrated for one time-step, $\Delta t$,
to give 
\begin{equation*}
P\left(\mathbf{r},\mathbf{p},{\bf s}_{1},{\bf s}_{2},t+\Delta t\right)=e^{-\Delta t\hat{\mathfrak{L}}_{FP}}P\left(\mathbf{r},\mathbf{p},{\bf s}_{1},{\bf s}_{2},t\right)
\end{equation*}
which can then be approximated using the second order (symmetrized)
Trotter expansion of the FP propagator
\cite{Bussi07} 
\begin{equation}
e^{-\Delta t\hat{\mathfrak{L}}_{FP}}=e^{-\frac{\Delta t}{2}\hat{\mathfrak{L}}_{diss}}e^{-\Delta t\hat{\mathfrak{L}}_{cons}}e^{-\frac{\Delta t}{2}\hat{\mathfrak{L}}_{diss}}+\mathcal{O}\left(\Delta t^{3}\right)
\label{eq:trotter}
\end{equation}
Although Eq. (\ref{eq:trotter}) gives a second order approximation
for the exact FP propagator, $P^{(eq)}\left(\mathbf{r},\mathbf{p},{\bf s}_{1},{\bf s}_{2}\right)$
is still a stationary solution of the approximate dynamics since $\hat{\mathfrak{L}}_{cons}P^{({\rm eq})}\left(\mathbf{r},\mathbf{p},{\bf s}_{1},{\bf s}_{2}\right)=0$
and $\hat{\mathfrak{L}}_{diss}P^{({\rm eq})}\left(\mathbf{r},\mathbf{p},{\bf s}_{1},{\bf s}_{2}\right)=0$
hold separately. 

To approximate the action of $e^{-\Delta t\mathcal{L}_{cons}}$, one
can split the conservative part of the Liouvillian into two contributions,
\cite{Donnelly05}
\begin{multline*}
\hat{\mathfrak{L}}_{r,s_{1}}=-\sum_{i\alpha}\frac{p_{i\alpha}}{m_{i}}\frac{\partial}{\partial r_{i\alpha}}-\sum_{k}\left(\omega_{k}s_{2}^{(k)}\right.\\
\left. -\sum_{i\alpha}\sum_{l\gamma}\frac{\sqrt{\mu_{l}\bar{\mu}}}{m_{i}}g_{i\alpha,l\gamma}\left(\mathbf{r}\right)c_{l\gamma}^{\left(k\right)}p_{i\alpha}\right)\frac{\partial}{\partial s_{1}^{(k)}}
\end{multline*}
and
\begin{multline*}
\hat{\mathfrak{L}}_{p,s_{2}}=\sum_{i\alpha}\left(\frac{\partial\bar{V}}{\partial r_{i\alpha}}-\sum_{l\gamma}\sum_{k}\sqrt{\frac{\mu_{l}}{\bar{\mu}}}g_{i\alpha,l\gamma}\left(\mathbf{r}\right)c_{l\gamma}^{\left(k\right)}s_{1}^{(k)}\right)\\
\frac{\partial}{\partial p_{i\alpha}}+\sum_{k}\omega_{k}s_{1}^{(k)}\frac{\partial}{\partial s_{2}^{(k)}}
\end{multline*}
and then use again the second order Trotter decomposition to obtain:
\begin{equation}
e^{-\Delta t\hat{\mathfrak{L}}_{cons}}=e^{-\frac{\Delta t}{2}\hat{\mathfrak{L}}_{p,s_{2}}}e^{-\Delta t\hat{\mathfrak{L}}_{r,s_{1}}}e^{-\frac{\Delta t}{2}\hat{\mathfrak{L}}_{p,s_{2}}}+\mathcal{O}\left(\Delta t^{3}\right)
\label{eq:lcons_2}
\end{equation}
Combining both decompositions, the following approximation for the
time-step propagation of the whole Liouvillian is finally obtained:
\cite{Bussi07}
\begin{multline}
e^{-\Delta t\hat{\mathfrak{L}}_{FP}}=e^{-\frac{\Delta t}{2}\hat{\mathfrak{L}}_{diss}}e^{-\frac{\Delta t}{2}\hat{\mathfrak{L}}_{p,s_{2}}}e^{-\Delta t\hat{\mathfrak{L}}_{r,s_{1}}}\\
e^{-\frac{\Delta t}{2}\hat{\mathfrak{L}}_{p,s_{2}}}e^{-\frac{\Delta t}{2}\hat{\mathfrak{L}}_{diss}}+\mathcal{O}\left(\Delta t^{3}\right)
\label{eq:trotter2}
\end{multline}

Each factor in Eq. (\ref{eq:trotter2}) (to be read from right to
left) corresponds to a single step in building up the action of $e^{-\Delta t\hat{\mathfrak{L}}_{FP}}$
on $P\left({\bf r},{\bf p},{\bf s}_{1},{\bf s}_{2},t\right)$, i.e.,
all of them in succession (from right to left) correspond to one time-step
propagation of the MD algorithm. The first and last steps are given
by $e^{-\frac{\Delta t}{2}\hat{\mathfrak{L}}_{diss}}$ which accounts
for the integration of the dissipative part of the dynamics related
to auxiliary DoFs, Eq. (\ref{eq:eoms_diss}). These equations describe
$K+1$ pairs of simple non-interacting Langevin equations corresponding
to the FP equation $\dot{P}\left({\bf s}_{1},{\bf s}_{2},t\right)=-\hat{\mathfrak{L}}_{diss}P\left({\bf s}_{1},{\bf s}_{2},t\right)$.
To integrate Eq. (\ref{eq:eoms_diss}) we then use a variant of a
well-known algorithm by Ermak and Buckholz
\cite{Ermak80,Allen}
\begin{equation}
s_{x}^{(k)}\left(t\right) \leftarrow a_{k}s_{x}^{(k)}\left(t\right)+b_{k}\xi_{x}^{(k)}\left(t\right)\;,
\label{eq:ss}
\end{equation}
where $x=1,2$ and $a_{k}=e^{-\Delta t/2\tau_{k}}$, $b_{k}=\sqrt{k_{B}T\bar{\mu}\left(1-a_{k}^{2}\right)}$,
while $\xi_{1}^{(k)}\left(t\right)$ and $\xi_{2}^{(k)}\left(t\right)$
comprise $K+1$ pairs of uncorrelated Wiener stochastic processes
with correlation functions $\left\langle \xi_{1,2}^{(k)}\left(t\right)\right\rangle =0$,
$\left\langle \xi_{x}^{(k)}\left(t\right)\xi_{x}^{(k^{\prime})}\left(t^{\prime}\right)\right\rangle =\delta_{kk^{\prime}}\delta\left(t-t^{\prime}\right)$,
and $\left\langle \xi_{1}^{(k)}\left(t\right)\xi_{2}^{(k^{\prime})}\left(t^{\prime}\right)\right\rangle =0$.
Note that Eq. (\ref{eq:ss}) reduces to 
\begin{equation*}
s_{x}^{(k)}\left(t\right)\leftarrow\sqrt{k_{B}T\bar{\mu}}\xi_{x}^{(k)}\left(t\right)\;,
\end{equation*}
in the strong friction limit, $\tau_{k}\to0$.

For the conservative part of the dynamics, Eqs. (\ref{eq:eoms_cons-r})-(\ref{eq:eoms_cons-s2}),
one can then work out a generalization of the Velocity-Verlet algorithm.
\cite{Tuckerman90,Donnelly05}
In particular, the action of the operator
$e^{-\Delta t\hat{\mathfrak{L}}_{p,s_{2}}/2}$ is equivalent to the
following step in the propagation algorithm:
\cite{Donnelly05}
\begin{multline}
p_{i\alpha}\leftarrow p_{i\alpha}+\left(-\frac{\partial\bar{V}(\mathbf{r})}{\partial r_{i\alpha}}\right.\\
\left.+\sum_{l\gamma}\sum_{k}\sqrt{\frac{\mu_{l}}{\bar{\mu}}}g_{i\alpha,l\gamma}\left(\mathbf{r}\right)c_{l\gamma}^{\left(k\right)}s_{1}^{(k)}\right)\frac{\Delta t}{2}
\label{eq:p-integrator}
\end{multline}
and
\begin{equation}
s_{2}^{(k)}\leftarrow s_{2}^{(k)}-\omega_{k}s_{1}^{(k)}\frac{\Delta t}{2}\;.
\label{eq:s2-integrator}
\end{equation}
These equations can also formally be obtained by integrating over
the same time Eqs. (\ref{eq:eoms_cons-p}) and (\ref{eq:eoms_cons-s2}).
Similarly, from the action of the operator $e^{-\Delta t\hat{\mathfrak{L}}_{r,s_{1}}}$
one obtains the following set of equations for the propagation dynamics:
\cite{Donnelly05}
\begin{equation}
r_{i\alpha}\leftarrow r_{i\alpha}+\frac{p_{i\alpha}}{m_{i}}\Delta t
\label{eq:r-integrator}
\end{equation}
and
\begin{multline}
s_{1}^{(k)}\leftarrow s_{1}^{(k)}+\left(\omega_{k}s_{2}^{(k)}\right.\\
\left.-\sum_{i\alpha}\sum_{l\gamma}\frac{\sqrt{\mu_{l}\bar{\mu}}}{m_{i}}g_{i\alpha,l\gamma}\left(\mathbf{r}\right)c_{l\gamma}^{\left(k\right)}p_{i\alpha}\right)\Delta t\;.
\label{eq:s1-integrator}
\end{multline}
Note that, in the limiting case of $g_{i\alpha,l\gamma}\left({\bf r}\right)=0$,
the equations above factorize into two independent Velocity-Verlet
steps for the physical and auxiliary DoFs.

Finally, combining Eqs. (\ref{eq:ss}) and (\ref{eq:p-integrator})-(\ref{eq:s1-integrator}),
the following algorithm for one time-step, $\Delta t$, integration
is found:
\begin{widetext}
\begin{equation}
\begin{aligned}s_{x}^{(k)} & \leftarrow a_{k}s_{x}^{(k)}+b_{k}\xi_{x}^{(k)}\;,\; x=1,2\\
p_{i\alpha} & \leftarrow p_{i\alpha}+\left(-\frac{\partial\bar{V}(\mathbf{r})}{\partial r_{i\alpha}}+\sum_{l\gamma}\sum_{k}\sqrt{\frac{\mu_{l}}{\bar{\mu}}}g_{i\alpha,l\gamma}\left(\mathbf{r}\right)c_{l\gamma}^{\left(k\right)}s_{1}^{(k)}\right)\frac{\Delta t}{2}\\
s_{2}^{(k)} & \leftarrow s_{2}^{(k)}-\omega_{k}s_{1}^{(k)}\frac{\Delta t}{2}\\
r_{i\alpha} & \leftarrow r_{i\alpha}+\frac{p_{i\alpha}}{m_{i}}\Delta t\\
s_{1}^{(k)} & \leftarrow s_{1}^{(k)}+\left(\omega_{k}s_{2}^{(k)}-\sum_{i\alpha}\sum_{l\gamma}\frac{\sqrt{\mu_{l}\bar{\mu}}}{m_{i}}g_{i\alpha,l\gamma}\left(\mathbf{r}\right)c_{l\gamma}^{\left(k\right)}p_{i\alpha}\right)\Delta t\\
p_{i\alpha} & \leftarrow p_{i\alpha}+\left(-\frac{\partial\bar{V}(\mathbf{r})}{\partial r_{i\alpha}}+\sum_{l\gamma}\sum_{k}\sqrt{\frac{\mu_{l}}{\bar{\mu}}}g_{i\alpha,l\gamma}\left(\mathbf{r}\right)c_{l\gamma}^{\left(k\right)}s_{1}^{(k)}\right)\frac{\Delta t}{2}\\
s_{2}^{(k)} & \leftarrow s_{2}^{(k)}-\omega_{k}s_{1}^{(k)}\frac{\Delta t}{2}\\
s_{x}^{(k)} & \leftarrow a_{k}s_{x}^{(k)}+b_{k}\xi_{x}^{(k)}\;,\; x=1,2
\end{aligned}
\label{eq:algo}
\end{equation}
\end{widetext}
It is essential that the equations above are executed in the given
order,\cite{Bussi07} as the accuracy and domain of applicability 
of the algorithm depend strongly on the ordering.\cite{Leimkuhler13a,Leimkuhler13b}
By iterating over the single time-step propagation defined
by Eq. (\ref{eq:algo}), it is possible to efficiently integrate our
original set of Eqs. (\ref{eq:EoM-physical-cvariable-final})-(\ref{eq:EoM-s2-final}).
Performing such simulations $W$ times, one obtains $W$ trajectories
$\left(\mathbf{r}^{w}\left(t\right),\mathbf{p}^{w}\left(t\right),{\bf s}_{1}^{w}\left(t\right),{\bf s}_{2}^{w}\left(t\right)\right)$
in the extended phase space, $w=1,\dots,W$. The evolution of any
physical observable $A\left({\bf r},{\bf p},t\right)$ is then
retrieved by taking the ensemble average 
\begin{equation}
\left\langle A\right\rangle \left(t\right)\equiv\frac{1}{W}\sum_{w}A\left({\bf r}^{w},{\bf p}^{w},t\right)
\label{eq:ave}
\end{equation}
As the observable $A$ does not depend of the auxiliary DoFs, while
the trajectories do, in the ensemble average defined above the auxiliary
DoFs are effectively traced out.

We finally note that the propagation algorithm used in this work
\cite{Bussi07} 
provides very accurate numerical averages of velocity depending functions, e.g. the velocity
autocorrelation function studied in Sec. \ref{sub:Analytical-solution} and Sec. \ref{sub:numerical}. 
However, even more accurate algorithms can be used if configurational averages need to be evaluated.
\cite{Leimkuhler13b}

\section{A single harmonic impurity in a Debye bath\label{sec:harmonic}}

The main objective of this paper is to demonstrate an efficient numerical
algorithm for solving the GLE defined in Eq. (\ref{eq:gle}) with
generic memory kernel and stochastic forces corresponding to a colored
noise. To this end, we need a simple, yet realistic, model of the
bath dynamics for which an analytic expression for the memory kernel
is available. This requirement is indeed crucial for a convincing
validation of the algorithm introduced in Sec. \ref{sub:integration}.
Therefore, we assume the bath to be a crystalline solid with the lattice
vectors $\mathbf{l}$. 

\subsection{The Debye bath\label{sub:debye_bath}}

For the sake of simplicity, we carry out the calculation for a 3D
cubic lattice, although the following ideas can be applied to non-orthogonal
lattices and low dimensional solids as well. In addition, we assume
there is a single atom of mass $\bar{\mu}$ in the unit cell. Then,
the vibration eigenproblem, Eq. (\ref{eq:vibr-problem}), is solved
analytically yielding eigenvectors $v_{{\bf l}\gamma}^{\left(\lambda{\bf q}\right)}=\delta_{\lambda\gamma}e^{i\mathbf{q}\cdot\mathbf{l}}/\sqrt{N_{l}}$,
where $\mathbf{q}$ is a vector in the Brillouin zone (BZ), $N_{l}$
the total number of $\mathbf{q}$ vectors, and $\mathbf{e}_{\gamma}^{(\lambda)}=\left(\delta_{\lambda\gamma}\right)$
are the three Cartesian vectors for the three acoustic branches labeled
by $\lambda$. 

\begin{figure}
\begin{centering}
\includegraphics[width=8cm]{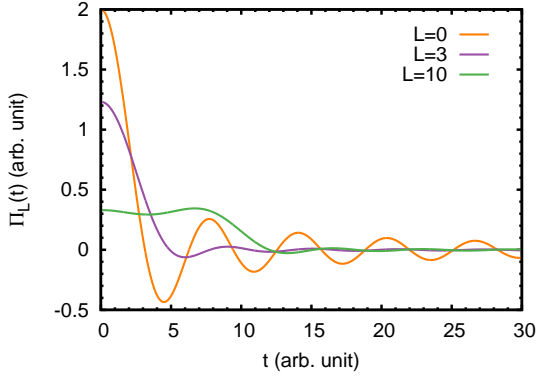}
\par\end{centering}
\caption{The polarization matrix of Eq. (\ref{eq:pol-function-debye-2}) as
a function of time $t$ for three values of $L=\left|\mathbf{l}-\mathbf{l}^{\prime}\right|$.
We used $\omega_{D}=v=1$ and $v_{c}=4\pi^{2}$. \label{fig:The-polarisation-function}}
\end{figure}

To provide an analytical expression for the memory kernel, it is convenient
to consider a Debye model in which the vibration frequencies depend
linearly on the modulus of the corresponding Brillouin vector, i.e.,
$\omega_{\mathbf{q}}=c\left|\mathbf{q}\right|$. In this case, the
bath polarization matrix in Eq. (\ref{eq:pol_ini}) reads
\begin{equation}
\Pi_{\mathbf{l}\gamma,\mathbf{l}^{\prime}\gamma^{\prime}}\left(t-t^{\prime}\right)=\delta_{\gamma\gamma^{\prime}}\frac{1}{N_{l}}\sum_{\mathbf{q}}\frac{e^{i\mathbf{q}\cdot\left(\mathbf{l}-\mathbf{l}^{\prime}\right)}}{\omega_{\mathbf{q}}^{2}}\cos\left(\omega_{\mathbf{q}}\left(t-t^{\prime}\right)\right)
\label{eq:stationary_part}
\end{equation}
In the thermodynamic limit, the sum in Eq. (\ref{eq:stationary_part})
can be replaced by an integral over a sphere of radius $q_{D}=\omega_{D}/c$,
where $\omega_{D}$ is the Debye frequency, and then
\begin{multline}
\Pi_{\mathbf{l}\gamma,\mathbf{l}^{\prime}\gamma^{\prime}}\left(t-t^{\prime}\right)=\delta_{\gamma\gamma^{\prime}}\frac{v_{c}}{\left(2\pi\right)^{3}}\int_{0}^{q_{D}}\cos\left(cq\left(t-t^{\prime}\right)\right)q^{2}\mbox{d}q\\
\int_{0}^{\pi}\frac{e^{iq\left|\mathbf{l}-\mathbf{l}^{\prime}\right|\cos\theta}}{c^{2}q^{2}}2\pi\sin\theta \mbox{d}\theta=\delta_{\gamma\gamma^{\prime}}\Pi_{\mathbf{l}-\mathbf{l}^{\prime}}(t-t^{\prime})
\label{eq:polar-mattix-debye}
\end{multline}
with the reduced bath polarization matrix defined as
\begin{widetext}
\begin{equation}
\Pi_{\mathbf{l}-\mathbf{l}^{\prime}}\left(t-t^{\prime}\right)=\frac{v_{c}}{4\pi^{2}c^{2}\left|\mathbf{l}-\mathbf{l}^{\prime}\right|}\left[\mbox{Si}\left(\omega_{D}\left(\left|t-t^{\prime}\right|+\frac{\left|\mathbf{l}-\mathbf{l}^{\prime}\right|}{c}\right)\right)-\mbox{Si}\left(\omega_{D}\left(\left|t-t^{\prime}\right|-\frac{\left|\mathbf{l}-\mathbf{l}^{\prime}\right|}{c}\right)\right)\right]
\label{eq:pol-function-debye-2}
\end{equation}
\end{widetext}
In Eq. (\ref{eq:pol-function-debye-2}), the function $\mbox{Si}(x)=\int_{0}^{x}\frac{\sin\left( x^{\prime}\right)}{x^{\prime}}\mbox{d}x^{\prime}$
is the integral sine function and $v_{c}$ is the volume of the unit
cell. The reduced polarization matrix, $\Pi_{\mathbf{l}-{\bf l^{\prime}}}\left(t-t^{\prime}\right)$,
demonstrates an oscillating character eventually decaying to zero
at the limit of $\left|t-t^{\prime}\right|\rightarrow\infty$ as is
shown in Fig. \ref{fig:The-polarisation-function}. This is the kind
of behavior which can be approximated by the expansion type of Eq.
(\ref{eq:pol}) by an appropriate choice of the free parameters.
In particular, for $\mathbf{l}=\mathbf{l}^{\prime}$, the bath polarization
matrix does not depend on the lattice vectors and is given by:
\begin{equation}
\begin{aligned}
\Pi_{\mathbf{l}\gamma,\mathbf{l}\gamma}\left(t-t^{\prime}\right)&=\frac{v_{c}}{2\pi c^{3}}\frac{\sin\left(\omega_{D}\left(t-t^{\prime}\right)\right)}{\pi\left(t-t^{\prime}\right)}\\
&=\frac{3\pi}{\omega_{D}^{3}}\frac{\sin\left(\omega_{D}\left(t-t^{\prime}\right)\right)}{\pi\left(t-t^{\prime}\right)}
\label{eq:diagoanl-pol-matrix-debye}
\end{aligned}
\end{equation}
In Eq. (\ref{eq:diagoanl-pol-matrix-debye}) we have made use of the
identity $\omega_{D}^{3}v_{c}=6\pi^{2}c^{3}$. Note that the polarization
matrix decays to zero when $\left|t-t^{\prime}\right|\rightarrow\infty$
and is an even function of its time argument, as required by Eq. (\ref{eq:pol_ini}).
By substituting Eq. (\ref{eq:diagoanl-pol-matrix-debye}) into Eq.
(\ref{eq:kernel-ini}), the following memory kernel for the Debye
model is finally obtained:
\begin{widetext}
\begin{equation}
K_{i\alpha,i^{\prime}\alpha^{\prime}}\left(t,t^{\prime};\mathbf{r}\right)=\bar{\mu}\sum_{\mathbf{l}\gamma}g_{i\alpha,{\bf l}\gamma}\left(\mathbf{r}\left(t\right)\right)\frac{3\pi}{\omega_{D}^{3}}\left[\frac{\sin\left(\omega_{D}\left(t-t^{\prime}\right)\right)}{\pi\left(t-t^{\prime}\right)}\right]g_{i^{\prime}\alpha^{\prime},{\bf l}\gamma}\left(\mathbf{r}\left(t^{\prime}\right)\right)
\label{eq:kernel-debye1}
\end{equation}
\end{widetext}

Without compromising the validation of the algorithm, we can still
devise an interesting test case (see Sec. \ref{sub:Analytical-solution})
by confining our attention to a model containing one atom moving along
a single Cartesian coordinate (say $z$) near the zero lattice site,
$\mathbf{l}=\mathbf{0}$. Assuming the atom-bath interaction to be
short-ranged, only the nearest-neighbor interactions must be included.
Finally, to simplify the model even further, we can adopt an approximation
$g_{i\alpha,\mathbf{l}\gamma}\left(\mathbf{r}\right)=g_{\mathbf{0}}\delta_{\mathbf{l}\mathbf{0}}\delta_{\alpha z}\delta_{\gamma z}$
in Eq. (\ref{eq:kernel-debye1}) to obtain 
\begin{equation}
K_{zz}\left(t,t^{\prime};\mathbf{r}\right)=\frac{3\pi}{\omega_{D}^{3}}\bar{\mu}g_{\mathbf{0}}^{2}\frac{\sin\left(\omega_{D}\left(t-t^{\prime}\right)\right)}{\pi\left(t-t^{\prime}\right)}
\label{eq:kernel-debye2}
\end{equation}
Hereafter, we will refer to any bath whose memory kernel can be expressed
as in Eq. (\ref{eq:kernel-debye2}) as a \emph{Debye bath}. Note that
in our actual calculations described below the factor $g_{\mathbf{0}}$
is a constant and does not depend on the atom position. 

In the following, we assume that one atomic impurity is coupled to the Debye bath. 
We also assume that, in the limit of vanishingly small coupling with the bath, 
this impurity can be modeled as a DoF with mass $\bar{\mu}$ 
subject to the harmonic potential $V(z)=\bar{\mu}\bar{\omega}_0^2z^2/2$.
Within the same model and according to the polaronic effect defined in Eq. (\ref{eq:potential_pol}), 
the coupling to the Debye bath causes a softening of this harmonic potential.
In particular, by substituting Eq. (\ref{eq:diagoanl-pol-matrix-debye}) 
and $f_{\mathbf{l}\gamma}(\mathbf{r})=g_{i\alpha,\mathbf{l}\gamma}\left(\mathbf{r}\right)z=g_{\mathbf{0}}\delta_{\mathbf{l0}}\delta_{\alpha z}\delta_{\gamma z}z$ 
into Eq. (\ref{eq:potential_pol}), one can write
\begin{equation}
\bar{V}\left(\mathbf{r}\right)=V\left(\mathbf{r}\right)-\frac{1}{2}\left(\frac{3\bar{\mu}g_{\mathbf{0}}^{2}}{\omega_{D}^{2}}\right)z^{2}=\frac{1}{2}\bar{\mu}\bar{\omega}_p^2z^2\;,\label{eq:potential_pol-1}
\end{equation}
where $\bar{\omega}_p$ is the effective harmonic frequency of the impurity.

As the coupling exceeds the critical value $g_0 = \omega_D \bar{\omega_0}/\sqrt{3}$,
$\bar{\omega}_p$ becomes negative leading to 
an artificial mechanical instability (the impurity ``falls down'' into the bath).\cite{Evstigneev10} 
However, in the next Section we shall see that, even before this critical value is hit,
the very distinction between bath and impurity is lost, as seen, e.g., in the FT of the velocity autocorrelation function 
(see Fig. \ref{fig:bath}).
In particular, for such a strong system-bath coupling, the linear model used in Eq. (\ref{eq:lagrangian-3}) might no longer be applicable and a non-linear
generalization should be considered.\cite{Evstigneev10}

Finally, we consider the limiting case of a Langevin dynamics with
memory kernel as in Eq. (\ref{eq:kernel_inf}). This case can be formally
considered by noticing that in the limit of $\omega_{D}\rightarrow\infty$
the function in the square brackets in the right hand side of Eq.
(\ref{eq:kernel-debye1}) tends to the delta function, so that one
can write: 
\begin{equation}
K_{zz}\left(t,t^{\prime}\right)=2\Gamma_{zz}\delta\left(t-t^{\prime}\right)
\label{eq:markov}
\end{equation}
where
\begin{equation}
\Gamma_{zz}\equiv\frac{3\pi}{2}\frac{\bar{\mu}}{\omega_{D}^{3}}g_{\mathbf{0}}^{2}
\label{eq:friction}
\end{equation}
As a characteristic ``memory time'' for the memory kernel in Eq.
(\ref{eq:kernel-debye2}) one may choose the time $\pi/2\omega_{D}$
when the memory kernel drops to zero. Therefore, for times $t\gg\pi/2\omega_{D}$
the Debye bath ``bears no memory''. In the limit of $\omega_{D}\rightarrow\infty$
this characteristic time becomes vanishing small, as expected.

\subsection{Analytic solution\label{sub:Analytical-solution}}

To test the integration algorithm explained in Sec. \ref{sub:integration},
we consider the following simple model in which a harmonic oscillator
is coupled to a Debye bath:
\begin{equation}
\bar{\mu}\ddot{r}=-\bar{\mu}\bar{\omega}_{p}^{2}r-\int_{-\infty}^{\infty}\widetilde{K}_{zz}\left(t-t^{\prime}\right)\dot{r}\left(t^{\prime}\right)\mbox{d}t^{\prime}+\eta_{1}(t)
\label{eq:EOM_harm}
\end{equation}
where the causal memory kernel $\widetilde{K}_{zz}\left(t-t^{\prime}\right)=\theta\left(t-t^{\prime}\right)K_{zz}\left(t-t^{\prime}\right)$
has been employed, $K_{zz}\left(t-t^{\prime}\right)$ is defined in
Eq. (\ref{eq:kernel-debye2}), and $\bar{\omega}_{p}$ is the frequency
of the harmonic oscillator reduced from its natural frequency $\bar{\omega}_{0}$
by the polaronic effect, see Eq. (\ref{eq:potential_pol-1}). The
FT of the memory kernel $K_{zz}\left(t-t^{\prime}\right)$ is calculated
easily as
\begin{equation}
K_{zz}\left(\omega\right)=\frac{3\pi\bar{\mu}g_{\mathbf{0}}^{2}}{\omega_{D}^{3}}\chi_{D}\left(\omega\right)=2\Gamma_{zz}\chi_{D}\left(\omega\right)
\label{eq:kernel_harm_ft}
\end{equation}
where the characteristic function, $\chi_{D}\left(\omega\right)$,
is defined so that $\chi_{D}(\omega)=1$ when $\omega\in\left[-\omega_{D},\omega_{D}\right]$
and zero otherwise, and $\Gamma_{zz}$ has been defined in Eq. (\ref{eq:friction}).
The FT of the causal memory kernel, $\widetilde{K}_{zz}\left(\omega\right)=\widetilde{K}_{1}\left(\omega\right)+i\widetilde{K}_{2}\left(\omega\right)$,
is calculated first by noticing that $K_{zz}\left(\omega\right)=2\mbox{Re}\left[\widetilde{K}_{zz}\left(\omega\right)\right]\equiv2\widetilde{K}_{1}\left(\omega\right)$
and then using the Kramers-Kronig relation to calculate its imaginary
part, $\widetilde{K}_{2}\left(\omega\right)$. By introducing the
bath ``self-energy'', 
\begin{equation}
\Sigma\left(\omega\right)=\frac{i}{\bar{\mu}}\widetilde{K}_{zz}\left(\omega\right)=\Sigma_{1}\left(\omega\right)+i\Sigma_{2}\left(\omega\right)
\label{eq:kernel2self}
\end{equation}
one obtains for it on the upper side of the real $\omega-$axis: 
\begin{equation}
\Sigma\left(\omega\right)=\frac{\Gamma_{zz}}{\pi\bar{\mu}}\left[\ln\left|\frac{\omega-\omega_{D}}{\omega+\omega_{D}}\right|+i\pi\chi_{D}\left(\omega\right)\right]
\label{eq:self_exact}
\end{equation}
so that $\Sigma_{2}\left(\omega\right)=\left(\Gamma_{zz}/\bar{\mu}\right)\chi_{D}\left(\omega\right)$,
while the expression 
\begin{equation}
\Sigma\left(\omega\right)=\frac{\Gamma_{zz}}{\pi\bar{\mu}}\ln\left(\frac{\omega-\omega_{D}}{\omega+\omega_{D}}\right)
\label{eq:self_exact-all-half-plane}
\end{equation}
is valid in the whole complex plane (a branch cut on the real axis
over the interval $\left[-\omega_{D},\omega_{D}\right]$ is assumed).
For $\left|\omega\right|>\omega_{D}$ the imaginary part for the self
energy is zero: $\Sigma_{2}\left(\omega\right)=0$. Note that in the
Markovian limit, $\omega_{D}\to\infty$, the self-energy becomes $\Sigma\left(\omega\right)=i\Gamma_{zz}/\bar{\mu}$,
as expected from Eqs. (\ref{eq:markov}) and (\ref{eq:kernel2self}).

The FT of the solution of Eq. (\ref{eq:EOM_harm}) reads:
\begin{equation}
\begin{aligned}
r\left(\omega\right)&=\bar{r}\left(\omega\right)+\frac{1}{\bar{\mu}}\frac{\eta_{1}\left(\omega\right)}{\bar{\omega}_{p}^{2}-\omega^{2}+\omega\Sigma\left(\omega\right)}\\
&=\bar{r}\left(\omega\right)+G\left(\omega\right)\eta_{1}\left(\omega\right)
\end{aligned}
\label{eq:EOM_harm_ft}
\end{equation}
where $\bar{r}\left(\omega\right)$ is a solution of the homogeneous
equation
\begin{equation}
\left[\bar{\omega}{}_{p}^{2}-\omega^{2}+\omega\Sigma(\omega)\right]\bar{r}\left(\omega\right)=0
\label{eq:homogenous}
\end{equation}
and $G(\omega)$ is the FT of the Green's function satisfying the
equation
\begin{equation}
\left[\bar{\omega}_{p}^{2}-\omega^{2}+\omega\Sigma(\omega)\right]G\left(\omega\right)=1\;.
\label{eq:inhomogenous}
\end{equation}
Eq. (\ref{eq:inhomogenous}) corresponds to the FT of Eq. (\ref{eq:EOM_harm})
in which the noise $\eta_{1}(t)$ has been replaced by the Dirac delta
function $\delta(t)$. 

To compute $\bar{r}\left(\omega\right)$, we use an exponential \emph{ansatz},
$r(t)\sim e^{i\bar{\omega}t}$, where $\bar{\omega}$ is a \emph{real}
frequency satisfying the equation:
\begin{equation}
\bar{\omega}_{p}^{2}-\bar{\omega}^{2}+\bar{\omega}\Sigma(\bar{\omega})=0
\label{eq:zeros}
\end{equation}
If such a solution exists, it yields persistent oscillations which
cannot be neglected as a transient phenomena. It can easily be seen
that if $\bar{\omega}$ is a root of this equation, then $-\bar{\omega}$
is also a root, i.e., the roots come in pairs $\pm\bar{\omega}$.
In addition, real roots of Eq. (\ref{eq:zeros}) are possible only
if $\left|\bar{\omega}\right|>\omega_{D}$, i.e., when $\Sigma_{2}(\bar{\omega})=0$.
Since the exponential solutions can be written in terms of delta functions
in the Fourier space, we can finally write:
\begin{multline}
r\left(\omega\right)=\sum_{j}\left[C_{j}\delta\left(\omega-\bar{\omega}_{j}\right)+C_{j}^{*}\delta\left(\omega+\bar{\omega}_{j}\right)\right]\\
+G(\omega)\eta_{1}(\omega)
\label{eq:qw}
\end{multline}
where $\pm\bar{\omega}_{j}$ are the roots of Eq. (\ref{eq:zeros})
and the arbitrary constants $C_{j}$ and $C_{j}^{*}$ are chosen to
satisfy the initial conditions of the problem. In the time domain
we obtain by taking the inverse FT of the expression in Eq. (\ref{eq:qw}):
\begin{equation*}
r(t)=2\sum_{j}\mbox{Re}\left[C_{j}e^{i\bar{\omega}_{j}t}\right]+\int_{-\infty}^{\infty}G\left(t-t^{\prime}\right)\eta_{1}\left(t^{\prime}\right)\mbox{d}t^{\prime}
\end{equation*}
Note that the solution of the homogeneous problem, $\bar{r}\left(t\right)=2\sum_{j}\mbox{Re}\left[C_{j}e^{i\bar{\omega}_{j}t}\right]$,
indeed describes persistent (i.e., undamped) oscillations of the system.
By graphical and numerical methods, one can find that there is just
one pair of roots, $\pm\bar{\omega}_{1}$, which satisfy the constraint
$\left|\bar{\omega}_{1}\right|>\omega_{D}$ for such a Debye bath,
i.e., there may only be one term in the sum over $j$:
\begin{equation}
r(t)=2C_{1}\cos\left(\bar{\omega}_{1}t\right)+\int_{-\infty}^{\infty}G\left(t-t^{\prime}\right)\eta_{1}\left(t^{\prime}\right)\mbox{d}t^{\prime}
\label{eq:r_sol}
\end{equation}

Persistent oscillations also appear in the velocity autocorrelation
function. From Eq. (\ref{eq:EOM_harm}) one obtains the equation satisfied
by the FT of the velocity $v\left(t\right)=\dot{r}\left(t\right)$,
namely
\begin{equation*}
\left[\bar{\omega}{}_{p}^{2}-\omega^{2}+\omega\Sigma(\omega)\right]v\left(\omega\right)=\frac{i\omega}{\bar{\mu}}\eta_{1}\left(\omega\right)
\end{equation*}
and then the equation satisfied by its square modulus
\begin{equation}
\left|\bar{\omega}{}_{p}^{2}-\omega^{2}+\omega\Sigma(\omega)\right|^{2}\left|v\left(\omega\right)\right|^{2}=\frac{\omega^{2}}{\bar{\mu}^{2}}\left|\eta_{1}\left(\omega\right)\right|^{2}
\label{eq:eom_vv_FT}
\end{equation}
By using the results reported in Appendix \ref{sec:auto} (in particular,
Eq. (\ref{eq:auto_xx_FT3})), one can see that Eq. (\ref{eq:eom_vv_FT})
provides a relation between the FT of the velocity autocorrelation
function, $\Phi_{vv}\left(\omega\right)$, and the FT of the noise
auto-correlation function, $\Phi_{\eta_{1}\eta_{1}}\left(\omega\right)$,
such that
\begin{equation}
\left|\bar{\omega}{}_{p}^{2}-\omega^{2}+\omega\Sigma(\omega)\right|^{2}\Phi_{vv}\left(\omega\right)=\frac{\omega^{2}}{\bar{\mu}^{2}}\Phi_{\eta_{1}\eta_{1}}\left(\omega\right)i
\label{eq:eom_vv_FT2}
\end{equation}
From Eq. (\ref{eq:Fluct-Diss-Theorem}), stating the (second) fluctuation-dissipation
theorem, we also know that $\Phi_{\eta_{1}\eta_{1}}\left(\omega\right)=k_{B}TK_{zz}\left(\omega\right)=2\bar{\mu}k_{B}T\Sigma_{2}\left(\omega\right)$.
Hence, the solution of Eq. (\ref{eq:eom_vv_FT2}) can be written as
\begin{multline}
\Phi_{vv}\left(\omega\right)=W_{1}\left[\delta\left(\omega-\bar{\omega}_{1}\right)+\delta\left(\omega+\bar{\omega}_{1}\right)\right]\\
+\left(\frac{k_{B}T}{\bar{\mu}}\right)\frac{2\omega^{2}\Sigma_{2}\left(\omega\right)}{\left[\bar{\omega}{}_{p}^{2}-\omega^{2}+\omega\Sigma_{1}(\omega)\right]^{2}+\omega^{2}\Sigma_{2}^{2}\left(\omega\right)}
\label{eq:vv_FT}
\end{multline}
where, as in Eq. (\ref{eq:r_sol}), a homogeneous term must also be
included. Note that the frequency of the persistent oscillation in
Eq. (\ref{eq:vv_FT}) is the same as in Eq. (\ref{eq:r_sol}) containing
the real solutions $\pm\bar{\omega}_{1}$ of Eq. (\ref{eq:zeros}).
To determine the (real) constant $W_{1}$, we note that the equipartition
theorem requires that 
\begin{equation*}
\lim_{t^{\prime}\to t}\left\langle v\left(t\right)v\left(t^{\prime}\right)\right\rangle =\frac{k_{B}T}{\bar{\mu}}
\end{equation*}
which yields the required condition for $W_{1}$:
\begin{widetext}
\begin{equation}
\int_{-\infty}^{+\infty}\Phi_{vv}\left(\omega\right)\mbox{d}\omega=2W_{1}+\left(\frac{k_{B}T}{\bar{\mu}}\right)\int_{-\omega_{D}}^{\omega_{D}}\frac{2\omega^{2}\Sigma_{2}\left(\omega\right)\mbox{d}\omega}{\left[\bar{\omega}{}_{p}^{2}-\omega^{2}+\omega\Sigma_{1}(\omega)\right]^{2}+\omega^{2}\Sigma_{2}^{2}\left(\omega\right)}=\frac{k_{B}T}{\bar{\mu}}
\label{eq:condition-for-W1}
\end{equation}
\end{widetext}
since for frequencies outside of the interval $\left(-\omega_{D},\omega_{D}\right)$,
the self-energy is real (i.e. $\Sigma_{2}(\omega)=0$).
\begin{figure}
\begin{centering}
\includegraphics[width=8cm]{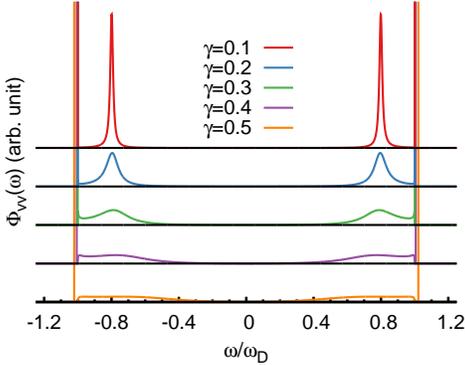}
\end{centering}
\caption{FT of the velocity autocorrelation function, $\Phi_{vv}\left(\omega\right)$
of Eq. (\ref{eq:vv_FT}), from GLE dynamics with a Debye bath (see
Eq. (\ref{eq:EOM_harm})) for different system-bath coupling strengths.
The natural frequency of the system is $\bar{\omega}_{0}=0.8\omega_{D}$
and the coupling $g_{\mathbf{0}}=\gamma\bar{\omega}_{0}^{2}$. The
contributions from the delta functions in Eq. (\ref{eq:vv_FT}) are
represented by vertical spikes.\label{fig:bath}}
\end{figure}

In Fig. \ref{fig:bath} we plot the FT of the velocity autocorrelation
function from the GLE dynamics defined in Eq. (\ref{eq:vv_FT}), having
set $\bar{\omega}_{0}=0.8\omega_{D}$ and $g_{\mathbf{0}}=\gamma\bar{\omega}_{0}^{2}$.
For a very weak coupling, i.e., $\gamma=0.1$, $\Phi_{vv}\left(\omega\right)$
presents two symmetric resonances centered at $\omega=\pm\bar{\omega}_{res}\approx\pm\bar{\omega}_{0}$
inside the interval $\omega\in\left(-\omega_{D},\omega_{D}\right)$
(see Table \ref{tab:table} for accurate numerical values of $\bar{\omega}_{res}$ obtained
by minimizing $\left|\bar{\omega}_{p}^{2}-\omega^{2}+\omega\Sigma_{1}(\omega)\right|$
for $\omega\in\left(-\omega_{D},\omega_{D}\right)$). The resonance
frequency, $\bar{\omega}_{res}$, decreases as the coupling increases
as a consequence of the polaronic correction discussed above. As the
coupling gets stronger, the two resonances broaden and they loose
their spectral weight as the integral $\int_{-\omega_{D}}^{+\omega_{D}}\Phi_{vv}\left(\omega\right)\mbox{d}\omega$
decreases. At the same time, the complementary spectral contribution
from the delta functions outside the interval $\omega\in\left(-\omega_{D},\omega_{D}\right)$,
i.e., the value of $W_{1}$, increases, as well as the frequency of
the persistent oscillation, $\bar{\omega}_{1}>\omega_{D}$ (see 
Table \ref{tab:table} for accurate numerical values obtained by solving Eq. (\ref{eq:zeros})
with respect to $\bar{\omega}$ with the constraint $\bar{\omega}>\omega_{D}$).

As discussed in the previous Section, a mechanical instability due the polaronic effect is predicted
for $g_0 > \omega_D \bar{\omega_0}/\sqrt{3}$, or $\gamma > 0.7217$ for our choice of the parameters. 
Note, however, that for any given value of $0<\gamma<0.7217$, $\bar{\omega}_{res}>\bar{\omega}_p$, i.e.,
the resonance in $\Phi_{vv}\left(\omega\right)$ is blue shifted with respect to the effective harmonic frequency of the impurity (see Table \ref{tab:table}).
This blue shift is an analogue of the so-called Lamb shift of quantum optics\cite{Petruccione} and does not appear in ordinary, i.e., Markovian,
Langevin dynamics\cite{Evstigneev10} for which $\Sigma_1(\omega)=0$ (see Eq. (\ref{eq:vv_harm_ft}) 
below and the discussion after Eq. (\ref{eq:self_exact-all-half-plane})). 

In practice, the blue shift caused by the real part of the bath ``self-energy'', $\Sigma_1(\omega)$, results in a slower convergence
of $\bar{\omega}_{res}$ to zero as $\gamma$ approaches the critical value for mechanical instability.
In other words, the interaction with the bath counteracts the polaronic effect so that, e.g.,
for $\gamma=0.5$, the renormalized harmonic frequency, $\bar{\omega}_{res}$ in Table \ref{tab:table}, is still noticeably larger 
than $\bar{\omega}_p$. Hence, although the analogue of the Lamb shift does not prevent an artificial mechanical instability 
for $\gamma > 0.7217$, within our GLE framework the softening caused by the linear approximation defined in Eq. (\ref{eq:lagrangian-3})
does not seem as severe as previously reported for ordinary, i.e., Markovian, Langevin dynamics.\cite{Evstigneev10}
\begin{table}
\begin{centering}
\begin{tabular}{|c|c|c|c|c|c|c|}
\hline 
$\gamma$ & $0.0$ & $0.1$ & $0.2$ & $0.3$ & $0.4$ & $0.5$\tabularnewline
\hline 
$\bar{\omega}_{p}/\omega_{D}$  & $0.8$ & $0.7923$ & $0.7687$ & $0.7276$ & $0.6659$ & $0.5769$\tabularnewline
\hline 
$\bar{\omega}_{res}/\omega_{D}$  & $0.8$ & $0.7990$ & $0.7958$ & $0.7891$ & $0.7751$ & $0.7419$\tabularnewline
\hline 
$\bar{\omega}_{1}/\omega_{D}$  & n/a & $1.0000$ & $1.0000$ & $1.0004$ & $1.0064$ & $1.0218$\tabularnewline
\hline 
$W_{1}/\left(\frac{k_{B}T}{\bar{\mu}}\right)$ & n/a & $0.0000$ & \textbf{$0.0000$} & $0.0071$ & $0.0593$ & $0.1205$\tabularnewline
\hline 
\end{tabular}
\end{centering}
\caption{Effective harmonic frequency $\bar{\omega}_p$, renormalized harmonic frequency $\bar{\omega}_{res}$, persistent
oscillation frequency $\bar{\omega}_{1}$, and weight $W_{1}$ (see
text) for several values of the dimensionless system-bath coupling,
$\gamma=g_{\mathbf{0}}/\bar{\omega}_{0}^{2}$.\label{tab:table}}
\end{table}

In the Markovian limit, $\omega_{D}\to\infty$, Eq. (\ref{eq:vv_FT})
simplifies to
\begin{equation}
\Phi_{vv}\left(\omega\right)=\left(\frac{k_{B}T}{\bar{\mu}}\right)\frac{2\bar{\mu}\omega^{2}\Gamma_{zz}}{\bar{\mu}^{2}\left(\bar{\omega}{}_{p}^{2}-\omega^{2}\right)^{2}+\omega^{2}\Gamma_{zz}^{2}}
\label{eq:vv_harm_ft}
\end{equation}
Note there are no solutions of Eq. (\ref{eq:zeros}) in this case
as $\Sigma_{2}=\Gamma_{zz}/\bar{\mu}\neq0$ everywhere on the whole
real axis and therefore there are no real solutions of Eq. (\ref{eq:zeros}),
i.e., persistent oscillations do not exist in the Markovian limit.
Taking the inverse FT from the $\Phi_{vv}\left(\omega\right)$ in
this case (the integration is most easily performed in the complex
plane) and after some tedious computations, one can work out analytically
the velocity autocorrelation function in the time domain as:
\begin{widetext}
\begin{equation}
\left\langle v\left(t\right)v\left(t^{\prime}\right)\right\rangle =\frac{k_{B}T}{\bar{\mu}}\begin{cases}
\left[\cos\left(\sqrt{D}\left(t-t^{\prime}\right)\right)-\frac{\sigma}{\sqrt{D}}\sin\left(\sqrt{D}\left|t-t^{\prime}\right|\right)\right]e^{-\sigma\left|t-t^{\prime}\right|} & {\rm if}\;\Gamma_{zz}<2\bar{\mu}\bar{\omega}_{p}\\
\left[\cosh\left(\sqrt{-D}\left(t-t^{\prime}\right)\right)-\frac{\sigma}{\sqrt{-D}}\sinh\left(\sqrt{-D}\left|t-t^{\prime}\right|\right)\right]e^{-\sigma\left|t-t^{\prime}\right|} & {\rm if}\;\Gamma_{zz}\ge2\bar{\mu}\bar{\omega}_{p}
\end{cases}
\label{eq:markov2}
\end{equation}
\end{widetext}
where, $D=\bar{\omega}_{p}^{2}-\left(\Gamma_{zz}/2\bar{\mu}\right)^{2}$
and $\sigma=\Gamma_{zz}/2\bar{\mu}$. Note that the equipartition
theorem is satisfied in both cases as $\left\langle v^{2}\left(t\right)\right\rangle =k_{B}T/\bar{\mu}$.
In the overdamped limit, when $\Gamma_{zz}\gg\bar{\mu}\bar{\omega}_{p}$,
the well-known Brownian motion result is also correctly retrieved:
\begin{equation*}
\left\langle v\left(t\right)v\left(t^{\prime}\right)\right\rangle \simeq\left(\frac{k_{B}T}{\bar{\mu}}\right)e^{-\Gamma_{zz}\left|t-t^{\prime}\right|/\bar{\mu}}
\end{equation*}

\subsection{Approximation of the memory kernel\label{sub:Approximation-to-memory}}

To perform MD simulations of GLE (\ref{eq:EOM_harm}), we map the
GLE into a set of complex Langevin equations, see Eq. (\ref{eq:algo}),
by introducing $K+1$ pairs of auxiliary DoFs, $s_{1}^{(k)}$ and
$s_{2}^{(k)}$, where $k=0,1,2,\ldots,K$. For this complex Langevin
dynamics to provide a faithful approximation to the actual GLE dynamics,
we have to make sure that our model of the polarization matrix in
Eq. (\ref{eq:pol}) faithfully approximates the actual polarization
matrix, Eq. (\ref{eq:diagoanl-pol-matrix-debye}). In other words,
we want the two functions of time to be approximately equal:
\begin{equation}
\frac{3\pi}{\omega_{D}^{3}}\frac{\sin\left(\omega_{D}\left(t-t^{\prime}\right)\right)}{\pi\left(t-t^{\prime}\right)}\approx\sum_{k=0}^{K}c_{k}^{2}e^{-\left|t-t^{\prime}\right|/\tau_{k}}\cos(\omega_{k}(t-t^{\prime}))
\label{eq:approximation}
\end{equation}
The nature of this approximation is better appreciated by comparing
the FT of both sides:
\begin{multline}
\frac{3\pi}{\omega_{D}^{3}}\chi_{D}\left(\omega\right)\approx\sum_{k=0}^{K}c_{k}^{2}\left[\frac{\tau_{k}}{1+(\omega-\omega_{k})^{2}\tau_{k}^{2}}\right.\\
\left. +\frac{\tau_{k}}{1+(\omega+\omega_{k})^{2}\tau_{k}^{2}}\right]
\label{eq:approximation_ft}
\end{multline}
As can be seen from Eq. (\ref{eq:approximation_ft}), the characteristic
function $\chi_{D}(\omega)$ is approximated as a weighted sum of
at most $K+1$ pairs of independent Lorentzian distributions centered
symmetrically about $\omega=\pm\omega_{k}$ and with the width at
half-height equal to $2/\tau_{k}$. In practice, a least square regression
\cite{Ceriotti2010a}
can be used to find an optimal set of parameters $\tau_{k}$, $\omega_{k}$,
and $c_{k}$ to provide the best approximation. Here we prefer a more
transparent analytic approximation which has the advantage to converge
as $K\to\infty$ (see Sec. \ref{sec:harmonic}).

In our method, the characteristic frequencies in Eqs. (\ref{eq:approximation})
and (\ref{eq:approximation_ft}) are chosen as $\omega_{k}=k\left(\omega_{D}/K\right)$
where $k=0,1,2,\ldots,K$. In this way, Eq. (\ref{eq:approximation_ft})
can be written as
\begin{widetext}
\begin{equation}
\frac{3\pi}{\omega_{D}^{3}}\chi_{D}\left(\omega\right)\approx\sum_{k=1}^{K}c_{k}^{2}\left[\frac{\tau_{k}}{1+(\omega-\omega_{k})^{2}\tau_{k}^{2}}+\frac{\tau_{k}}{1+(\omega+\omega_{k})^{2}\tau_{k}^{2}}\right]+c_{0}^{2}\frac{2\tau_{0}}{1+\omega^{2}\tau_{0}^{2}}
\label{eq:approximation_ft2}
\end{equation}
\end{widetext}
where in the right hand side of Eq. (\ref{eq:approximation_ft2})
we have discriminated between the cases $k\neq0$ and $k=0$ in the
original summation. Note that in the last case the pair of Lorentzian
is degenerate, i.e., they coincide. Eq. (\ref{eq:approximation_ft2})
gives a weighted expansion of the FT of the polarization matrix in
terms of equally spaced (over the frequency interval $\omega\in\left[-\omega_{D},\omega_{D}\right]$)
Lorentzians. To have an uniform expansion, we also require the Lorentzians
to have the same width, i.e., $\tau_{k}=\tau$, and to be equally
weighted, i.e., $c_{k}=c$ for $k>1$ and $c_{0}=c/\sqrt{2}$.

Finally, to fix the parameters $c$ and $\tau$, we require that:
(i) the left and right hand sides of Eq. (\ref{eq:approximation})
are strictly equal for $t=t^{\prime}$; (ii) the left and right hand
sides of Eq. (\ref{eq:approximation_ft2}) are strictly equal for
$\omega=0$. In practice, these two conditions correspond to the short
and long time behaviors of the bath polarization matrix, respectively.
It is easy to see that these requirements are satisfied by choosing
$\tau=\lambda(2K+1)/2\omega_{D}$ and $c=\sqrt{6/(2K+1)}/\omega_{D}$,
where the dimensionless constant $\lambda$ is determined self-consistently
from
\begin{equation*}
\lambda=\pi\left(1+2\sum_{k=1}^{K}\frac{1}{1+k^{2}\lambda^{2}\left(1+\frac{1}{2K}\right)^{2}}\right)^{-1}
\end{equation*}

It is worth noting that, after fixing $K$, in the Markovian limit
$\omega_{D}\to\infty$ we have that $\tau\to0$ , i.e., the characteristic
time of the polarization matrix over which it is greater than zero
is tending to zero, i.e., the polarization matrix ``bears no memory''
as explained at the end of Sec. \ref{sub:debye_bath}.

\subsection{\textcolor{black}{Numerical results}}\label{sub:numerical}

In this Section we present the results of our MD simulations of the
GLE equation (\ref{eq:EOM_harm}) describing a single harmonic oscillator
embedded in the Debye bath. Using the general theory described in
Sections \ref{sub:Mapping-of-GLE}-\ref{sub:integration}, $K+1$
pairs of the auxiliary DoFs are introduced with the parameters as
explained in Section \ref{sub:Approximation-to-memory}, which allow
a mapping of the GLE onto a set of white noise Langevin type equations. 

\begin{figure}
\begin{centering}
\includegraphics[width=8cm]{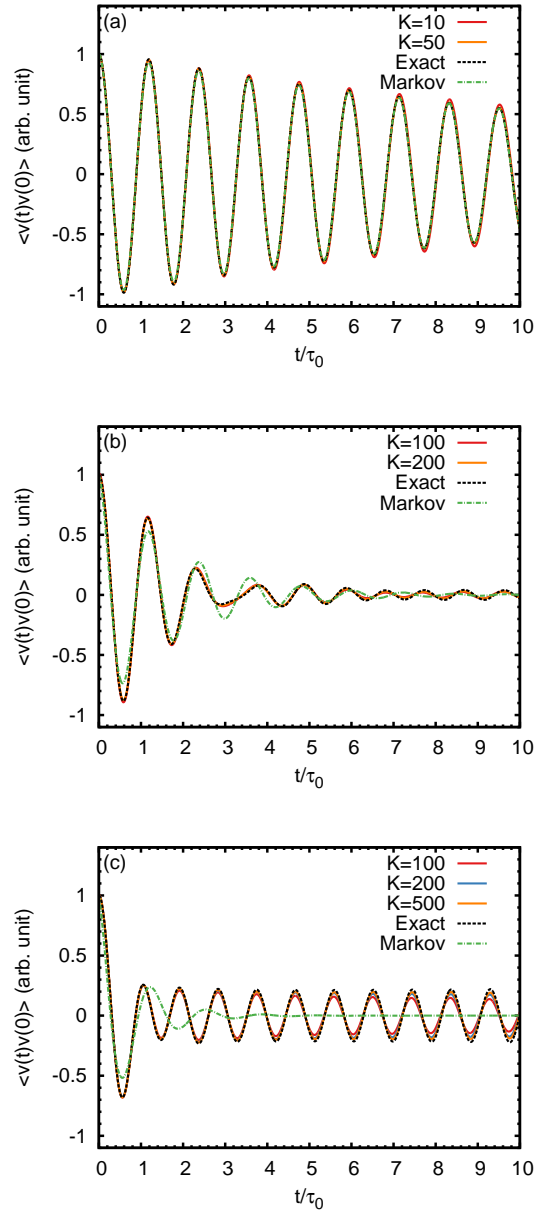}
\par\end{centering}
\caption{Comparison of the (rescaled) velocity autocorrelation functions from
the numerical simulation of Eq. (\ref{eq:EOM_harm}) (solid colored
curves) with the exact results from the inverse FT of Eq. (\ref{eq:vv_FT})
(black dotted curve), and the Markovian limit (green dashed curve)
defined in Eq. (\ref{eq:markov2}). In the Markovian limit instead
of the bare frequency $\bar{\omega}_{p}$, we used the renormalized
frequency, $\bar{\omega}_{res}$, reported in Table \ref{tab:table}.
The integer parameter $K$ sets the accuracy of the numerical approximation,
see Sec. \ref{sub:Approximation-to-memory}. Panels report results
for different system-bath coupling: (a) weak coupling, $\gamma=g_{\mathbf{0}}/\bar{\omega}_{0}^{2}=0.1$;
(b) intermediate coupling, $\gamma=0.3$; (c) strong coupling, $\gamma=0.5$.
\label{fig:results}}
\end{figure}

In Fig. \ref{fig:results} we show the velocity autocorrelation function
obtained by numerically evaluating the GLE dynamics of the harmonic
oscillator. The purpose of these simulations is to demonstrate the
convergence of the numerical algorithm based on the mapping we developed.
The accuracy of our MD simulations is verified by comparing the computed
correlation function with the exact result obtained by the inverse
FT of Eq. (\ref{eq:vv_FT}); we can also compare our correlation function
with the exact prediction of Eq. (\ref{eq:markov2}) in the simple
Markovian limit.

In Fig. \ref{fig:results}(a) we show results for a weak system-bath
coupling, when $\gamma=g_{\mathbf{0}}/\bar{\omega}_{0}^{2}=0.1$.
As the exact velocity correlation function is obtained from the inverse
FT of $\Phi_{vv}\left(\omega\right)$ in Eq. (\ref{eq:vv_FT}), it
is worth recalling that, in the case of $\gamma=0.1$ the function
$\Phi_{vv}\left(\omega\right)$ presents two very strong resonances
in the interval $\omega\in\left[-\omega_{D},\omega_{D}\right]$, see
Fig. \ref{fig:bath}. In addition, the weight $W_{1}$ of the persistent
oscillations at $\omega=\bar{\omega}_{1}\approx\omega_{D}$ is negligible
in this case, see Table \ref{tab:table}. For all these reasons, it
is justified to approximate the FT of the velocity autocorrelation
function with the second term in Eq. (\ref{eq:vv_FT}). This is the
same expression as in Eq. (\ref{eq:vv_harm_ft}) obtained in the Markovian
limit, but with the renormalized harmonic frequency, $\bar{\omega}_{res}$
(for its numerical value, see Table \ref{tab:table}), instead of
the natural one, $\bar{\omega}_{0}$. In fact, using this renormalized
Markovian limit yields a very good agreement with the exact results.
At the same time, the approximate GLE integration algorithm presented
in Sec. \ref{sub:integration} with a limited number of auxiliary
DoFs ($K<50$), which has been supplemented by the analytic fitting
procedure described in Sec. \ref{sub:Approximation-to-memory}, gives
a very good agreement with the exact result as well.

In Fig. \ref{fig:results}(b) we show results for an intermediate
system-bath coupling of $\gamma=0.3$. In this case, the two symmetric
resonances of $\Phi_{vv}\left(\omega\right)$ in the interval $\omega\in\left[-\omega_{D},\omega_{D}\right]$
are rather broadened, see Fig. \ref{fig:bath}. As a consequence,
by using Eq. (\ref{eq:vv_harm_ft}) with the appropriate $\bar{\omega}_{res}$
(see Table \ref{tab:table}), we no longer obtain a good agreement
with the exact velocity autocorrelation function. On the other hand,
our approximate GLE numerical integration still gives an excellent
agreement with the exact result, provided the number of auxiliary
DoFs is large enough, i.e., $K\sim100$.

Finally, in Fig. \ref{fig:results}(c) we show results for a strong
system-bath coupling, i.e. for $\gamma=$0.5. In this case, $\Phi_{vv}\left(\omega\right)$
does not show any resonant features within the interval $\omega\in\left[-\omega_{D},\omega_{D}\right]$,
(see Fig. \ref{fig:bath}) and the weight of the persistent oscillations,
$W_{1}/\left(\frac{k_{B}T}{\bar{\mu}}\right)\approx12\%$ is non-negligible,
see Table \ref{tab:table}. As a consequence, the renormalized Markovian
limit completely fails in the asymptotic limit, i.e., it does not
give persistent oscillations at all. On the other hand, our approximate
GLE numerical integration still provides a convergent approximation
when a sufficient number of auxiliary DoFs is selected.

\section{Discussion and conclusions\label{sec:conclusions}}

In summary, we have devised a very general integration scheme for
conducting GLE dynamics on realistic systems. This scheme considers
two parts of the simulated system: the environment and the real system.
The first step of our algorithm is to calculate the polarization matrix,
see Eq. (\ref{eq:FT-pol}) which does not need to be positive definite.
\cite{Baczewski13}
In principle, in order to do this, one has to conduct a separate simulation
to determine the vibration frequencies of the environment alone, i.e.,
uncoupled from the real system. Then the auxiliary DoFs required by
our integration scheme are determined, e.g., using an analytic approach,
as described in \ref{sub:Approximation-to-memory}. Finally, these
auxiliary DoFs are propagated via our integration scheme, which we
have outlined in Sec. \ref{sub:integration}. Our solution bears many
similarities to the algorithm previously presented by Ceriotti \emph{et
al.}
\cite{Ceriotti2009a,Ceriotti2011,Morrone2011} 
which provides
an optimal thermostat for equilibrium MD simulations. However, the
integration scheme presented in this article conforms to the physical
response of the bath by taking proper consideration of its characteristic
time scales and  is, in principle, better suited for out-of-equilibrium
MD simulations.

We have demonstrated the convergence of our approximate GLE integration
algorithm for the non-trivial case of a single harmonic oscillator
embedded in a Debye bath. In doing so, we have used a simplified representation
of the polarization matrix. In this system, we observed convergence
to the exact velocity autocorrelation function even in the strong
system-bath coupling limit, i.e., when there are no resonant features
in the FT of the velocity autocorrelation function, $\Phi_{vv}\left(\omega\right)$,
and the weight of the persistent oscillations is not negligible. The
reason for such a good agreement, which occurs regardless of the strength
of the system-bath coupling, can be traced back to the specific functional
form of the memory kernel in Eq. (\ref{eq:kernel-ini}). There, the
dependence on the system-bath coupling through the terms $g_{i\alpha,l\gamma}\left(\mathbf{r}\right)$
appears factorized. Hence, one has to fit only the polarization matrix
of the bath, which in fact does not depend on the system-bath coupling
strength. 

Regarding the rate of convergence, it is important to note that the
aim of this work is not to optimize the numerical performance of the
fitting algorithm. However, our analytical approach yields a more
transparent demonstration of the algorithm convergence for the selected
test case. For more realistic systems, we expect a smaller number
of auxiliary DoFs would be needed to achieve convergence by numerically
fitting the polarization matrix in Eq. (\ref{eq:pol}) to the exact
one in Eq. (\ref{eq:diagoanl-pol-matrix-debye}), e.g., by the least
square regression.

\begin{acknowledgments}
LS would like to acknowledge useful conversations with Roberto D'Agosta,
Michele Ceriotti, and Ian Ford, as well as the financial support from
EPSRC, grant EP/J019259/1. 
\end{acknowledgments}

\appendix

\section{Derivation of the Fokker-Plank equation\label{sec:Derivation-of-the-FP}}

It is known
\cite{Gillespie96a,Gillespie96b,Risken}
that the system
of stochastic differential equations 
\begin{equation}
\dot{X}_{a}=h_{a}\left(\mathbf{X},t\right)+\sum_{b}G_{ab}\left(\mathbf{X},t\right)\xi_{b}(t)\;,
\label{eq:gen-stoch-DE}
\end{equation}
for the stochastic variables $\mathbf{X}=\left\{ X_{a}(t)\right\} $,
with $\xi_{a}(t)$ being the Wiener processes defined by 
\begin{equation*}
\left\langle \xi_{a}(t)\right\rangle =0\;,\quad\left\langle \xi_{a}\left(t\right)\xi_{a^{\prime}}\left(t^{\prime}\right)\right\rangle =\delta_{aa^{\prime}}\delta\left(t-t^{\prime}\right)\;,
\end{equation*}
is equivalent to the following Fokker-Plank equation for the probability
distribution function $P\left(\mathbf{X},t\right)$: 
\begin{multline}
\frac{\partial P}{\partial t}\left(\mathbf{X},t\right)=-\sum_{a}\frac{\partial}{\partial X_{a}}\left[\left(h_{a}\left(\mathbf{X},t\right)P\left(\mathbf{X},t\right)\right)\right.\\
\left.-\frac{1}{2}\sum_{b}\frac{\partial}{\partial X_{b}}\left(D_{ab}\left(\mathbf{X},t\right)P\left(\mathbf{X},t\right)\right)\right]
\label{eq:gen-FP-eq}
\end{multline}
where $D_{ab}\left(\mathbf{X},t\right)=\sum_{c}G_{ac}\left(\mathbf{X},t\right)G_{cb}\left(\mathbf{X},t\right)$.

In our case, the set $\mathbf{X}$ is formed by the stochastic variables
$\left\{ r_{i\alpha},p_{i\alpha},s_{1}^{(k)},s_{2}^{(k)}\right\} $.
The quantities $h_{a}\left(\mathbf{X},t\right)$ are given in the
right hand sides of Eq. (\ref{eq:complex_LEs}), excluding the terms
containing the noise, i.e.,
\begin{widetext}
\begin{equation*}
\begin{aligned}
&h_{r_{i\alpha}}=\frac{p_{i\alpha}}{m_{i}}\;,\\
&h_{p_{i\alpha}}=-\frac{\partial\bar{V}}{\partial r_{i\alpha}}+\sum_{l\gamma}\sqrt{\frac{\mu_{l}}{\bar{\mu}}}g_{i\alpha,l\gamma}\left(\mathbf{r}\right)\sum_{k}c_{l\gamma}^{\left(k\right)}s_{1}^{(k)}\;,\\
&h_{s_{1}^{(k)}}=-s_{1}^{(k)}/\tau_{k}+\omega_{k}s_{2}^{(k)}+\sum_{l\gamma}\sqrt{\bar{\mu}\mu_{l}}c_{l\gamma}^{(k)}\sum_{i\alpha}g_{ia,l\gamma}\left(\mathbf{r}(t)\right)\frac{p_{i\alpha}}{m_{i}}\;,\\
&h_{s_{2}^{(k)}}=-s_{2}^{(k)}/\tau_{k}-\omega_{k}s_{1}^{(k)}\;,
\end{aligned}
\end{equation*}
\end{widetext}
while the only non-zero coefficients of $G_{ab}$ are $G_{s_{1}^{(k)},s_{1}^{(k)}}=G_{s_{2}^{(k)},s_{2}^{(k)}}=\sqrt{2k_{B}T\bar{\mu}/\tau_{k}}$.
Since in our case the matrices $G_{ab}\left(\mathbf{X},t\right)$
are constant and diagonal, the matrix $D_{ab}^{(2)}\left(\mathbf{X},t\right)$
is diagonal as well with the only non-zero elements being $D_{s_{1}^{(k)},s_{1}^{(k)}}^{(2)}=D_{s_{2}^{(k)},s_{2}^{(k)}}^{(2)}=2k_{B}T\bar{\mu}/\tau_{k}$.
Substitution of these matrices into Eq. (\ref{eq:gen-FP-eq}) yields
the equations reported in Sec. \ref{sec:fp}.

\section{Equilibrium solution of the Fokker-Plank equation\label{sec:equilibrium}}

In this appendix, we show that $P^{({\rm eq})}\left(\mathbf{r},\mathbf{p},{\bf s}_{1},{\bf s}_{2}\right)$
defined in Eq. (\ref{eq:pdf_eq}) is an equilibrium PDF, i.e., that
\begin{equation*}
P^{({\rm eq})}\left(\mathbf{r},\mathbf{p},{\bf s}_{1},{\bf s}_{2}\right)=\lim_{t\to\infty}P\left(\mathbf{r},\mathbf{p},{\bf s}_{1},{\bf s}_{2},t\right)
\end{equation*}
under the hypothesis that $\hat{\mathfrak{L}}_{cons}P^{({\rm eq})}=0$
and $\hat{\mathfrak{L}}_{diss}P^{({\rm eq})}=0$ hold separately (see
Sec. \ref{sec:fp}). To this end, we proceed by constructing an appropriate
Lyapunov functional.
\cite{Arnold}
Let us take
\begin{equation*}
L\left[P\left(\mathbf{r},\mathbf{p},{\bf s}_{1},{\bf s}_{2}\right)\right]=\int\frac{\left[P\left(\mathbf{r},\mathbf{p},{\bf s}_{1},{\bf s}_{2}\right)-P^{({\rm eq})}\right]^{2}}{P^{({\rm eq})}\left(\mathbf{r},\mathbf{p},{\bf s}_{1},{\bf s}_{2}\right)}\mbox{d}\mathbf{v}\;,
\end{equation*}
where $\mbox{d}\mathbf{v}=\mbox{d}\mathbf{r}\mbox{d}\mathbf{p}\mbox{d}\mathbf{s}_{1}\mbox{d}\mathbf{s}_{2}$, as a candidate functional and show that: (i) $L\left[P\left(\mathbf{r},\mathbf{p},{\bf s}_{1},{\bf s}_{2}\right)\right]\ge L\left[P^{({\rm eq})}\left(\mathbf{r},\mathbf{p},{\bf s}_{1},{\bf s}_{2}\right)\right]$
and (ii) $\frac{d}{dt}L\left[P\left(\mathbf{r},\mathbf{p},{\bf s}_{1},{\bf s}_{2},t\right)\right]<0$
if $P\left(\mathbf{r},\mathbf{p},{\bf s}_{1},{\bf s}_{2},t\right)\neq P^{({\rm eq})}\left(\mathbf{r},\mathbf{p},{\bf s}_{1},{\bf s}_{2}\right)$. 

We first note that
\begin{equation}
L\left[P\left(\mathbf{r},\mathbf{p},{\bf s}_{1},{\bf s}_{2}\right)\right]=\left\Vert \frac{P\left(\mathbf{r},\mathbf{p},{\bf s}_{1},{\bf s}_{2}\right)-P^{({\rm eq})}}{\sqrt{P^{({\rm eq})}\left(\mathbf{r},\mathbf{p},{\bf s}_{1},{\bf s}_{2}\right)}}\right\Vert ^{2}\;,
\label{eq:lyapunov2}
\end{equation}
i.e., the candidate Lyapunov functional corresponds to the square
of the Euclidean distance between the two (square integrable) functions
$P\left(\mathbf{r},\mathbf{p},{\bf s}_{1},{\bf s}_{2}\right)/\sqrt{P^{({\rm eq})}\left(\mathbf{r},\mathbf{p},{\bf s}_{1},{\bf s}_{2}\right)}$
and $\sqrt{P^{({\rm eq})}\left(\mathbf{r},\mathbf{p},{\bf s}_{1},{\bf s}_{2}\right)}.$
Hence, property (i) follows from the properties of the Euclidean norm.
In particular,
\begin{equation*}
L\left[P\left(\mathbf{r},\mathbf{p},{\bf s}_{1},{\bf s}_{2}\right)\right]=0\quad\Leftrightarrow\quad P\left(\mathbf{r},\mathbf{p},{\bf s}_{1},{\bf s}_{2}\right)=P^{({\rm eq})}\;.
\end{equation*}
At this point, we can also define the neighborhood of $P^{({\rm eq})}\left(\mathbf{r},\mathbf{p},{\bf s}_{1},{\bf s}_{2}\right)$
with radius $\varepsilon$ as the set of all the PDFs $P\left(\mathbf{r},\mathbf{p},{\bf s}_{1},{\bf s}_{2}\right)$
such that $L\left[P\left(\mathbf{r},\mathbf{p},{\bf s}_{1},{\bf s}_{2}\right)\right]<\varepsilon$.
Therefore proving property (ii) is the same as proving that the FP
dynamics in Eq. (\ref{eq:fp}) maps a PDF in the neighborhood of
$P^{({\rm eq})}\left(\mathbf{r},\mathbf{p},{\bf s}_{1},{\bf s}_{2}\right)$
with radius $\varepsilon$ to a PDF in the neighborhood of $P^{({\rm eq})}\left(\mathbf{r},\mathbf{p},{\bf s}_{1},{\bf s}_{2}\right)$
of radius $\varepsilon^{\prime}$, with $\varepsilon^{\prime}<\varepsilon$.
In other words, by proving property (ii) we want to show that the
FP dynamics in Eq. (\ref{eq:fp}) provides a contraction and that
$P^{({\rm eq})}\left(\mathbf{r},\mathbf{p},{\bf s}_{1},{\bf s}_{2}\right)$
is the fixed point of this contraction.

Therefore, in the next step, we note that
\begin{equation*}
L\left[e^{t\hat{\mathfrak{L}}_{{\rm cons}}}P\left(\mathbf{r},\mathbf{p},{\bf s}_{1},{\bf s}_{2}\right)\right]=L\left[P\left(\mathbf{r},\mathbf{p},{\bf s}_{1},{\bf s}_{2}\right)\right]
\end{equation*}
as $\hat{\mathfrak{L}}_{{\rm cons}}\sqrt{P^{({\rm eq})}\left(\mathbf{r},\mathbf{p},{\bf s}_{1},{\bf s}_{2}\right)}=0$.
In fact, $e^{t\hat{\mathfrak{L}}_{{\rm cons}}}$ is an isometry, i.e.,
$\left\Vert e^{t\hat{\mathfrak{L}}_{{\rm cons}}}\Psi\left(\mathbf{r},\mathbf{p},{\bf s}_{1},{\bf s}_{2}\right)\right\Vert =\left\Vert \Psi\left(\mathbf{r},\mathbf{p},{\bf s}_{1},{\bf s}_{2}\right)\right\Vert $
for any square integrable $\Psi\left(\mathbf{r},\mathbf{p},{\bf s}_{1},{\bf s}_{2}\right)$,
which leaves the equilibrium solution invariant. One can think of
this isometry as a rotation of the space of  $\Psi\left(\mathbf{r},\mathbf{p},{\bf s}_{1},{\bf s}_{2}\right)$
centered at $\Psi^{({\rm eq})}\left(\mathbf{r},\mathbf{p},{\bf s}_{1},{\bf s}_{2}\right)=\sqrt{P^{({\rm eq})}\left(\mathbf{r},\mathbf{p},{\bf s}_{1},{\bf s}_{2}\right)}$.
Hence, if we define
\begin{equation}
\Psi\left(\mathbf{r},\mathbf{p},{\bf s}_{1},{\bf s}_{2},t\right)=\frac{e^{t\hat{\mathfrak{L}}_{{\rm cons}}}P\left(\mathbf{r},\mathbf{p},{\bf s}_{1},{\bf s}_{2},t\right)}{\sqrt{P^{({\rm eq})}\left(\mathbf{r},\mathbf{p},{\bf s}_{1},{\bf s}_{2}\right)}}\;,
\label{eq:wf_FP}
\end{equation}
we can also rewrite Eq. (\ref{eq:lyapunov2}) as
\begin{equation*}
L\left[P\left(\mathbf{r},\mathbf{p},{\bf s}_{1},{\bf s}_{2},t\right)\right]=\left\Vert \Psi\left(\mathbf{r},\mathbf{p},{\bf s}_{1},{\bf s}_{2},t\right)-\Psi^{({\rm eq})}\right\Vert ^{2}
\label{eq:E2_FP}
\end{equation*}

By taking the time derivative of Eq. (\ref{eq:wf_FP}), we first find that
\begin{widetext}
\begin{equation*}
\dot{\Psi}\left(\mathbf{r},\mathbf{p},{\bf s}_{1},{\bf s}_{2},t\right)=\frac{e^{t\hat{\mathfrak{L}}_{{\rm cons}}}\left[\hat{\mathfrak{L}}_{{\rm cons}}P\left(\mathbf{r},\mathbf{p},{\bf s}_{1},{\bf s}_{2},t\right)\right]+e^{t\hat{\mathfrak{L}}_{{\rm cons}}}\dot{P}\left(\mathbf{r},\mathbf{p},{\bf s}_{1},{\bf s}_{2},t\right)}{\sqrt{P^{({\rm eq})}\left(\mathbf{r},\mathbf{p},{\bf s}_{1},{\bf s}_{2}\right)}}
\end{equation*}
\end{widetext}
and then, by using Eq. (\ref{eq:fp}), that
\begin{equation}
\dot{\Psi}\left(\mathbf{r},\mathbf{p},{\bf s}_{1},{\bf s}_{2},t\right)=-\frac{e^{t\hat{\mathfrak{L}}_{{\rm cons}}}\hat{\mathfrak{L}}_{diss}P\left(\mathbf{r},\mathbf{p},{\bf s}_{1},{\bf s}_{2},t\right)}{\sqrt{P^{({\rm eq})}\left(\mathbf{r},\mathbf{p},{\bf s}_{1},{\bf s}_{2}\right)}}\;.
\label{eq:step}
\end{equation}
Hence, one can use Eq. (\ref{eq:wf_FP}) to write $P\left(\mathbf{r},\mathbf{p},{\bf s}_{1},{\bf s}_{2},t\right)$
as a function of $\Psi\left(\mathbf{r},\mathbf{p},{\bf s}_{1},{\bf s}_{2},t\right)$,
i.e.,
\begin{equation*}
\begin{aligned}
P\left(\mathbf{r},\mathbf{p},{\bf s}_{1},{\bf s}_{2},t\right) & = & e^{-t\hat{\mathfrak{L}}_{{\rm cons}}}\left[\sqrt{P^{({\rm eq})}}\Psi\left(\mathbf{r},\mathbf{p},{\bf s}_{1},{\bf s}_{2},t\right)\right]\\
 & = & \sqrt{P^{({\rm eq})}}e^{-t\hat{\mathfrak{L}}_{{\rm cons}}}\Psi\left(\mathbf{r},\mathbf{p},{\bf s}_{1},{\bf s}_{2},t\right)\;,
\end{aligned}
\end{equation*}
and substitute into Eq. (\ref{eq:step}) to obtain
\begin{equation}
\dot{\Psi}=-\frac{e^{t\hat{\mathfrak{L}}_{{\rm cons}}}\hat{\mathfrak{L}}_{diss}\left[\sqrt{P^{({\rm eq})}}e^{-t\hat{\mathfrak{L}}_{{\rm cons}}}\Psi\right]}{\sqrt{P^{({\rm eq})}}}\;.
\label{eq:step2}
\end{equation}
Due to the peculiar functional form of $P^{({\rm eq})}\left(\mathbf{r},\mathbf{p},{\bf s}_{1},{\bf s}_{2}\right)$,
one can also derive the following equation:
\begin{multline}
\hat{\mathfrak{L}}_{diss}\left[\sqrt{P^{({\rm eq})}}e^{-t\hat{\mathfrak{L}}_{{\rm cons}}}\Psi\left(\mathbf{r},\mathbf{p},{\bf s}_{1},{\bf s}_{2},t\right)\right]\\
=\sqrt{P^{({\rm eq})}}\hat{\mathcal{H}}_{{\rm diss}}e^{-t\hat{\mathfrak{L}}_{{\rm cons}}}\Psi\left(\mathbf{r},\mathbf{p},{\bf s}_{1},{\bf s}_{2},t\right)
\label{eq:step3}
\end{multline}
where
\begin{widetext}
\begin{equation}
\hat{\mathcal{H}}_{{\rm diss}}=\left(2k_{B}T\bar{\mu}\right)\sum_{k}\left\{ -\frac{1}{2}\left[\frac{\partial^{2}}{\partial\left(s_{1}^{(k)}\right)^{2}}+\frac{\partial^{2}}{\partial\left(s_{2}^{(k)}\right)^{2}}\right]+\frac{1}{\left(2k_{B}T\bar{\mu}\right)^{2}}\left[\left(s_{1}^{(k)}\right)^{2}+\left(s_{2}^{(k)}\right)^{2}\right]-\frac{1}{\left(2k_{B}T\bar{\mu}\right)}\right\} \;.
\label{eq:h_diss1}
\end{equation}
\end{widetext}
An effective EoM for $\Psi\left(\mathbf{r},\mathbf{p},{\bf s}_{1},{\bf s}_{2},t\right)$
is eventually found by substituting Eq. (\ref{eq:step3}) into Eq.
(\ref{eq:step2})
\begin{multline}
\dot{\Psi} = -\frac{e^{t\hat{\mathfrak{L}}_{{\rm cons}}}\left[\sqrt{P^{({\rm eq})}}\hat{\mathcal{H}}_{{\rm diss}}e^{-t\hat{\mathfrak{L}}_{{\rm cons}}}\Psi\right]}{\sqrt{P^{({\rm eq})}}}\\
=-e^{t\hat{\mathfrak{L}}_{{\rm cons}}}\hat{\mathcal{H}}_{{\rm diss}}e^{-t\hat{\mathfrak{L}}_{{\rm cons}}}\Psi
\label{eq:schro_imag1}
\end{multline}
Note that the effective Hamiltonian defined in Eq. (\ref{eq:h_diss1})
describes a collection of $2(K+1)$ independent two-dimensional quantum
harmonic oscillators and its spectrum can be easily computed. In particular,
the ``energy'' of the ground state, $\varepsilon_{{\bf 0}}$, turns
out to be exactly zero. 

By using Eq. (\ref{eq:schro_imag1}), one can compute the time derivative
of $L\left[P\left(\mathbf{r},\mathbf{p},{\bf s}_{1},{\bf s}_{2},t\right)\right]$
through Eq. (\ref{eq:E2_FP}) as
\begin{equation}
\frac{d}{dt}L\left[P\left(\mathbf{r},\mathbf{p},{\bf s}_{1},{\bf s}_{2},t\right)\right]=-\left\langle e^{-t\hat{\mathfrak{L}}_{{\rm cons}}}\Psi\left|\hat{\mathcal{H}}_{{\rm diss}}\right|e^{-t\hat{\mathfrak{L}}_{{\rm cons}}}\Psi\right\rangle 
\label{eq:contraction}
\end{equation}
where we have employed the usual inner product of the square integrable
functions. Finally, because of the variational inequality, we have
that
\begin{equation}
\left\langle e^{-t\hat{\mathfrak{L}}_{{\rm cons}}}\Psi\left|\hat{\mathcal{H}}_{{\rm diss}}\right|e^{-t\hat{\mathfrak{L}}_{{\rm cons}}}\Psi\right\rangle \ge\varepsilon_{{\bf 0}}=0
\label{eq:inequality}
\end{equation}
and, by substituting Eq. (\ref{eq:inequality}) into Eq. (\ref{eq:contraction}),
we find that 
\begin{equation}
\frac{d}{dt}L\left[P\left(\mathbf{r},\mathbf{p},{\bf s}_{1},{\bf s}_{2},t\right)\right]\le0
\label{eq:contraction2}
\end{equation}
which proves property (ii). In particular, 
\begin{equation*}
\frac{d}{dt}L\left[P\left(\mathbf{r},\mathbf{p},{\bf s}_{1},{\bf s}_{2},t\right)\right]=0\quad\Leftrightarrow\quad P\left(\mathbf{r},\mathbf{p},{\bf s}_{1},{\bf s}_{2},t\right)=P^{({\rm eq})}\;,
\end{equation*}
as the equality in Eq. (\ref{eq:inequality}) holds just for the (non-degenerate)
ground state of $\hat{\mathcal{H}}_{{\rm diss}}$, i.e., $\Psi\left(\mathbf{r},\mathbf{p},{\bf s}_{1},{\bf s}_{2}\right)=\Psi^{({\rm eq})}\left(\mathbf{r},\mathbf{p},{\bf s}_{1},{\bf s}_{2}\right)=\sqrt{P^{({\rm eq})}\left(\mathbf{r},\mathbf{p},{\bf s}_{1},{\bf s}_{2}\right)}$.
This also proves the uniqueness of the equilibrium solution under
the hypothesis assumed in Sec. \ref{sec:fp}.

\section{Autocorrelation functions\label{sec:auto}}

Let $x\left(t\right)$ be a dynamic observable, e.g., an atomic velocity,
defined in the time interval $t\in\left[-T/2,T/2\right]$. Assuming
the dynamics to be ergodic, one can substitute an ensemble averages
with a time average and then compute the autocorrelation function
of $x\left(t\right)$ as follows
\cite{Allen}: 
\begin{widetext}
\begin{equation}
\left\langle x\left(t\right)x\left(t^{\prime}\right)\right\rangle =\lim_{T\to\infty}\frac{1}{T}\int_{-\infty}^{+\infty}\chi_{T}\left(t+s\right)x\left(t+s\right)\chi_{T}\left(t^{\prime}+s\right)x\left(t^{\prime}+s\right)\mbox{d}s
\label{eq:auto_xx}
\end{equation}
\end{widetext}
where the characteristic function, $\chi_{T}\left(t\right)$, is defined
so that $\chi_{T}(t)=1$ when $t\in\left[-T/2,T/2\right]$ and zero
otherwise. The FT of $\left\langle x\left(t\right)x\left(t^{\prime}\right)\right\rangle $
is taken as
\begin{equation}
\Phi_{xx}\left(\omega\right)=\int_{-\infty}^{+\infty}e^{-i\omega t}\left\langle x\left(t\right)x\left(0\right)\right\rangle\mbox{d}t 
\label{eq:auto_xx_FT}
\end{equation}
where we have further assumed that the dynamics reaches a stationary
(i.e., time-translation invariant) state. 

By substituting Eq. (\ref{eq:auto_xx}) into Eq. (\ref{eq:auto_xx_FT}),
one obtains a relation between the FT of the autocorrelation function
of $x\left(t\right)$ and the modulus square of the FT of $x\left(t\right)$:
\begin{multline}
\Phi_{xx}\left(\omega\right)=\lim_{T\to\infty}\frac{1}{T}\left|\int_{-\infty}^{+\infty}e^{-i\omega t}\chi_{T}\left(t\right)x\left(t\right)\mbox{d}t\right|^{2}\\
=\lim_{T\to\infty}\frac{1}{T}\left|\int_{-T/2}^{T/2}e^{-i\omega t}x\left(t\right)\mbox{d}t\right|^{2}
\label{eq:auto_xx_FT2}
\end{multline}
Finally, Eq. (\ref{eq:auto_xx_FT2}) tells us that, given any two
dynamic observables, say $x\left(t\right)$ and $y\left(t\right)$,
the following equation holds
\begin{equation}
\frac{\Phi_{xx}\left(\omega\right)}{\Phi_{yy}\left(\omega\right)}=\left|\frac{\int_{-\infty}^{+\infty}e^{-i\omega t}x\left(t\right)\mbox{d}t}{\int_{-\infty}^{+\infty}e^{-i\omega t}y\left(t\right)\mbox{d}t}\right|^{2}
\label{eq:auto_xx_FT3}
\end{equation}
whenever $x\left(t\right)$ and $y\left(t\right)$ are defined over
the same time interval. In practice, Eq. (\ref{eq:auto_xx_FT2}) is
also the starting point of a very efficient numerical algorithm to
compute an autocorrelation function by means of the fast Fourier transform
(FFT).
\cite{Allen}

\section{GLE not constrained by the fluctuation-dissipation theorem \label{sec:quantum}}

The mapping scheme proposed in Section \ref{sub:Mapping-of-GLE} was
based on an assumption that the (second) fluctuation-dissipation theorem,
Eq. (\ref{eq:Fluct-Diss-Theorem}), must hold, whereby the correlation
function of the colored noise is exactly proportional to the memory
kernel in the GLE (\ref{eq:gle}). We shall briefly state here a simple
generalization of the method which allows one going beyond this assumption. 

The equations given below establish a complex Langevin dynamics that
is equivalent to a GLE (\ref{eq:gle}) in which the correlation function
of the stochastic forces is a decaying function of the time difference,
$\left|t-t^{\prime}\right|$, but it is no longer required to be proportional
to the memory kernel. In this new scheme, Eqs. (\ref{eq:EoM-s1-final})
and (\ref{eq:EoM-s2-final}) for the auxiliary DoFs are modified as
follows:
\begin{widetext}
\begin{equation}
\dot{s}_{1}^{(k)}=-s_{1}^{(k)}/\tau_{k}+\omega_{k}s_{2}^{(k)}-\sum_{l\gamma}\sqrt{\bar{\mu}\mu_{l}}c_{l\gamma}^{(k)}\sum_{i\alpha}g_{ia,l\gamma}\left(\mathbf{r}(t)\right)\dot{r}_{i\alpha}(t)+\sqrt{\frac{2k_{B}T\bar{\mu}Q\left(\omega_{k}\right)}{\tau_{k}}}{\bf \xi}_{1}^{(k)}
\label{eq:EoM-s1-final_alt}
\end{equation}
\end{widetext}
and
\begin{equation}
\dot{s}_{2}^{(k)}=-s_{2}^{(k)}/\tau_{k}-\omega_{k}s_{1}^{(k)}+\sqrt{\frac{2k_{B}T\bar{\mu}Q\left(\omega_{k}\right)}{\tau_{k}}}{\bf \xi}_{2}^{(k)}
\label{eq:EoM-s2-final_alt}
\end{equation}
A calculation similar to that performed in Section \ref{sub:Mapping-of-GLE}
yields the same expression (\ref{eq:pol}) for the polarization matrix,,
while the correlation of the stochastic forces changes to: 
\begin{widetext}
\begin{equation*}
\left\langle \eta_{i\alpha}(t)\eta_{i^{\prime}\alpha^{\prime}}\left(t^{\prime}\right)\right\rangle =k_{B}T\sum_{l\gamma}\sum_{l^{\prime}\gamma^{\prime}}\sqrt{\mu_{l}\mu_{l^{\prime}}}g_{i\alpha,l\gamma}\left(\mathbf{r}(t)\right)\left[\sum_{k}Q\left(\omega_{k}\right)c_{l\gamma}^{(k)}c_{l^{\prime}\gamma^{\prime}}^{(k)}\phi_{k}\left(t-t^{\prime}\right)\right]g_{i^{\prime}\alpha^{\prime},l^{\prime}\gamma^{\prime}}\left(\mathbf{r}\left(t^{\prime}\right)\right)
\end{equation*}
\end{widetext}
By appropriately choosing the frequency weight function $Q(\omega)$,
one can simulate a GLE dynamics with a colored noise which is no
longer proportional to the memory kernel.
\bibliographystyle{apsrev4-1}

\begin{thebibliography}{10}%
\makeatletter
\providecommand \@ifxundefined [1]{%
 \ifx #1\undefined \expandafter \@firstoftwo
 \else \expandafter \@secondoftwo
\fi
}%
\providecommand \@ifnum [1]{%
 \ifnum #1\expandafter \@firstoftwo
 \else \expandafter \@secondoftwo
\fi
}%
\providecommand \enquote [1]{``#1''}%
\providecommand \bibnamefont  [1]{#1}%
\providecommand \bibfnamefont [1]{#1}%
\providecommand \citenamefont [1]{#1}%
\providecommand\href[0]{\@sanitize\@href}%
\providecommand\@href[1]{\endgroup\@@startlink{#1}\endgroup\@@href}%
\providecommand\@@href[1]{#1\@@endlink}%
\providecommand \@sanitize [0]{\begingroup\catcode`\&12\catcode`\#12\relax}%
\@ifxundefined \pdfoutput {\@firstoftwo}{%
 \@ifnum{\z@=\pdfoutput}{\@firstoftwo}{\@secondoftwo}%
}{%
 \providecommand\@@startlink[1]{\leavevmode}%
 \providecommand\@@endlink[0]{}%
}{%
 \providecommand\@@startlink[1]{%
  \leavevmode
  \pdfstartlink
   attr{/Border[0 0 1 ]/H/I/C[0 1 1]}%
   user{/Subtype/Link/A<</Type/Action/S/URI/URI(#1)>>}%
  \relax
 }%
 \providecommand\@@endlink[0]{\pdfendlink}%
}%
\providecommand \url  [0]{\begingroup\@sanitize \@url }%
\providecommand \@url [1]{\endgroup\@href {#1}{\urlprefix}}%
\providecommand \urlprefix [0]{URL }%
\providecommand \Eprint[0]{\href }%
\@ifxundefined \urlstyle {%
  \providecommand \doi [1]{doi:\discretionary{}{}{}#1}%
}{%
  \providecommand \doi [0]{doi:\discretionary{}{}{}\begingroup
  \urlstyle{rm}\Url }%
}%
\providecommand \doibase [0]{http://dx.doi.org/}%
\providecommand \Doi[1]{\href{\doibase#1}}%
\providecommand \bibAnnote [3]{%
  \BibitemShut{#1}%
  \begin{quotation}\noindent
    \textsc{Key:}\ #2\\\textsc{Annotation:}\ #3%
  \end{quotation}%
}%
\providecommand \bibAnnoteFile [2]{%
  \IfFileExists{#2}{\bibAnnote {#1} {#2} {\input{#2}}}{}%
}%
\providecommand \typeout [0]{\immediate \write \m@ne }%
\providecommand \selectlanguage [0]{\@gobble}%
\providecommand \bibinfo [0]{\@secondoftwo}%
\providecommand \bibfield [0]{\@secondoftwo}%
\providecommand \translation [1]{[#1]}%
\providecommand \BibitemOpen[0]{}%
\providecommand \bibitemStop [0]{}%
\providecommand \bibitemNoStop [0]{.\EOS\space}%
\providecommand \EOS [0]{\spacefactor3000\relax}%
\providecommand \BibitemShut [1]{\csname bibitem#1\endcsname}%
\bibitem{Segal2002}%
  \BibitemOpen
  \bibfield{author}{%
  \bibinfo {author} {\bibfnamefont{D.}~\bibnamefont{Segal}}\ and\ \bibinfo
  {author} {\bibfnamefont{A.}~\bibnamefont{Nitzan}},\ }%
  \bibfield{journal}{%
  \bibinfo {journal} {J. Chem. Phys.}\ }%
  \textbf{\bibinfo {volume} {117}},\ \bibinfo {pages} {3915} (\bibinfo {year}
  {2002})%
  \bibAnnoteFile{NoStop}{Segal2002}%
\bibitem{Dubi2011}%
  \BibitemOpen
  \bibfield{author}{%
  \bibinfo {author} {\bibfnamefont{Y.}~\bibnamefont{Dubi}}\ and\ \bibinfo
  {author} {\bibfnamefont{M.}~\bibnamefont{Di~Ventra}},\ }%
  \bibfield{journal}{%
  \bibinfo {journal} {Rev. Mod. Phys.}\ }%
  \textbf{\bibinfo {volume} {83}},\ \bibinfo {pages} {131} (\bibinfo {year}
  {2011})%
  \bibAnnoteFile{NoStop}{Dubi2011}%
\bibitem{Berber2000}%
  \BibitemOpen
  \bibfield{author}{%
  \bibinfo {author} {\bibfnamefont{S.}~\bibnamefont{Berber}}, \bibinfo {author}
  {\bibfnamefont{Y.-K.}\ \bibnamefont{Kwon}},\ and\ \bibinfo {author}
  {\bibfnamefont{D.}~\bibnamefont{Tom\'{a}nek}},\ }%
  \bibfield{journal}{%
  \bibinfo {journal} {Phys. Rev. Lett.}\ }%
  \textbf{\bibinfo {volume} {84}},\ \bibinfo {pages} {4613} (\bibinfo {year}
  {2000})%
  \bibAnnoteFile{NoStop}{Berber2000}%
\bibitem{Kim2001}%
  \BibitemOpen
  \bibfield{author}{%
  \bibinfo {author} {\bibfnamefont{P.}~\bibnamefont{Kim}}, \bibinfo {author}
  {\bibfnamefont{L.}~\bibnamefont{Shi}}, \bibinfo {author}
  {\bibfnamefont{A.}~\bibnamefont{Majumdar}},\ and\ \bibinfo {author}
  {\bibfnamefont{P.~L.}\ \bibnamefont{McEuen}},\ }%
  \bibfield{journal}{%
  \bibinfo {journal} {Phys. Rev. Lett.}\ }%
  \textbf{\bibinfo {volume} {87}},\ \bibinfo {pages} {215502} (\bibinfo {year}
  {2001})%
  \bibAnnoteFile{NoStop}{Kim2001}%
\bibitem{Shi2002}%
  \BibitemOpen
  \bibfield{author}{%
  \bibinfo {author} {\bibfnamefont{L.}~\bibnamefont{Shi}}\ and\ \bibinfo
  {author} {\bibfnamefont{A.}~\bibnamefont{Majumdar}},\ }%
  \bibfield{journal}{%
  \bibinfo {journal} {J. Heat Trans. - T. ASME}\ }%
  \textbf{\bibinfo {volume} {124}},\ \bibinfo {pages} {329} (\bibinfo {year}
  {2002})%
  \bibAnnoteFile{NoStop}{Shi2002}%
\bibitem{Padgett2004}%
  \BibitemOpen
  \bibfield{author}{%
  \bibinfo {author} {\bibfnamefont{C.~W.}\ \bibnamefont{Padgett}}\ and\
  \bibinfo {author} {\bibfnamefont{D.~W.}\ \bibnamefont{Brenner}},\ }%
  \bibfield{journal}{%
  \bibinfo {journal} {Nano Letters}\ }%
  \textbf{\bibinfo {volume} {4}},\ \bibinfo {pages} {1051} (\bibinfo {year}
  {2004})%
  \bibAnnoteFile{NoStop}{Padgett2004}%
\bibitem{Hu2008}%
  \BibitemOpen
  \bibfield{author}{%
  \bibinfo {author} {\bibfnamefont{M.}~\bibnamefont{Hu}}, \bibinfo {author}
  {\bibfnamefont{P.}~\bibnamefont{Keblinski}}, \bibinfo {author}
  {\bibfnamefont{J.-S.}\ \bibnamefont{Wang}},\ and\ \bibinfo {author}
  {\bibfnamefont{N.}~\bibnamefont{Raravikar}},\ }%
  \bibfield{journal}{%
  \bibinfo {journal} {Journal of Applied Physics}\ }%
  \textbf{\bibinfo {volume} {104}},\ \bibinfo {pages} {083503} (\bibinfo {year}
  {2008})%
  \bibAnnoteFile{NoStop}{Hu2008}%
\bibitem{Padgett2006}%
  \BibitemOpen
  \bibfield{author}{%
  \bibinfo {author} {\bibfnamefont{C.~W.}\ \bibnamefont{Padgett}}, \bibinfo
  {author} {\bibfnamefont{O.}~\bibnamefont{Shenderova}},\ and\ \bibinfo
  {author} {\bibfnamefont{D.~W.}\ \bibnamefont{Brenner}},\ }%
  \bibfield{journal}{%
  \bibinfo {journal} {Nano Lett.}\ }%
  \textbf{\bibinfo {volume} {6}},\ \bibinfo {pages} {1827} (\bibinfo {year}
  {2006})%
  \bibAnnoteFile{NoStop}{Padgett2006}%
\bibitem{Yang2008}%
  \BibitemOpen
  \bibfield{author}{%
  \bibinfo {author} {\bibfnamefont{N.}~\bibnamefont{Yang}}, \bibinfo {author}
  {\bibfnamefont{G.}~\bibnamefont{Zhang}},\ and\ \bibinfo {author}
  {\bibfnamefont{B.}~\bibnamefont{Li}},\ }%
  \bibfield{journal}{%
  \bibinfo {journal} {Nano Lett.}\ }%
  \textbf{\bibinfo {volume} {8}},\ \bibinfo {pages} {276} (\bibinfo {year}
  {2008})%
  \bibAnnoteFile{NoStop}{Yang2008}%
\bibitem{Estreicher2009}%
  \BibitemOpen
  \bibfield{author}{%
  \bibinfo {author} {\bibfnamefont{S.~K.}\ \bibnamefont{Estreicher}}\ and\
  \bibinfo {author} {\bibfnamefont{T.~M.}\ \bibnamefont{Gibbons}},\ }%
  \bibfield{journal}{%
  \bibinfo {journal} {Physica B}\ }%
  \textbf{\bibinfo {volume} {404}},\ \bibinfo {pages} {4509} (\bibinfo {year}
  {2009})%
  \bibAnnoteFile{NoStop}{Estreicher2009}%
\bibitem{Cahill2002}%
  \BibitemOpen
  \bibfield{author}{%
  \bibinfo {author} {\bibfnamefont{D.~G.}\ \bibnamefont{Cahill}}, \bibinfo
  {author} {\bibfnamefont{K.}~\bibnamefont{Goodson}},\ and\ \bibinfo {author}
  {\bibfnamefont{A.}~\bibnamefont{Majumdar}},\ }%
  \bibfield{journal}{%
  \bibinfo {journal} {J. Heat Trans. - T. ASME}\ }%
  \textbf{\bibinfo {volume} {124}},\ \bibinfo {pages} {223} (\bibinfo {year}
  {2002})%
  \bibAnnoteFile{NoStop}{Cahill2002}%
\bibitem{Pop2010}%
  \BibitemOpen
  \bibfield{author}{%
  \bibinfo {author} {\bibfnamefont{E.}~\bibnamefont{Pop}},\ }%
  \bibfield{journal}{%
  \bibinfo {journal} {Nano. Res.}\ }%
  \textbf{\bibinfo {volume} {3}},\ \bibinfo {pages} {147} (\bibinfo {year}
  {2010})%
  \bibAnnoteFile{NoStop}{Pop2010}%
\bibitem{Zebarjadi2012}%
  \BibitemOpen
  \bibfield{author}{%
  \bibinfo {author} {\bibfnamefont{M.}~\bibnamefont{Zebarjadi}}, \bibinfo
  {author} {\bibfnamefont{K.}~\bibnamefont{Esfarjani}}, \bibinfo {author}
  {\bibfnamefont{M.~S.}\ \bibnamefont{Dresselhaus}}, \bibinfo {author}
  {\bibfnamefont{Z.~F.}\ \bibnamefont{Ren}},\ and\ \bibinfo {author}
  {\bibfnamefont{G.}~\bibnamefont{Chen}},\ }%
  \bibfield{journal}{%
  \bibinfo {journal} {Energy Environ. Sci.}\ }%
  \textbf{\bibinfo {volume} {5}},\ \bibinfo {pages} {5147} (\bibinfo {year}
  {2012})%
  \bibAnnoteFile{NoStop}{Zebarjadi2012}%
\bibitem{West2006}%
  \BibitemOpen
  \bibfield{author}{%
  \bibinfo {author} {\bibfnamefont{D.}~\bibnamefont{West}}\ and\ \bibinfo
  {author} {\bibfnamefont{S.~K.}\ \bibnamefont{Estreicher}},\ }%
  \bibfield{journal}{%
  \bibinfo {journal} {Phys. Rev. Lett.}\ }%
  \textbf{\bibinfo {volume} {96}},\ \bibinfo {pages} {115504} (\bibinfo {year}
  {2006})%
  \bibAnnoteFile{NoStop}{West2006}%
\bibitem{Kermode2008}%
  \BibitemOpen
  \bibfield{author}{%
  \bibinfo {author} {\bibfnamefont{J.~R.}\ \bibnamefont{Kermode}}, \bibinfo
  {author} {\bibfnamefont{T.}~\bibnamefont{Albaret}}, \bibinfo {author}
  {\bibfnamefont{D.}~\bibnamefont{Sherman}}, \bibinfo {author}
  {\bibfnamefont{N.}~\bibnamefont{Bernstein}}, \bibinfo {author}
  {\bibfnamefont{P.}~\bibnamefont{Gumbsch}}, \bibinfo {author}
  {\bibfnamefont{M.~C.}\ \bibnamefont{Payne}}, \bibinfo {author}
  {\bibfnamefont{G.}~\bibnamefont{Cs\'{a}nyi}},\ and\ \bibinfo {author}
  {\bibfnamefont{A.}~\bibnamefont{De~Vita}},\ }%
  \bibfield{journal}{%
  \bibinfo {journal} {Nature}\ }%
  \textbf{\bibinfo {volume} {455}},\ \bibinfo {pages} {1224} (\bibinfo {year}
  {2008})%
  \bibAnnoteFile{NoStop}{Kermode2008}%
\bibitem{Backus2005}%
  \BibitemOpen
  \bibfield{author}{%
  \bibinfo {author} {\bibfnamefont{E.~H.~G.}\ \bibnamefont{Backus}}, \bibinfo
  {author} {\bibfnamefont{A.}~\bibnamefont{Eichler}}, \bibinfo {author}
  {\bibfnamefont{A.~W.}\ \bibnamefont{Kleyn}},\ and\ \bibinfo {author}
  {\bibfnamefont{M.}~\bibnamefont{Bonn}},\ }%
  \bibfield{journal}{%
  \bibinfo {journal} {Science}\ }%
  \textbf{\bibinfo {volume} {310}},\ \bibinfo {pages} {1790} (\bibinfo {year}
  {2005})%
  \bibAnnoteFile{NoStop}{Backus2005}%
\bibitem{Ueba2005}%
  \BibitemOpen
  \bibfield{author}{%
  \bibinfo {author} {\bibfnamefont{H.}~\bibnamefont{Ueba}}\ and\ \bibinfo
  {author} {\bibfnamefont{M.}~\bibnamefont{Wolf}},\ }%
  \bibfield{journal}{%
  \bibinfo {journal} {Science}\ }%
  \textbf{\bibinfo {volume} {310}},\ \bibinfo {pages} {1774} (\bibinfo {year}
  {2005})%
  \bibAnnoteFile{NoStop}{Ueba2005}%
\bibitem{Mak1987}%
  \BibitemOpen
  \bibfield{author}{%
  \bibinfo {author} {\bibfnamefont{C.~H.}\ \bibnamefont{Mak}}, \bibinfo
  {author} {\bibfnamefont{B.~G.}\ \bibnamefont{Koehler}}, \bibinfo {author}
  {\bibfnamefont{J.~L.}\ \bibnamefont{Brand}},\ and\ \bibinfo {author}
  {\bibfnamefont{S.~M.}\ \bibnamefont{George}},\ }%
  \bibfield{journal}{%
  \bibinfo {journal} {J. Chem. Phys.}\ }%
  \textbf{\bibinfo {volume} {87}},\ \bibinfo {pages} {2340} (\bibinfo {year}
  {1987})%
  \bibAnnoteFile{NoStop}{Mak1987}%
\bibitem{Chan2007}%
  \BibitemOpen
  \bibfield{author}{%
  \bibinfo {author} {\bibfnamefont{W.~L.}\ \bibnamefont{Chan}}\ and\ \bibinfo
  {author} {\bibfnamefont{E.}~\bibnamefont{Chason}},\ }%
  \bibfield{journal}{%
  \bibinfo {journal} {J. Appl. Phys.}\ }%
  \textbf{\bibinfo {volume} {101}},\ \bibinfo {pages} {121301} (\bibinfo {year}
  {2007})%
  \bibAnnoteFile{NoStop}{Chan2007}%
\bibitem{Toussaint2009}%
  \BibitemOpen
  \bibfield{author}{%
  \bibinfo {author} {\bibfnamefont{U.}~\bibnamefont{von Toussaint}}, \bibinfo
  {author} {\bibfnamefont{P.~N.}\ \bibnamefont{Maya}},\ and\ \bibinfo {author}
  {\bibfnamefont{C.}~\bibnamefont{Hopf}},\ }%
  \bibfield{journal}{%
  \bibinfo {journal} {J. Nucl. Mater.}\ }%
  \textbf{\bibinfo {volume} {386-388}},\ \bibinfo {pages} {353} (\bibinfo
  {year} {2009})%
  \bibAnnoteFile{NoStop}{Toussaint2009}%
\bibitem{Wucher2010}%
  \BibitemOpen
  \bibfield{author}{%
  \bibinfo {author} {\bibfnamefont{A.}~\bibnamefont{Wucher}}\ and\ \bibinfo
  {author} {\bibfnamefont{N.}~\bibnamefont{Winograd}},\ }%
  \bibfield{journal}{%
  \bibinfo {journal} {Anal. Bioanal. Chem.}\ }%
  \textbf{\bibinfo {volume} {396}},\ \bibinfo {pages} {105} (\bibinfo {year}
  {2010})%
  \bibAnnoteFile{NoStop}{Wucher2010}%
\bibitem{Szlufarska2008}%
  \BibitemOpen
  \bibfield{author}{%
  \bibinfo {author} {\bibfnamefont{I.}~\bibnamefont{Szlufarska}}, \bibinfo
  {author} {\bibfnamefont{M.}~\bibnamefont{Chandross}},\ and\ \bibinfo {author}
  {\bibfnamefont{R.~W.}\ \bibnamefont{Carpick}},\ }%
  \bibfield{journal}{%
  \bibinfo {journal} {J. Phys. D: Appl. Phys.}\ }%
  \textbf{\bibinfo {volume} {41}},\ \bibinfo {pages} {123001} (\bibinfo {year}
  {2008})%
  \bibAnnoteFile{NoStop}{Szlufarska2008}%
\bibitem{Benassi10}%
  \BibitemOpen
  \bibfield{author}{%
  \bibinfo {author} {\bibfnamefont{A.}~\bibnamefont{Benassi}}, \bibinfo
  {author} {\bibfnamefont{A.}~\bibnamefont{Vanossi}}, \bibinfo {author}
  {\bibfnamefont{G.~E.}\ \bibnamefont{Santoro}},\ and\ \bibinfo {author}
  {\bibfnamefont{E.}~\bibnamefont{Tosatti}},\ }%
  \bibfield{journal}{%
  \bibinfo {journal} {Phys. Rev. B}\ }%
  \textbf{\bibinfo {volume} {82}},\ \bibinfo {pages} {081401} (\bibinfo {year}
  {2010})%
  \bibAnnoteFile{NoStop}{Benassi10}%
\bibitem{Benassi12}%
  \BibitemOpen
  \bibfield{author}{%
  \bibinfo {author} {\bibfnamefont{A.}~\bibnamefont{Benassi}}, \bibinfo
  {author} {\bibfnamefont{A.}~\bibnamefont{Vanossi}}, \bibinfo {author}
  {\bibfnamefont{G.}~\bibnamefont{Santoro}},\ and\ \bibinfo {author}
  {\bibfnamefont{E.}~\bibnamefont{Tosatti}},\ }%
  \bibfield{journal}{%
  \bibinfo {journal} {Tribology Letters}\ }%
  \textbf{\bibinfo {volume} {48}},\ \bibinfo {pages} {41} (\bibinfo {year}
  {2012})%
  \bibAnnoteFile{NoStop}{Benassi12}%
\bibitem{Lohrasebi2011}%
  \BibitemOpen
  \bibfield{author}{%
  \bibinfo {author} {\bibfnamefont{A.}~\bibnamefont{Lohrasebi}}, \bibinfo
  {author} {\bibfnamefont{M.}~\bibnamefont{Neek-Amal}},\ and\ \bibinfo {author}
  {\bibfnamefont{M.~R.}\ \bibnamefont{Ejtehadi}},\ }%
  \bibfield{journal}{%
  \bibinfo {journal} {Physical Review E}\ }%
  \textbf{\bibinfo {volume} {83}},\ \bibinfo {pages} {042601} (\bibinfo {month}
  {Apr}\ \bibinfo {year} {2011})%
  \bibAnnoteFile{NoStop}{Lohrasebi2011}%
\bibitem{Toton2010}%
  \BibitemOpen
  \bibfield{author}{%
  \bibinfo {author} {\bibfnamefont{D.}~\bibnamefont{Toton}}, \bibinfo {author}
  {\bibfnamefont{C.~D.}\ \bibnamefont{Lorenz}}, \bibinfo {author}
  {\bibfnamefont{N.}~\bibnamefont{Rompotis}}, \bibinfo {author}
  {\bibfnamefont{N.}~\bibnamefont{Martsinovich}},\ and\ \bibinfo {author}
  {\bibfnamefont{L.}~\bibnamefont{Kantorovich}},\ }%
  \bibfield{journal}{%
  \bibinfo {journal} {J. Phys.: Condens. Matter}\ }%
  \textbf{\bibinfo {volume} {22}},\ \bibinfo {pages} {074205} (\bibinfo {year}
  {2010})%
  \bibAnnoteFile{NoStop}{Toton2010}%
\bibitem{Andersen1980}%
  \BibitemOpen
  \bibfield{author}{%
  \bibinfo {author} {\bibfnamefont{H.~C.}\ \bibnamefont{Andersen}},\ }%
  \bibfield{journal}{%
  \bibinfo {journal} {J. Chem. Phys.}\ }%
  \textbf{\bibinfo {volume} {72}},\ \bibinfo {pages} {2384} (\bibinfo {year}
  {1980})%
  \bibAnnoteFile{NoStop}{Andersen1980}%
\bibitem{Nose1984}%
  \BibitemOpen
  \bibfield{author}{%
  \bibinfo {author} {\bibfnamefont{S.}~\bibnamefont{Nos\'{e}}},\ }%
  \bibfield{journal}{%
  \bibinfo {journal} {Mol. Phys.}\ }%
  \textbf{\bibinfo {volume} {52}},\ \bibinfo {pages} {255} (\bibinfo {year}
  {1984})%
  \bibAnnoteFile{NoStop}{Nose1984}%
\bibitem{Nose1984a}%
  \BibitemOpen
  \bibfield{author}{%
  \bibinfo {author} {\bibfnamefont{S.}~\bibnamefont{Nos\'{e}}},\ }%
  \bibfield{journal}{%
  \bibinfo {journal} {J. Chem. Phys.}\ }%
  \textbf{\bibinfo {volume} {81}},\ \bibinfo {pages} {511} (\bibinfo {year}
  {1984})%
  \bibAnnoteFile{NoStop}{Nose1984a}%
\bibitem{Hoover1985}%
  \BibitemOpen
  \bibfield{author}{%
  \bibinfo {author} {\bibfnamefont{W.~G.}\ \bibnamefont{Hoover}},\ }%
  \bibfield{journal}{%
  \bibinfo {journal} {Phys. Rev. A}\ }%
  \textbf{\bibinfo {volume} {31}},\ \bibinfo {pages} {1695} (\bibinfo {year}
  {1985})%
  \bibAnnoteFile{NoStop}{Hoover1985}%
\bibitem{Schneider1978}%
  \BibitemOpen
  \bibfield{author}{%
  \bibinfo {author} {\bibfnamefont{T.}~\bibnamefont{Schneider}}\ and\ \bibinfo
  {author} {\bibfnamefont{E.}~\bibnamefont{Stoll}},\ }%
  \bibfield{journal}{%
  \bibinfo {journal} {Phys. Rev. B}\ }%
  \textbf{\bibinfo {volume} {17}},\ \bibinfo {pages} {1302} (\bibinfo {year}
  {1978})%
  \bibAnnoteFile{NoStop}{Schneider1978}%
\bibitem{Bussi2007}%
  \BibitemOpen
  \bibfield{author}{%
  \bibinfo {author} {\bibfnamefont{G.}~\bibnamefont{Bussi}}, \bibinfo {author}
  {\bibfnamefont{D.}~\bibnamefont{Donadio}},\ and\ \bibinfo {author}
  {\bibfnamefont{M.}~\bibnamefont{Parrinello}},\ }%
  \bibfield{journal}{%
  \bibinfo {journal} {J. Chem. Phys.}\ }%
  \textbf{\bibinfo {volume} {126}},\ \bibinfo {pages} {014101} (\bibinfo {year}
  {2007})%
  \bibAnnoteFile{NoStop}{Bussi2007}%
\bibitem{Szlufarska2007}%
  \BibitemOpen
  \bibfield{author}{%
  \bibinfo {author} {\bibfnamefont{I.}~\bibnamefont{Szlufarska}}, \bibinfo
  {author} {\bibfnamefont{R.~K.}\ \bibnamefont{Kalia}}, \bibinfo {author}
  {\bibfnamefont{A.}~\bibnamefont{Nakano}},\ and\ \bibinfo {author}
  {\bibfnamefont{P.}~\bibnamefont{Vashishta}},\ }%
  \bibfield{journal}{%
  \bibinfo {journal} {Journal of Applied Physics}\ }%
  \textbf{\bibinfo {volume} {102}},\ \bibinfo {pages} {023509} (\bibinfo {year}
  {2007})%
  \bibAnnoteFile{NoStop}{Szlufarska2007}%
\bibitem{Barry2009}%
  \BibitemOpen
  \bibfield{author}{%
  \bibinfo {author} {\bibfnamefont{P.~R.}\ \bibnamefont{Barry}}, \bibinfo
  {author} {\bibfnamefont{P.~Y.}\ \bibnamefont{Chiu}}, \bibinfo {author}
  {\bibfnamefont{S.~S.}\ \bibnamefont{Perry}}, \bibinfo {author}
  {\bibfnamefont{W.~G.}\ \bibnamefont{Sawyer}}, \bibinfo {author}
  {\bibfnamefont{S.~R.}\ \bibnamefont{Phillpot}},\ and\ \bibinfo {author}
  {\bibfnamefont{S.~B.}\ \bibnamefont{Sinnott}},\ }%
  \bibfield{journal}{%
  \bibinfo {journal} {J.Phys.: Condens. Matter}\ }%
  \textbf{\bibinfo {volume} {21}},\ \bibinfo {pages} {144201} (\bibinfo {year}
  {2009})%
  \bibAnnoteFile{NoStop}{Barry2009}%
\bibitem{Trevethan2004}%
  \BibitemOpen
  \bibfield{author}{%
  \bibinfo {author} {\bibfnamefont{T.}~\bibnamefont{Trevethan}}\ and\ \bibinfo
  {author} {\bibfnamefont{L.}~\bibnamefont{Kantorovich}},\ }%
  \bibfield{journal}{%
  \bibinfo {journal} {Physical Review B}\ }%
  \textbf{\bibinfo {volume} {70}},\ \bibinfo {pages} {115411} (\bibinfo {year}
  {2004})%
  \bibAnnoteFile{NoStop}{Trevethan2004}%
\bibitem{Mazyar2006}%
  \BibitemOpen
  \bibfield{author}{%
  \bibinfo {author} {\bibfnamefont{O.~A.}\ \bibnamefont{Mazyar}}\ and\ \bibinfo
  {author} {\bibfnamefont{W.~L.}\ \bibnamefont{Hase}},\ }%
  \bibfield{journal}{%
  \bibinfo {journal} {The Journal of Physical Chemistry A}\ }%
  \textbf{\bibinfo {volume} {110}},\ \bibinfo {pages} {526} (\bibinfo {year}
  {2006})%
  \bibAnnoteFile{NoStop}{Mazyar2006}%
\bibitem{Hu2009}%
  \BibitemOpen
  \bibfield{author}{%
  \bibinfo {author} {\bibfnamefont{J.}~\bibnamefont{Hu}}, \bibinfo {author}
  {\bibfnamefont{X.}~\bibnamefont{Ruan}},\ and\ \bibinfo {author}
  {\bibfnamefont{Y.~P.}\ \bibnamefont{Chen}},\ }%
  \bibfield{journal}{%
  \bibinfo {journal} {Nano Letters}\ }%
  \textbf{\bibinfo {volume} {9}},\ \bibinfo {pages} {2730} (\bibinfo {year}
  {2009})%
  \bibAnnoteFile{NoStop}{Hu2009}%
\bibitem{Guo2010}%
  \BibitemOpen
  \bibfield{author}{%
  \bibinfo {author} {\bibfnamefont{J.}~\bibnamefont{Guo}}, \bibinfo {author}
  {\bibfnamefont{B.}~\bibnamefont{Wen}}, \bibinfo {author}
  {\bibfnamefont{R.}~\bibnamefont{Melnik}}, \bibinfo {author}
  {\bibfnamefont{S.}~\bibnamefont{Yao}},\ and\ \bibinfo {author}
  {\bibfnamefont{T.}~\bibnamefont{Li}},\ }%
  \bibfield{journal}{%
  \bibinfo {journal} {Physica E: Low-dimensional Systems and Nanostructures}\
  }%
  \textbf{\bibinfo {volume} {43}},\ \bibinfo {pages} {155} (\bibinfo {year}
  {2010})%
  \bibAnnoteFile{NoStop}{Guo2010}%
\bibitem{Hu2011}%
  \BibitemOpen
  \bibfield{author}{%
  \bibinfo {author} {\bibfnamefont{L.}~\bibnamefont{Hu}}, \bibinfo {author}
  {\bibfnamefont{T.}~\bibnamefont{Desai}},\ and\ \bibinfo {author}
  {\bibfnamefont{P.}~\bibnamefont{Keblinski}},\ }%
  \bibfield{journal}{%
  \bibinfo {journal} {Physical Review B}\ }%
  \textbf{\bibinfo {volume} {83}},\ \bibinfo {pages} {195423} (\bibinfo {year}
  {2011})%
  \bibAnnoteFile{NoStop}{Hu2011}%
\bibitem{Manikandan2011}%
  \BibitemOpen
  \bibfield{author}{%
  \bibinfo {author} {\bibfnamefont{P.}~\bibnamefont{Manikandan}}, \bibinfo
  {author} {\bibfnamefont{J.~A.}\ \bibnamefont{Carter}}, \bibinfo {author}
  {\bibfnamefont{D.~D.}\ \bibnamefont{Dlott}},\ and\ \bibinfo {author}
  {\bibfnamefont{W.~L.}\ \bibnamefont{Hase}},\ }%
  \bibfield{journal}{%
  \bibinfo {journal} {The Journal of Physical Chemistry C}\ }%
  \textbf{\bibinfo {volume} {115}},\ \bibinfo {pages} {9622} (\bibinfo {year}
  {2011})%
  \bibAnnoteFile{NoStop}{Manikandan2011}%
\bibitem{Hsu2007}%
  \BibitemOpen
  \bibfield{author}{%
  \bibinfo {author} {\bibfnamefont{W.-D.}\ \bibnamefont{Hsu}}, \bibinfo
  {author} {\bibfnamefont{S.}~\bibnamefont{Tepavcevic}}, \bibinfo {author}
  {\bibfnamefont{L.}~\bibnamefont{Hanley}},\ and\ \bibinfo {author}
  {\bibfnamefont{S.}~\bibnamefont{Sinnott}},\ }%
  \bibfield{journal}{%
  \bibinfo {journal} {Journal of Physical Chemistry C}\ }%
  \textbf{\bibinfo {volume} {111}},\ \bibinfo {pages} {4199} (\bibinfo {year}
  {2007})%
  \bibAnnoteFile{NoStop}{Hsu2007}%
\bibitem{Berendsen1984}%
  \BibitemOpen
  \bibfield{author}{%
  \bibinfo {author} {\bibfnamefont{H.~J.~C.}\ \bibnamefont{Berendsen}},
  \bibinfo {author} {\bibfnamefont{J.~P.~M.}\ \bibnamefont{Postma}}, \bibinfo
  {author} {\bibfnamefont{W.~F.}\ \bibnamefont{van Gunsteren}}, \bibinfo
  {author} {\bibfnamefont{A.}~\bibnamefont{DiNola}},\ and\ \bibinfo {author}
  {\bibfnamefont{J.~R.}\ \bibnamefont{Haak}},\ }%
  \bibfield{journal}{%
  \bibinfo {journal} {The Journal of Chemical Physics}\ }%
  \textbf{\bibinfo {volume} {81}},\ \bibinfo {pages} {3684} (\bibinfo {year}
  {1984})%
  \bibAnnoteFile{NoStop}{Berendsen1984}%
\bibitem{Thompson1990}%
  \BibitemOpen
  \bibfield{author}{%
  \bibinfo {author} {\bibfnamefont{P.~A.}\ \bibnamefont{Thompson}}\ and\
  \bibinfo {author} {\bibfnamefont{M.~O.}\ \bibnamefont{Robbins}},\ }%
  \bibfield{journal}{%
  \bibinfo {journal} {Phys.}\ }%
  \textbf{\bibinfo {volume} {41}},\ \bibinfo {pages} {6830} (\bibinfo {year}
  {1990})%
  \bibAnnoteFile{NoStop}{Thompson1990}%
\bibitem{Lorenz2010}%
  \BibitemOpen
  \bibfield{author}{%
  \bibinfo {author} {\bibfnamefont{C.~D.}\ \bibnamefont{Lorenz}}, \bibinfo
  {author} {\bibfnamefont{M.}~\bibnamefont{Chandross}},\ and\ \bibinfo {author}
  {\bibfnamefont{G.~S.}\ \bibnamefont{Grest}},\ }%
  \bibfield{journal}{%
  \bibinfo {journal} {Journal of Adhesion Science and Technology}\ }%
  \textbf{\bibinfo {volume} {24}},\ \bibinfo {pages} {2453} (\bibinfo {year}
  {2010})%
  \bibAnnoteFile{NoStop}{Lorenz2010}%
\bibitem{Zwanzig2001}%
  \BibitemOpen
  \bibfield{author}{%
  \bibinfo {author} {\bibfnamefont{R.}~\bibnamefont{Zwanzig}},\ }%
  \emph{\bibinfo {title} {Nonequilibrium Statistical Mechanics}}\ (\bibinfo
  {publisher} {Oxford University Press},\ \bibinfo {year} {2001})%
  \bibAnnoteFile{NoStop}{Zwanzig2001}%
\bibitem{Kantorovich2008a}%
  \BibitemOpen
  \bibfield{author}{%
  \bibinfo {author} {\bibfnamefont{L.}~\bibnamefont{Kantorovich}},\ }%
  \bibfield{journal}{%
  \bibinfo {journal} {Physical Review B}\ }%
  \textbf{\bibinfo {volume} {78}},\ \bibinfo {pages} {094304} (\bibinfo {year}
  {2008})%
  \bibAnnoteFile{NoStop}{Kantorovich2008a}%
\bibitem{Ceriotti2009a}%
  \BibitemOpen
  \bibfield{author}{%
  \bibinfo {author} {\bibfnamefont{M.}~\bibnamefont{Ceriotti}}, \bibinfo
  {author} {\bibfnamefont{G.}~\bibnamefont{Bussi}},\ and\ \bibinfo {author}
  {\bibfnamefont{M.}~\bibnamefont{Parrinello}},\ }%
  \bibfield{journal}{%
  \bibinfo {journal} {Physical Review Letters}\ }%
  \textbf{\bibinfo {volume} {102}},\ \bibinfo {pages} {020601} (\bibinfo {year}
  {2009})%
  \bibAnnoteFile{NoStop}{Ceriotti2009a}%
\bibitem{Ceriotti2011}%
  \BibitemOpen
  \bibfield{author}{%
  \bibinfo {author} {\bibfnamefont{M.}~\bibnamefont{Ceriotti}}, \bibinfo
  {author} {\bibfnamefont{D.~E.}\ \bibnamefont{Manolopoulos}},\ and\ \bibinfo
  {author} {\bibfnamefont{M.}~\bibnamefont{Parrinello}},\ }%
  \bibfield{journal}{%
  \bibinfo {journal} {The Journal of Chemical Physics}\ }%
  \textbf{\bibinfo {volume} {134}},\ \bibinfo {pages} {084104} (\bibinfo {year}
  {2011})%
  \bibAnnoteFile{NoStop}{Ceriotti2011}%
\bibitem{Morrone2011}%
  \BibitemOpen
  \bibfield{author}{%
  \bibinfo {author} {\bibfnamefont{J.~A.}\ \bibnamefont{Morrone}}, \bibinfo
  {author} {\bibfnamefont{T.~E.}\ \bibnamefont{Markland}}, \bibinfo {author}
  {\bibfnamefont{M.}~\bibnamefont{Ceriotti}},\ and\ \bibinfo {author}
  {\bibfnamefont{B.~J.}\ \bibnamefont{Berne}},\ }%
  \bibfield{journal}{%
  \bibinfo {journal} {The Journal of Chemical Physics}\ }%
  \textbf{\bibinfo {volume} {134}},\ \bibinfo {pages} {014103} (\bibinfo {year}
  {2011})%
  \bibAnnoteFile{NoStop}{Morrone2011}%
\bibitem{Dammak2009}%
  \BibitemOpen
  \bibfield{author}{%
  \bibinfo {author} {\bibfnamefont{H.}~\bibnamefont{Dammak}}, \bibinfo {author}
  {\bibfnamefont{Y.}~\bibnamefont{Chalopin}}, \bibinfo {author}
  {\bibfnamefont{M.}~\bibnamefont{Laroche}}, \bibinfo {author}
  {\bibfnamefont{M.}~\bibnamefont{Hayoun}},\ and\ \bibinfo {author}
  {\bibfnamefont{J.-J.}\ \bibnamefont{Greffet}},\ }%
  \bibfield{journal}{%
  \bibinfo {journal} {Physical Review Letters}\ }%
  \textbf{\bibinfo {volume} {103}},\ \bibinfo {pages} {190601} (\bibinfo {year}
  {2009})%
  \bibAnnoteFile{NoStop}{Dammak2009}%
\bibitem{Barrat2011}%
  \BibitemOpen
  \bibfield{author}{%
  \bibinfo {author} {\bibfnamefont{J.-L.}\ \bibnamefont{Barrat}}\ and\ \bibinfo
  {author} {\bibfnamefont{D.}~\bibnamefont{Rodney}},\ }%
  \bibfield{journal}{%
  \bibinfo {journal} {Journal of Statistical Physics}\ }%
  \textbf{\bibinfo {volume} {144}},\ \bibinfo {pages} {679} (\bibinfo {month}
  {Aug}\ \bibinfo {year} {2011})%
  \bibAnnoteFile{NoStop}{Barrat2011}%
\bibitem{Lu12}%
  \BibitemOpen
  \bibfield{author}{%
  \bibinfo {author} {\bibfnamefont{J.-T.}\ \bibnamefont{L\"u}}, \bibinfo
  {author} {\bibfnamefont{M.}~\bibnamefont{Brandbyge}}, \bibinfo {author}
  {\bibfnamefont{P.}~\bibnamefont{Hedeg\aa{}rd}}, \bibinfo {author}
  {\bibfnamefont{T.~N.}\ \bibnamefont{Todorov}},\ and\ \bibinfo {author}
  {\bibfnamefont{D.}~\bibnamefont{Dundas}},\ }%
  \bibfield{journal}{%
  \bibinfo {journal} {Phys. Rev. B}\ }%
  \textbf{\bibinfo {volume} {85}},\ \bibinfo {pages} {245444} (\bibinfo {year}
  {2012})%
  \bibAnnoteFile{NoStop}{Lu12}%
\bibitem{Biele13}%
  \BibitemOpen
  \bibfield{author}{%
  \bibinfo {author} {\bibfnamefont{R.}~\bibnamefont{Biele}}, \bibinfo {author}
  {\bibfnamefont{C.}~\bibnamefont{Tim}},\ and\ \bibinfo {author}
  {\bibfnamefont{R.}~\bibnamefont{D'Agosta}},\ }%
  \enquote{\bibinfo {title} {{Time-convolutionless stochastic Schr\"{o}dinger
  equation for open quantum systems: application to thermal transport and
  relaxation}},}\  (\bibinfo {year} {2013}),\
  \Eprint{http://arxiv.org/abs/1203.3785}{arXiv:1203.3785}%
  \bibAnnoteFile{NoStop}{Biele13}%
\bibitem{Baczewski13}%
  \BibitemOpen
  \bibfield{author}{%
  \bibinfo {author} {\bibfnamefont{A.~D.}\ \bibnamefont{Baczewski}}\ and\
  \bibinfo {author} {\bibfnamefont{S.~D.}\ \bibnamefont{Bond}},\ }%
  \bibfield{journal}{%
  \bibinfo {journal} {The Journal of Chemical Physics}\ }%
  \textbf{\bibinfo {volume} {139}},\ \bibinfo {eid} {044107} (\bibinfo {year}
  {2013})%
  \bibAnnoteFile{NoStop}{Baczewski13}%
\bibitem{Luczka05}%
  \BibitemOpen
  \bibfield{author}{%
  \bibinfo {author} {\bibfnamefont{J.}~\bibnamefont{Luczka}},\ }%
  \bibfield{journal}{%
  \bibinfo {journal} {Chaos: An Interdisciplinary Journal of Nonlinear
  Science}\ }%
  \textbf{\bibinfo {volume} {15}},\ \bibinfo {eid} {026107} (\bibinfo {year}
  {2005})%
  \bibAnnoteFile{NoStop}{Luczka05}%
\bibitem{Rice44}%
  \BibitemOpen
  \bibfield{author}{%
  \bibinfo {author} {\bibfnamefont{O.}~\bibnamefont{Rice}},\ }%
  \bibfield{journal}{%
  \bibinfo {journal} {Bell Systems Tech. J.}\ }%
  \textbf{\bibinfo {volume} {23}},\ \bibinfo {pages} {282} (\bibinfo {year}
  {1944})%
  \bibAnnoteFile{NoStop}{Rice44}%
\bibitem{Billah90}%
  \BibitemOpen
  \bibfield{author}{%
  \bibinfo {author} {\bibfnamefont{K.~Y.~R.}\ \bibnamefont{Billah}}\ and\
  \bibinfo {author} {\bibfnamefont{M.}~\bibnamefont{Shinozuka}},\ }%
  \bibfield{journal}{%
  \bibinfo {journal} {Phys. Rev. A}\ }%
  \textbf{\bibinfo {volume} {42}},\ \bibinfo {pages} {7492} (\bibinfo {year}
  {1990})%
  \bibAnnoteFile{NoStop}{Billah90}%
\bibitem{Mannella92}%
  \BibitemOpen
  \bibfield{author}{%
  \bibinfo {author} {\bibfnamefont{R.}~\bibnamefont{Mannella}}\ and\ \bibinfo
  {author} {\bibfnamefont{V.}~\bibnamefont{Palleschi}},\ }%
  \bibfield{journal}{%
  \bibinfo {journal} {Phys. Rev. A}\ }%
  \textbf{\bibinfo {volume} {46}},\ \bibinfo {pages} {8028} (\bibinfo {year}
  {1992})%
  \bibAnnoteFile{NoStop}{Mannella92}%
\bibitem{Allen}%
  \BibitemOpen
  \bibfield{author}{%
  \bibinfo {author} {\bibfnamefont{M.}~\bibnamefont{Allen}}\ and\ \bibinfo
  {author} {\bibfnamefont{D.}~\bibnamefont{Tildesley}},\ }%
  \emph{\bibinfo {title} {{Computer Simulation of Liquids}}},\ \bibinfo
  {edition} {{New}}\ ed.\ (\bibinfo {publisher} {{Clarendon Press, Oxford}},\
  \bibinfo {year} {1989})%
  \bibAnnoteFile{NoStop}{Allen}%
\bibitem{Ceriotti2009b}%
  \BibitemOpen
  \bibfield{author}{%
  \bibinfo {author} {\bibfnamefont{M.}~\bibnamefont{Ceriotti}}, \bibinfo
  {author} {\bibfnamefont{G.}~\bibnamefont{Bussi}},\ and\ \bibinfo {author}
  {\bibfnamefont{M.}~\bibnamefont{Parrinello}},\ }%
  \bibfield{journal}{%
  \bibinfo {journal} {Phys. Rev. Lett.}\ }%
  \textbf{\bibinfo {volume} {103}},\ \bibinfo {pages} {030603} (\bibinfo {year}
  {2009})%
  \bibAnnoteFile{NoStop}{Ceriotti2009b}%
\bibitem{Ceriotti2010b}%
  \BibitemOpen
  \bibfield{author}{%
  \bibinfo {author} {\bibfnamefont{M.}~\bibnamefont{Ceriotti}}\ and\ \bibinfo
  {author} {\bibfnamefont{M.}~\bibnamefont{Parrinello}},\ }%
  \bibfield{journal}{%
  \bibinfo {journal} {Procedia Computer Science}\ }%
  \textbf{\bibinfo {volume} {1}},\ \bibinfo {pages} {1607} (\bibinfo {year}
  {2010})%
  \bibAnnoteFile{NoStop}{Ceriotti2010b}%
\bibitem{Gillespie96a}%
  \BibitemOpen
  \bibfield{author}{%
  \bibinfo {author} {\bibfnamefont{D.~T.}\ \bibnamefont{Gillespie}},\ }%
  \bibfield{journal}{%
  \bibinfo {journal} {American Journal of Physics}\ }%
  \textbf{\bibinfo {volume} {64}},\ \bibinfo {pages} {225} (\bibinfo {year}
  {1996})%
  \bibAnnoteFile{NoStop}{Gillespie96a}%
\bibitem{Gillespie96b}%
  \BibitemOpen
  \bibfield{author}{%
  \bibinfo {author} {\bibfnamefont{D.~T.}\ \bibnamefont{Gillespie}},\ }%
  \bibfield{journal}{%
  \bibinfo {journal} {American Journal of Physics}\ }%
  \textbf{\bibinfo {volume} {64}},\ \bibinfo {pages} {1246} (\bibinfo {year}
  {1996})%
  \bibAnnoteFile{NoStop}{Gillespie96b}%
\bibitem{Ceriotti2010a}%
  \BibitemOpen
  \bibfield{author}{%
  \bibinfo {author} {\bibfnamefont{M.}~\bibnamefont{Ceriotti}}, \bibinfo
  {author} {\bibfnamefont{G.}~\bibnamefont{Bussi}},\ and\ \bibinfo {author}
  {\bibfnamefont{M.}~\bibnamefont{Parrinello}},\ }%
  \bibfield{journal}{%
  \bibinfo {journal} {Journal of Chemical Theory and Computation}\ }%
  \textbf{\bibinfo {volume} {6}},\ \bibinfo {pages} {1170} (\bibinfo {year}
  {2010})%
  \bibAnnoteFile{NoStop}{Ceriotti2010a}%
\bibitem{Tuckerman90}%
  \BibitemOpen
  \bibfield{author}{%
  \bibinfo {author} {\bibfnamefont{M.}~\bibnamefont{Tuckerman}}, \bibinfo
  {author} {\bibfnamefont{B.~J.}\ \bibnamefont{Berne}},\ and\ \bibinfo {author}
  {\bibfnamefont{G.~J.}\ \bibnamefont{Martyna}},\ }%
  \bibfield{journal}{%
  \bibinfo {journal} {The Journal of Chemical Physics}\ }%
  \textbf{\bibinfo {volume} {97}},\ \bibinfo {pages} {1990} (\bibinfo {year}
  {1992})%
  \bibAnnoteFile{NoStop}{Tuckerman90}%
\bibitem{Donnelly05}%
  \BibitemOpen
  \bibfield{author}{%
  \bibinfo {author} {\bibfnamefont{D.}~\bibnamefont{Donnelly}}\ and\ \bibinfo
  {author} {\bibfnamefont{E.}~\bibnamefont{Rogers}},\ }%
  \bibfield{journal}{%
  \bibinfo {journal} {American Journal of Physics}\ }%
  \textbf{\bibinfo {volume} {73}},\ \bibinfo {pages} {938} (\bibinfo {year}
  {2005})%
  \bibAnnoteFile{NoStop}{Donnelly05}%
\bibitem{Bussi07}%
  \BibitemOpen
  \bibfield{author}{%
  \bibinfo {author} {\bibfnamefont{G.}~\bibnamefont{Bussi}}\ and\ \bibinfo
  {author} {\bibfnamefont{M.}~\bibnamefont{Parrinello}},\ }%
  \bibfield{journal}{%
  \bibinfo {journal} {Phys. Rev. E}\ }%
  \textbf{\bibinfo {volume} {75}},\ \bibinfo {pages} {056707} (\bibinfo {year}
  {2007})%
  \bibAnnoteFile{NoStop}{Bussi07}%
\bibitem{Ermak80}%
  \BibitemOpen
  \bibfield{author}{%
  \bibinfo {author} {\bibfnamefont{D.~L.}\ \bibnamefont{Ermak}}\ and\ \bibinfo
  {author} {\bibfnamefont{H.}~\bibnamefont{Buckholz}},\ }%
  \bibfield{journal}{%
  \bibinfo {journal} {Journal of Computational Physics}\ }%
  \textbf{\bibinfo {volume} {35}},\ \bibinfo {pages} {169 } (\bibinfo {year}
  {1980})%
  \bibAnnoteFile{NoStop}{Ermak80}%
\bibitem{Leimkuhler13a}%
  \BibitemOpen
  \bibfield{author}{%
  \bibinfo {author} {\bibfnamefont{B.}~\bibnamefont{Leimkuhler}}\ and\ \bibinfo
  {author} {\bibfnamefont{C.}~\bibnamefont{Matthews}},\ }%
  \bibfield{journal}{%
  \bibinfo {journal} {Applied Mathematics Research eXpress}\ }%
  \textbf{\bibinfo {volume} {2013}},\ \bibinfo {pages} {34} (\bibinfo {year}
  {2013})%
  \bibAnnoteFile{NoStop}{Leimkuhler13a}%
\bibitem{Leimkuhler13b}%
  \BibitemOpen
  \bibfield{author}{%
  \bibinfo {author} {\bibfnamefont{B.}~\bibnamefont{Leimkuhler}}\ and\ \bibinfo
  {author} {\bibfnamefont{C.}~\bibnamefont{Matthews}},\ }%
  \bibfield{journal}{%
  \bibinfo {journal} {The Journal of Chemical Physics}\ }%
  \textbf{\bibinfo {volume} {138}},\ \bibinfo {eid} {174102} (\bibinfo {year}
  {2013})%
  \bibAnnoteFile{NoStop}{Leimkuhler13b}%
\bibitem{Evstigneev10}%
  \BibitemOpen
  \bibfield{author}{%
  \bibinfo {author} {\bibfnamefont{M.}~\bibnamefont{Evstigneev}}\ and\ \bibinfo
  {author} {\bibfnamefont{P.}~\bibnamefont{Reimann}},\ }%
  \bibfield{journal}{%
  \bibinfo {journal} {Phys. Rev. B}\ }%
  \textbf{\bibinfo {volume} {82}},\ \bibinfo {pages} {224303} (\bibinfo {year}
  {2010})%
  \bibAnnoteFile{NoStop}{Evstigneev10}%
\bibitem{Petruccione}%
  \BibitemOpen
  \bibfield{author}{%
  \bibinfo {author} {\bibfnamefont{H.-P.}\ \bibnamefont{Breuer}}\ and\ \bibinfo
  {author} {\bibfnamefont{F.}~\bibnamefont{Petruccione}},\ }%
  \emph{\bibinfo {title} {{The Theory of Open Quantum Systems}}},\ \bibinfo
  {edition} {{New}}\ ed.\ (\bibinfo {publisher} {{Oxford University Press,
  Oxford}},\ \bibinfo {year} {2002})%
  \bibAnnoteFile{NoStop}{Petruccione}%
\bibitem{Risken}%
  \BibitemOpen
  \bibfield{author}{%
  \bibinfo {author} {\bibfnamefont{H.}~\bibnamefont{Risken}},\ }%
  \emph{\bibinfo {title} {{The Fokker-Planck Equation: Methods of Solutions and
  Applications}}},\ \bibinfo {edition} {{Second}}\ ed.\ (\bibinfo {publisher}
  {{Springer, Berlin}},\ \bibinfo {year} {1996})%
  \bibAnnoteFile{NoStop}{Risken}%
\bibitem{Arnold}%
  \BibitemOpen
  \bibfield{author}{%
  \bibinfo {author} {\bibfnamefont{V.}~\bibnamefont{Arnold}},\ }%
  \emph{\bibinfo {title} {{Ordinary Differential Equations}}},\ \bibinfo
  {edition} {{3rd}}\ ed.\ (\bibinfo {publisher} {{Springer, Berlin}},\ \bibinfo
  {year} {2006})%
  \bibAnnoteFile{NoStop}{Arnold}%
\end{thebibliography}
\end{document}